\documentclass[3p]{elsarticle}
\journal{Computational Physics}

\usepackage{xr}

\usepackage{wrapfig}
\usepackage{amssymb}
\usepackage{amsmath}
\usepackage{tikz}
\usepackage{pgfplots}

\usepackage[colorlinks,linkcolor=blue,urlcolor=blue,citecolor=blue]{hyperref}
\usepackage{todonotes}
\usepackage{listings}
\usepackage[capitalise]{cleveref}

\usepackage{placeins}
\usepackage{comment}
\usepackage{booktabs}
\usepackage{caption}
\usepackage{soul}
\usepackage[nosuper,acronym]{glossaries}
\newacronym{UQ}{UQ}{Uncertainty Quantification}
\newacronym{MC}{MC}{Monte Carlo}
\newacronym{MCMC}{MCMC}{Markov chain Monte Carlo}
\newacronym{MLMCMC}{MLMCMC}{Multilevel Markov chain Monte Carlo}
\newacronym{MLMC}{MLMC}{Multilevel Monte Carlo}
\newacronym{HPC}{HPC}{High Performance Computing}
\newacronym{DUNE}{DUNE}{Distributed and Unified Numerics Environment}
\newacronym{FE}{FE}{finite element}
\newacronym{PDE}{PDE}{partial differential equation}
\newacronym{MUQ}{MUQ}{MIT UQ Library}
\newacronym{UMBridge}{UM-Bridge}{UQ and Modeling Bridge}
\newacronym{HTTP}{HTTP}{Hypertext Transfer Protocol}
\newacronym{MPI}{MPI}{Message Passing Interface}
\newacronym{MAP}{MAP}{maximum a posteriori probability}
\newacronym{pdf}{PDF}{probability density function}
\newacronym{sgmk}{SGMK}{Sparse Grids Matlab Kit}
\newacronym{GKE}{GKE}{Google Kubernetes Engine}
\newacronym{SMT}{SMT}{Simultaneous Multithreading}
\newacronym{QMC}{QMC}{Quasi-Monte Carlo}
\newacronym{GCP}{GCP}{Google Cloud Platform}
\newacronym{GP}{GP}{Gaussian Process}
\newacronym{HI}{HI}{Hydrogen Isotopes}
\newacronym{KL}{KL}{Karhunen–Loève}

\glsdisablehyper

\usepackage{xtab}

\usepackage{pifont}
\newcommand{\cmark}{\ding{51}}%
\newcommand{\xmark}{\ding{55}}%

\newcommand{\Frou}{\operatorname{\mathit{F\kern-.15em r}}} % http://de.wikipedia.org/wiki/Froude-Zahl
\newcommand{\Dr}{\operatorname{\mathit{T}}} 
\newcommand{\thetashape}{s}

\makeatletter

\newcommand*{\addFileDependency}[1]{% argument=file name and extension
\typeout{(#1)}% latexmk will find this if $recorder=0
\@addtofilelist{#1}
\IfFileExists{#1}{}{\typeout{No file #1.}}
}\makeatother

\begin{document}

\begin{frontmatter}
\title{Democratizing Uncertainty Quantification}

\author[KIT]{Linus Seelinger\corref{cor1}}
\ead{linus.seelinger@kit.edu}
\cortext[cor1]{Corresponding author}

\author[DU]{Anne Reinarz}
%\ead{anne.k.reinarz@durham.ac.uk} %Second address breaks formatting...

\author[DL]{Mikkel B. Lykkegaard}
\author[UKAEA]{Robert Akers}
\author[DTU]{Amal M. A. Alghamdi}
\author[CSUa]{David Aristoff}
\author[CSUb]{Wolfgang Bangerth}
\author[UB]{Jean Bénézech}
\author[INM]{Matteo Diez}
\author[BMGF]{Kurt Frey}
\author[SNL]{John D. Jakeman}
\author[DTU]{Jakob S. Jørgensen}
\author[UCM]{Ki-Tae Kim}
\author[IMATI]{Benjamin M. Kent}
\author[IMATI]{Massimiliano Martinelli}
\author[SOL]{Matthew Parno}
\author[INM]{Riccardo Pellegrini}
\author[UCM]{Noemi Petra}
\author[CI,DTU]{Nicolai A. B. Riis}
\author[BMGF]{Katherine Rosenfeld}
\author[INM]{Andrea Serani}
\author[IMATI]{Lorenzo Tamellini}
\author[UTA]{Umberto Villa}
\author[DL]{Tim J. Dodwell}
\author[HU]{Robert Scheichl}

%\cortext[author] {Corresponding author.\\\textit{E-mail address:} linus.seelinger@iwr.uni-heidelberg.de}
\affiliation[KIT]{organization={Scientific Computing Center, Karlsruhe Institute of Technology}, city={Karlsruhe}, country={Germany}}
\affiliation[DU]{organization={Department of Computer Science, Durham University},            city={Durham}, country={United Kingdom}}
%\affiliation[1]{organization={Scientific Computing Center, Karlsruhe Institute of Technology}, city={Karlsruhe}, country={Germany}}{}

\affiliation[DL]{organization={digiLab}, city={Exeter}, country={United Kingdom}}
\affiliation[UKAEA]{organization={UK Atomic Energy Authority}, city={Oxford}, country={United Kingdom}}
\affiliation[DTU]{organization={Department of Applied Mathematics and Computer Science, Technical University of Denmark (DTU)}, city={Lyngby}, country={Denmark}}
\affiliation[CSUa]{organization={Department of Mathematics, Colorado State University}, city={Fort Collins}, state={CO}, country={USA}}
\affiliation[CSUb]{organization={Department of Mathematics, Department of Geosciences, Colorado State University}, city={Fort Collins}, state={CO}, country={USA}}
\affiliation[UB]{organization={Centre for Integrated Materials, Processes and Structures, Department of Mechanical Engineering, University of Bath}, city={Bath}, country={United Kingdom}}
\affiliation[INM]{organization={National Research Council-Institute of Marine Engineering}, city={Rome}, country={Italy}}
\affiliation[BMGF]{organization={Institute for Disease Modeling, Global Health Division, Bill \& Melinda Gates Foundation}}
\affiliation[SNL]{organization={Optimization and Uncertainty Quantification, Sandia National Laboratories}, city={Albuquerque}, state={NM}, country={USA}}
\affiliation[UCM]{organization={University of California, Merced}, country={USA}}
\affiliation[IMATI]{organization={National Research Council-Institute for Applied Mathematics and Information Technologies ``E. Magenes''}, city={Pavia}, country={Italy}}
\affiliation[SOL]{organization={Solea Energy}, city={Thetford}, state={VT}, country={USA}}
\affiliation[CI]{organization={Copenhagen Imaging ApS}, city={Herlev}, country={Denmark}}
\affiliation[UTA]{organization={The University of Texas at Austin}, country={USA}}
\affiliation[HU]{organization={Institute for Mathematics / Interdisciplinary Center for Scientific Computing (IWR), Heidelberg University}, %city={Heidelberg}, 
country={Germany}}

%\begin{refsection}

%\maketitle

\begin{abstract}
%
%\gls{UQ} is vital to safety-critical model-based analyses, but the widespread adoption of sophisticated \gls{UQ} methods is limited by high technical complexity of the associated computational routines. In this paper, we introduce UM-Bridge, a software protocol that breaks down the technical complexity of advanced UQ applications. At its core, UM-Bridge is a generic interface between the key components of a \gls{UQ} application. It facilitates the interoperability of generic \gls{UQ} software with arbitrary simulation codes, and enables portability of simulation models.  This separation of concerns enables the accelerated development of \gls{UQ} methods and effective interdisciplinary collaboration, forgoing user lock-in and making it easy to test new \gls{UQ} algorithms and perform complex \gls{UQ} analyses. Further, UM-Bridge enables effortless scalability on large compute clusters.
%
%In addition, we present a library of \gls{UQ} benchmark problems, distributed alongside UM-Bridge and accessible through its generic interface. These benchmarks support ongoing \gls{UQ} methodology research and enable robust and reproducible performance comparisons. We demonstrate the benefits of UM-Bridge to \gls{UQ} workflows with several scientific applications, harnessing large-scale \gls{HPC} resources for massively parallel \gls{UQ}, even using \gls{UQ} codes not originally designed for \gls{HPC}.
%
% Shortened to meet word count
\gls{UQ} is vital to safety-critical model-based analyses, but the widespread adoption of sophisticated \gls{UQ} methods is limited by technical complexity. In this paper, we introduce UM-Bridge (the \gls{UQ} and Modeling Bridge), a high-level abstraction and software protocol that facilitates universal interoperability of \gls{UQ} software with simulation codes. It breaks down the technical complexity of advanced \gls{UQ} applications and enables separation of concerns between experts. UM-Bridge democratizes \gls{UQ} by allowing effective interdisciplinary collaboration, accelerating the development of advanced \gls{UQ} methods, and making it easy to perform \gls{UQ} analyses from prototype to \gls{HPC} scale.

In addition, we present a library of ready-to-run \gls{UQ} benchmark problems, all easily accessible through UM-Bridge. These benchmarks support \gls{UQ} methodology research, enabling reproducible performance comparisons. We demonstrate UM-Bridge with several scientific applications, harnessing \gls{HPC} resources even using \gls{UQ} codes not designed with \gls{HPC} support.
\end{abstract}


\end{frontmatter}

\begin{figure*}
\centering    
\includegraphics[width=.8\textwidth]{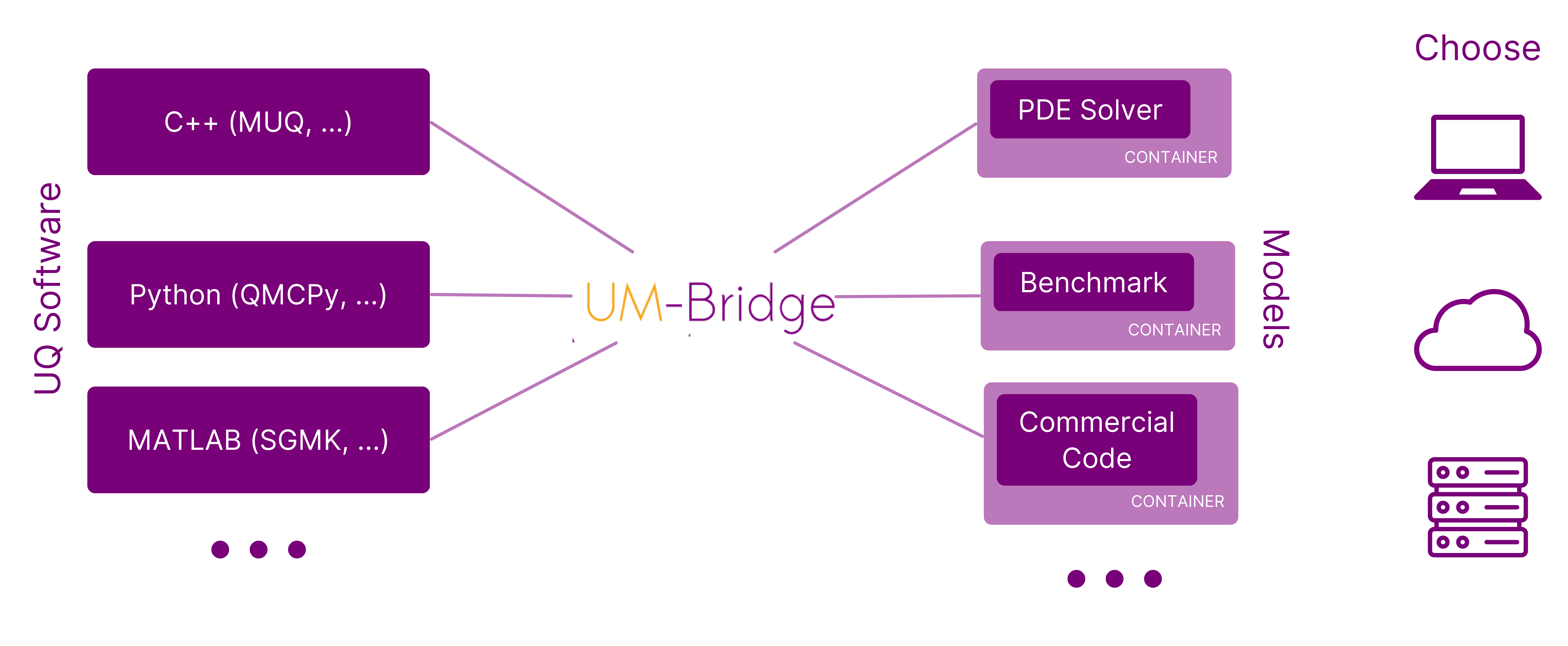}
\caption{\it UM-Bridge connecting UQ and numerical models through a universal interface.}
\end{figure*}

\section{Introduction}

The quantification of uncertainties is crucial for safety-critical applications where an understanding of the associated risks is integral to a broader analysis, such as in engineering design, clinical trials, weather and extreme event forecasting, environmental risk assessment, and infectious disease control, to name but a few. Such risk analyses are typically model-driven since they involve extrapolating complex processes into the future. Accurate quantitative information on uncertainties in predictions underlying decisions can lead to deeper insight into the problem and, consequently, better decisions.

The last couple of decades have seen remarkable growth in the use of \gls{UQ}, including uncertainty propagation and Bayesian inference \cite{bay_inverse}. This is partly due to efficient and robust techniques such as Latin hypercube, quasi-Monte Carlo, ADVI, HMC, and NUTS \cite{kucukelbir2017automatic, duane_hybrid_1987, hoffman2014no}, and partly due to flexible, open-source implementations of those algorithms, e.g. SciPy stats \cite{scipy}, Stan \cite{Stan}, PyMC \cite{pyMC3}, and Pyro \cite{bingham2018pyro}.

The success of these software packages can be attributed to the high level of abstraction with which they allow a user to define their statistical model. The subtle details of the underlying algorithms are not exposed to the user, who can simply specify their statistical model using high-level representations of the relevant components (prior, likelihood, observation model, etc.) and solve it using existing general-purpose algorithms.
% This allows statisticians and domain experts to solve complex problems without having detailed knowledge about the (often sophisticated) modelling software. %
This arrangement is a form of \textit{separation of concerns}: The algorithm developer does not require detailed knowledge of the domain expert's application and vice versa.

Notwithstanding those successes, advanced statistical inference tools 
are not yet employed extensively in conjunction with computationally demanding models. Such models appear, for example, in engineering design, in geophysics, or in medical imaging. We believe that this is in part due to a chicken-and-egg problem: UQ has not been demonstrated to be useful in those areas, but demonstration requires advanced knowledge of both application and \gls{UQ} techniques that few research groups have. Our work addresses both of those issues by reducing the required level of knowledge via generic interfaces.

We rely on an existing and largely universal theoretical vocabulary for the interaction between models and \gls{UQ}: The application model (or data-generating process) is represented as a \emph{forward operator} $F(\theta)$, which is a map from parameter space to data space. In forward \gls{UQ}, the parameter $\theta$ is uncertain, and we want to infer the probability distribution of the outputs $F(\theta)$; in inverse problems, we specify a likelihood model for the observed data based on $F(\theta)$ and aim to infer the probability distribution of $\theta$. Many \gls{UQ} algorithms exploit knowledge of derivatives of the forward map to increase computational efficiency; as a consequence, application models needs to be able to expose the computation of gradients (Jacobians) or Hessians of $F$, provided they are available within a particular application. This short list of shared components -- evaluating $F(\theta)$ and, perhaps, derivatives -- can be understood as a form of weak coupling between the statistical and the application model: Only a small part of the machinery inside the application model needs to be exposed to the statistical model, while the application model does not need to see the statistical model at all. This simple fact implies that it is -- in theory -- relatively straightforward to apply a wide variety of \gls{UQ} methods to arbitrary application models. 

%promises straightforward application of \gls{UQ} to many models, 
%This mathematical ``interface'' appears naturally in many \gls{UQ} software packages in some form or another.
Despite that, in practice, the software packages mentioned above are typically only used to solve problems with relatively simple data-generating processes, such as basic numerical integration or hierarchical Bayesian regression.
This dichotomy can at least in part be attributed to the high technical complexity of combining advanced software packages for statistical inference with state-of-the-art implementations of sophisticated application models. The technical complexity arises since there is no equally universal counterpart to the mathematical interface in software, and becomes particularly challenging in \gls{HPC} applications.

To address these issues, we introduce UM-Bridge: A universal link between \emph{any} \gls{UQ} package and \emph{any} forward model that breaks down complex software stacks into manageable components. Language-specific integrations make UM-Bridge models appear as native entities (classes, function calls etc.) in the respective programming language or even as native model classes in specific \gls{UQ} packages. 

Its unified interface allows linking arbitrary tools and makes it easy to swap out components on either side, thus reducing user lock-in. UM-Bridge further offers portable models through optional containerization, which offers a high degree of separation of concerns between \gls{UQ} and model experts.
In order to strengthen rigorous performance comparisons in \gls{UQ}, we present what we believe to be the first library of \gls{UQ} benchmark problems, developed by the authors of this paper and provided as part of an open-source software repository. UM-Bridge support makes the benchmarks available to virtually any \gls{UQ} software and ensures portability and reproducibility via containerization. We describe these \gls{UQ} benchmarks in detail in \Cref{sec:benchmarks}.

In addition, UM-Bridge offers a universal approach to scale-up \gls{UQ} applications on parallel compute clusters that is fully model agnostic. In a black-box fashion, UM-Bridge %even
allows for \gls{HPC}-scale computations with %existing 
\gls{UQ} codes that had never been intended for that purpose. In \Cref{sec:applications}, we describe in detail the workflow for three complex \gls{UQ} applications, where \gls{UQ} software is successfully coupled with an \gls{HPC} code.

Thus, by seperating concerns of UQ, model and HPC experts, UM-Bridge democratizes \gls{UQ} and paves the way to its widespread adoption, providing transparent scalability and facilitating reproducible benchmarking. %lowering the entry bar for non-specialists wishing to apply cutting-edge \gls{UQ} or statistical tools to complex computer models.

%\todo[inline]{Provide an outline of the paper here.}

%\paragraph{Impact}
%\todo[inline]{Suggestion: Show what impact we have already had: UQ package integrations, successful projects, large number of contributors to benchmarks, early industry adoption, open source peer bonus, google research credits, blog posts/siam news}
%\todo[inline]{AR: Not sure if this is the right place for such an advertisement. Maybe we need a news section on the documentation? Or a website for the project?}

\section{UM-Bridge -- a new paradigm for UQ software integration}

%\subsection{Rapid development of UQ applications}

% Accelerating development
% Breaking down complexity
The aim of UM-Bridge is to establish a new paradigm for \gls{UQ} software integration, enabling the analysis of previously intractable problems, development of better methods, and streamlining the workflow for every stakeholder. At its core, UM-Bridge links arbitrary \gls{UQ} and model applications through a network protocol, which leads to an entire range of new opportunities for \gls{UQ}. Stakeholders include applied scientists and engineers who employ \gls{UQ} methodologies to interrogate challenging problems, developers of novel \gls{UQ} methods who require objective and reproducible benchmarks to validate their methods, and application modellers who wish to integrate their model into a \gls{UQ} workflow. Here, we outline the benefits of the UM-Bridge paradigm to the various stakeholders.

\begin{itemize}
\item \textbf{Benefits for scientists and engineers.}
While \gls{UQ} techniques are critical for making trustworthy predictions, particularly in safety-critical contexts, rigorous \gls{UQ} is often held back by perceived complexity and cost of setting up a UQ workflow. UM-Bridge contributes to the establishment of ``UQ-as-a-Service'' by providing an environment to quantify uncertainties through a streamlined interface for all stages of the UQ workflow. This allows researchers with varying levels of expertise to engage in the rigorous analysis of uncertainties.

\item \textbf{Benefits for \gls{UQ} method developers.}
The development of efficient algorithms for \gls{UQ} requires access to objective and challenging problems that can be used to adequately benchmark the algorithms in question. UM-Bridge enables \gls{UQ} method developers to separate the development of \gls{UQ} methods, such as the design of experiments, surrogate model construction, and uncertainty propagation, from the model or application being studied. Each UM-Bridge model and benchmark can be accessed from any \gls{UQ} platform through the respective UM-Bridge integration, promoting flexibility in \gls{UQ} workflows and enabling method developers to easily test their methods on new problems.

\item \textbf{Benefits for model code developers.}
The coupling of \gls{UQ} software with an application model currently requires intricate knowledge of both software packages.
% the application model and the specifics of a particular \gls{UQ} software. 
The UM-Bridge model interaction interface (\Cref{sec:architecture}) allows for application modellers to simply equip their model with UM-Bridge support and otherwise focus on the performance of the application model itself. They can thus create performant models that are both fully compatible with virtually any \gls{UQ} software and easy to share, enabling effective collaboration with %between the application modellers and 
their \gls{UQ} counterparts.
\end{itemize}

In the following, we outline some of the mechanisms by which UM-Bridge accelerates development and simplifies adoption of advanced UQ methods, contributing to the democratization of \gls{UQ}.

\subsection{Making tools accessible}

\begin{table}
  \center
    \caption{\it Currently available UM-Bridge integrations. Ticks indicate support for UM-Bridge clients (\gls{UQ} algorithms) and UM-Bridge servers (models) in the respective language or \gls{UQ} package. }
  \label{tab:lang_support}
\begin{center}
  \begin{tabular}{c c c}
   \toprule
   \bf Language & \bf Client & \bf Server \\ [0.5ex]
    \midrule\midrule
    Python & \cmark & \cmark \\
    C++ & \cmark & \cmark \\
    Matlab & \cmark & \xmark \\
    R & \cmark & \xmark \\
    Julia & \cmark & \cmark \\
    \bottomrule
\end{tabular} \qquad\qquad
\begin{tabular}{c c c}
\toprule
    \bf \gls{UQ} package & \bf Client & \bf Server \\ [0.5ex]
    \midrule\midrule
    CUQIpy \cite{alghamdi2023cuqipy,riis2023cuqipy} & \cmark & \cmark \\
    emcee \cite{emcee} & \cmark & \xmark \\
    MUQ \cite{MUQ} & \cmark & \cmark \\
    PyApprox \cite{PyApprox} & \cmark & \xmark \\
    PyMC \cite{pyMC3} & \cmark & \xmark \\
    QMCPy \cite{QMCPy} & \cmark & \xmark \\
    SGMK \cite{sparse_grids_matlab_kit} & \cmark & \xmark \\
    tinyDA \cite{tinyDA} & \cmark & \xmark \\
    TT-Toolbox \cite{tt-toolbox} & \cmark & \xmark \\
    UQPy \cite{UQPy} & \cmark & \xmark \\
   \bottomrule
  \end{tabular}
\end{center}
\end{table}

UM-Bridge makes advanced UQ tools accessible to a wide range of applications by providing a unified and language-agnostic interface. 
This is especially important since each community has specific requirements and therefore tends to use specific languages and software packages. For example, efficient numerical solvers for \gls{PDE} models are often limited by computational performance and consequently need fine-grained control over memory operations, sometimes use low-level hardware features, and often interact with many other software libraries providing specialized algorithms (see, for example, \cite{arndt2019dealii}); they are therefore often implemented in lower-level languages like C++ or Fortran. On the other hand, \gls{UQ} methods themselves are often far less compute-intensive, are relatively self-contained, and so allow for fast development in high-level languages like Python or Matlab. The UM-Bridge interface allows these communities to continue using their preferred tools, while still being able to collaborate.

Native integrations for various languages and \gls{UQ} packages (see \Cref{tab:lang_support}) implement the UM-Bridge network protocol both for client side (e.g. \gls{UQ} software) and server side (numerical models). Many \gls{UQ} packages such as PyMC \cite{pyMC3} or QMCPy \cite{QMCPy} have a specific model interface, which the respective integrations implement. UM-Bridge thus acts as a cross-language and cross-platform translation layer.

%\begin{wrapfigure}{R}{.4\linewidth}
%\centering
%   \includegraphics[width=\linewidth]{Images/l2sea/l2sea_example}
%    \caption{\it Fortran code for modelling the pressure distribution and the wave elevation generated by a vessel. UM-Bridge support is achieved through a simple wrapper, enabling parallelized \gls{UQ} from, e.g., Matlab.}
%  \label{fig:l2sea_results_section}
%\end{wrapfigure}

\begin{figure}[t]
\centering
   \includegraphics[width=0.25\linewidth]{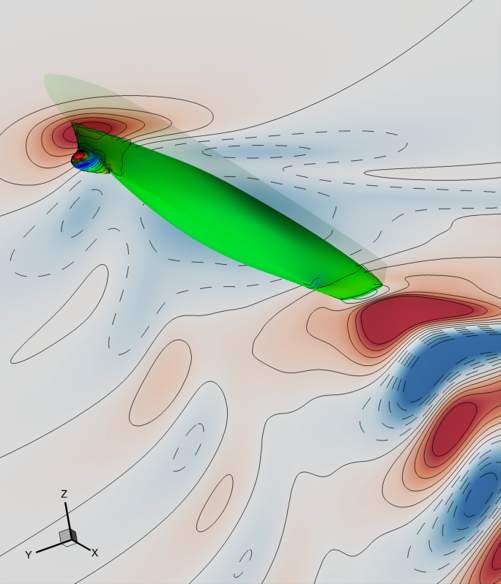} \qquad\includegraphics[width=.7\linewidth]{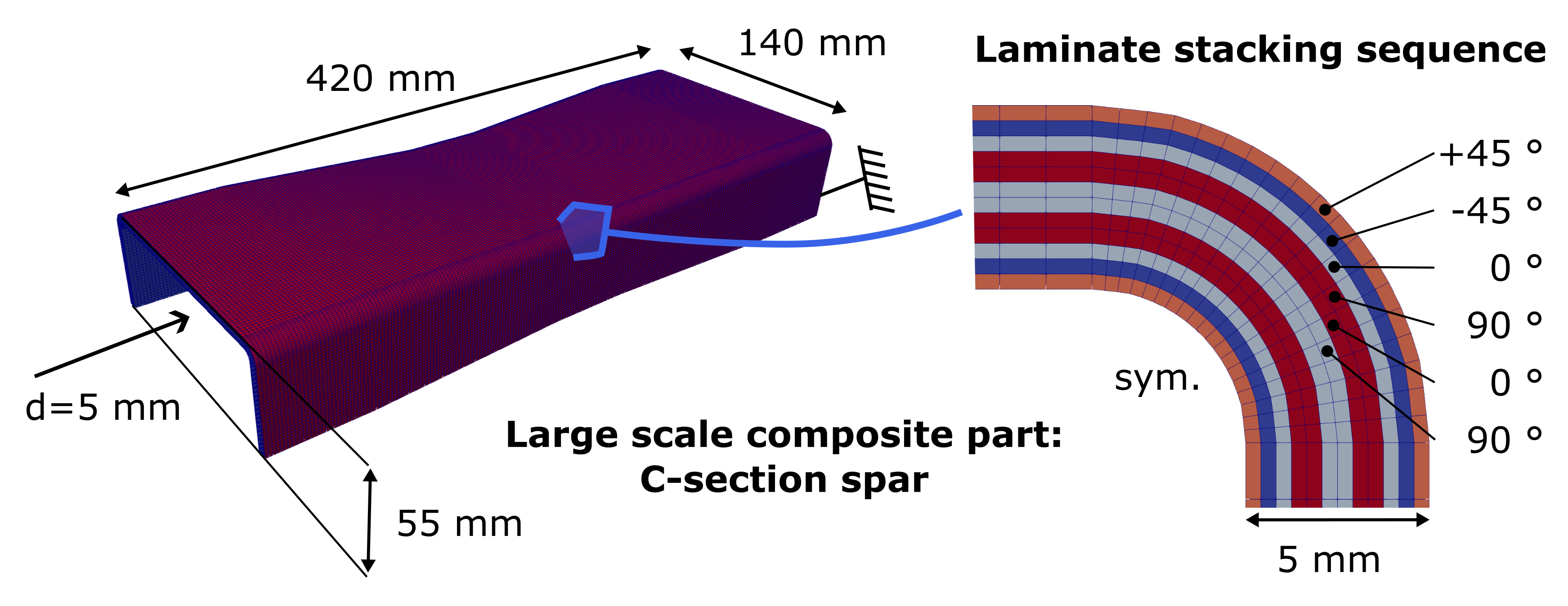}    
   \caption{\it Fortran code for modelling the pressure distribution and the wave elevation generated by a vessel. UM-Bridge support is achieved through a simple wrapper, enabling parallelized \gls{UQ} from, e.g., Matlab (left figure). FE model of a carbon fibre composite aerospace component implemented in a complex \gls{PDE} solver in C++. Through UM-Bridge and containerization, a \gls{UQ} expert could solve uncertainty propagation without technical knowledge of the model code (center \& right figures).}
  \label{fig:l2sea_results_section}
\end{figure}

More generally, any language supporting HTTP communication can implement UM-Bridge support. Models that are written in unsupported languages or that cannot be embedded in a higher-level code may still reap most benefits of UM-Bridge when called from a simple UM-Bridge wrapper.
As an example that demonstrates how previously incompatible applications are now readily available to high-level \gls{UQ} tools, in \Cref{sec:application_sparse_grids} we seamlessly apply a Matlab \gls{UQ} code to a Fortran-based model of a ship hull \cite{pellegrini2022mathematics}, as shown in \Cref{fig:l2sea_results_section} (left).

\subsection{Separating concerns}

Building \gls{UQ} applications requires expertise in both \gls{UQ} and numerical modelling. In addition to a theoretical understanding, it also calls for experience with software packages in both domains. It is often time-consuming or outright infeasible to achieve uch an overarching expertise  the constraints of a project.

Since UM-Bridge makes the mathematical model available through a universal software interface, it now becomes possible to properly separate concerns between the model and \gls{UQ} experts. The model expert can focus on implementing the map $F$ and (if available) its derivatives, while the \gls{UQ} expert can focus on \gls{UQ} aspects. This is especially important when applying state-of-the-art \gls{UQ} methods to complex models. %, where each expert likely has little knowledge of the intricacies of the respective other field. 
%UM-Bridge allows these experts to collaborate regardless, 
%removing a significant obstacle hindering widespread adoption of \gls{UQ}.

In addition, since the UM-Bridge interface is network-based, it becomes straightforward to containerize models in \gls{UQ} workflows. This allows sharing models among collaborators and running them on any machine or cluster without per-machine installation effort. The model expert may now pass their implementation of the model as a ready-to-use container to the \gls{UQ} expert, which further contributes to separation of concerns.

%\begin{wrapfigure}{R}{.5\columnwidth}
%    \centering
%    \includegraphics[width=.5\columnwidth]{Images/Composite_application/composite_model_description.png}
%    \caption{\it FE model of a carbon fibre composite aerospace component implemented in a complex \gls{PDE} solver in C++. Through UM-Bridge and containerization, a \gls{UQ} expert could solve uncertainty propagation without technical knowledge of the model code.}
%    \label{fig:composites_results_section}
%\end{wrapfigure}

A further benefit of containerization is reproducibility of the model outputs. Besides being integral to the scientific method, it is helpful for developing \gls{UQ} applications across teams. Model container images can be published alongside the scientific results, enabling the exact reproducibility of scientific work. We exploit this feature in the benchmark library below.

Illustrating separation of concerns, in application \Cref{sec:application_qmc} a simulation expert developed an advanced modelling application in C++, %{fig}:composites_results_section}), 
containerized it, and shared it with a \gls{UQ} expert, see \Cref{fig:l2sea_results_section} (right). The \gls{UQ} expert was able to quantify the parametric uncertainty using a high-level \gls{UQ} package written in Python, without further integration work or any deeper technical knowledge of the model implementation.

\subsection{Providing transparent scalability}

A major roadblock to \gls{UQ} on large-scale models is the need for \gls{HPC} capability. To achieve that, both the underlying model and the UQ method need support for parallelization in theory and implementation. Additionally, the parallelization techniques involved have to be compatible and tightly linked. % (\Cref{fig:monolithic_architecture}).

\begin{comment}
\begin{figure}[h]
\centering
\includegraphics[width=\columnwidth]{Images/monolithic.pdf}

\todo[inline]{I'm not sure the picture provides sufficient value
  compared to the space it takes up. I think everything you say with
  the picture is already said more concisely in the text. The only non-generic part of it
  is the image on the left, which is too small to read. I vote for
  removal. (The figure is also somewhat duplicative of Fig. 1.)}

\caption{\label{fig:monolithic}\it Monolithic coupling of UQ, model and HPC capabilities within a single application. SuperMUC image from \cite{SuperMUCImage}.}
\label{fig:monolithic_architecture}
\end{figure}
\end{comment}

In practice, many UQ packages do support parallelization in some
capacity but are often limited to a single machine or a specific
way of setting up a parallel simulation. On the other hand, many
\gls{PDE} solvers are able to use HPC resources, typically through \gls{MPI}. However, they often come with specific
assumptions about the nature of the machine they are running on.
As part of UM-Bridge, we introduce an architecture that separates parallelization of UQ and model: %(\Cref{fig:umbridge_architecture_high_level})
A cluster-side load balancer accepts parallel model evaluation requests from the UQ code, without making any assumptions on what parallelization technique the UQ code is employing. It then distributes those requests among independent instances of the model, each of which may now also employ any parallelization strategy.

\begin{comment}
\begin{figure}[h]
\centering
\includegraphics[width=\columnwidth]{Images/umbridge.pdf}

\todo[inline]{Same here as for the previous picture. I vote to remove.}

\caption{\it UM-Bridge providing HPC support as a separate, general-purpose component.
SuperMUC image from \cite{SuperMUCImage}.
}
\label{fig:umbridge_architecture_high_level}
\end{figure}
\end{comment}

With such an architecture, even thread-parallel UQ codes can offload
costly numerical simulation runs to a (potentially remote)
cluster. Many existing UQ codes do not require any modification to
scale up via UM-Bridge, and neither do typical models. 
%UM-Bridge supports 
Both cloud-clusters and \gls{HPC} systems are supported.

We illustrate these concepts with the application in Appendix
\ref{sec:application_mlda}. It shows how an existing \gls{UQ} code can transparently offload costly model evaluations to a remote cluster of 2800 processor cores, running 100 instances of a sophisticated, parallel tsunami model.

\subsection{Facilitating reproducible benchmarking} \label{sec:results_benchmarks}

The current lack of available and reproducible UQ benchmarks is severely limiting quantitative comparisons of competing UQ
methods. The few existing initiatives %' implementations 
are typically limited to a single programming language (e.g., \cite{PyApprox}). Support for more languages then requires
reimplementation (e.g., \cite{AristoffBangerth}), but this is clearly not viable for complex models.

A universal interface like UM-Bridge can address this
problem. Thus, we have collected a large number of benchmark
problems (described in more detail in \Cref{sec:uq-benchmark-library} below), 
with the goal of being:

\begin{itemize}
 \item \textbf{Representative:} The benchmarks are a curated collection, aiming to cover all properties of models relevant to \gls{UQ}. That includes different choices of parameter dimension, smoothness of model output, model run time etc. Testing methods across several benchmarks in our collection can be used to demonstrate their behaviour on problems ranging from simple artificial tests all the way to realistic large-scale simulations.
 \item \textbf{Easy to use:} All benchmark problems are implemented with UM-Bridge support and have easily accessible documentation. %Thus, 
 They can readily be used to benchmark any \gls{UQ} package or prototype code.
 \item \textbf{Reproducible:} Containerization ensures that benchmark problems are fully defined in portable software, avoiding potential ambiguities in definition or implementation details.
\end{itemize}

The library aims to establish a neutral foundation enabling subsequent studies. We intentionally do not include benchmark results of competing methods, since specific metrics may depend strongly on the goals of such a comparison, and results will change as future methods are developed.

\begin{comment}
We include the following three types:

\begin{enumerate}
  \item \textbf{Models} implement a numerical model of a physical phenomenon, such as a PDE solvers. They are intended as general-purpose models that can be used in a wide range of user-defined UQ applications, and typically expose a number of configuration options for a high level of control. You can find the models in Appendix \ref{sec:models}.
  \item \textbf{Inverse problems} define a complete Bayesian posterior density, including prior, likelihood and observations. The goal is then to determine the posterior distribution or quantities derived from it. They too are intentionally restricted in terms of configuration options. However, for methods needing finer-grained control beyond the posterior density and e.g. direct access to the likelihood itself, the forward model is provided as well. You can find the inverse benchmarks in Appendix \ref{sec:inference}.
  \item \textbf{Forward problems} that are defined by a model and a specific probability distribution over the model inputs. The goal is then to find the corresponding distribution of the model output, or a quantity derived therefrom. They are intentionally restricted in terms of configuration options to avoid potential ambiguities. You can find the propagation benchmarks in Appendix \ref{sec:propagation}.
\end{enumerate}
\end{comment}

\Cref{sec:benchmarks} gives detailed documentation of all
currently available models and benchmark problems. We intend for this library to
be a growing collection and encourage future contributions to it. This
collection can also serve related areas, like optimization, and an
example in this direction can be found in \Cref{sec:optimisation}.

\section{UM-Bridge architecture and usage}

\gls{UQ} problems typically revolve around a deterministic model,
which is operationalized as a \emph{forward operator} $F :
\mathbb{R}^n \rightarrow \mathbb{R}^m$. The forward operator takes a
parameter vector $\theta$ of dimensions $n$ (containing uncertain
forcing terms and/or coefficients in the model) to some model output, a vector of dimension $m$.

Uncertainty propagation problems are then defined by considering
$\theta$ as a random vector following some distribution $\pi(\theta)$. The goal is to find properties of the distribution of $F(\theta)$,
such as moments, quantiles, the \gls{pdf} or the cumulative distribution function.

On the other hand, Bayesian inference defines the probability of a parameter $\theta$ given observed data $y$ in terms of a prior distribution $\pi_{\text{prior}}$ and a likelihood function $L$:
\[ 
\pi_{\text{posterior}}(\theta | y) \propto \pi_{\text{prior}}(\theta) L(y | \theta).
\]
The first factor on the right represents prior
information about the parameter distribution we seek, while the
likelihood $L(y | \theta)$ provides a %is typically defined based on a 
measure of distance between model
prediction $F(\theta)$ and data $y$. The goal is then to characterize $\pi_{\text{posterior}}$, for example through sampling or by estimating its moments.

These concepts constitute the typical theoretical context of \gls{UQ} methods to which UM-Bridge lends itself. In the following, we explore how UM-Bridge implements these concepts.

\subsection{UM-Bridge architecture}
\label{sec:architecture}

Many forward \gls{UQ} methods (e.g. many \gls{MC} samplers, or stochastic collocation \cite{SC_Survey})
only require model evaluations $\{F(\theta_i)\}$ at a %an (at least in practice) 
finite number
of points $\theta_1, \ldots, \theta_N$. Other methods may additionally need derivatives of $F$ at
$\theta_i$. Similarly, inversion problems are often solved by pointwise evaluations of $\pi_{post}$, for which different methods can
be employed (Metropolis-Hastings MCMC \cite{MHMCMC}, Hamiltonian MCMC
\cite{HamiltonianMCMC}, NUTS \cite{hoffman2014no}, etc.). Again, these approaches require only forward model evaluations $F(\theta_i)$ and, in some cases, derivatives at a finite number of points $\theta_1, \ldots, \theta_N$.

In general, across most \gls{UQ} algorithms, the model is typically assumed to provide (some of) the following pointwise mathematical operations:
\begin{itemize}
    \item Model evaluation $F(\theta)$, \vspace{-0.5ex}
    \item Gradient $v^\top J_F(\theta)$,\vspace{-0.5ex}
    \item Jacobian action $J_F(\theta) v$, \vspace{-0.5ex}
    \item Hessian action.
\end{itemize}

%\todo[inline]{Hessian action is a bit more complex, see \url{https://mituq.bitbucket.io/source/_site/latest/group__modpieces.html} at the bottom. Not sure how to write this up briefly}

Any \gls{UQ} method whose requirements are limited to this list could therefore -- in principle -- be applied to any model supporting the appropriate operations.

\begin{figure}[t]
\centering

\includegraphics[width=0.48\columnwidth]{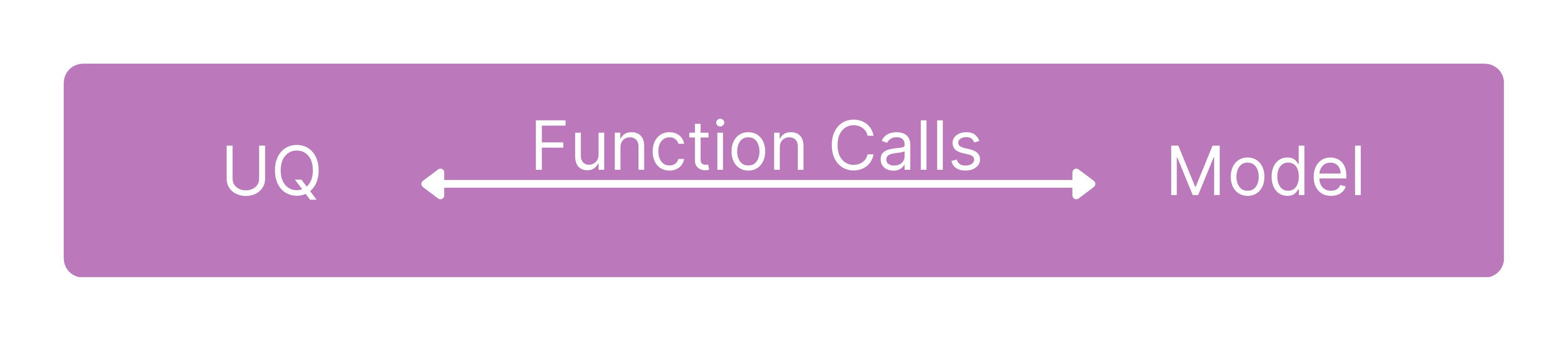} \hspace{1em}
\includegraphics[width=0.48\columnwidth]{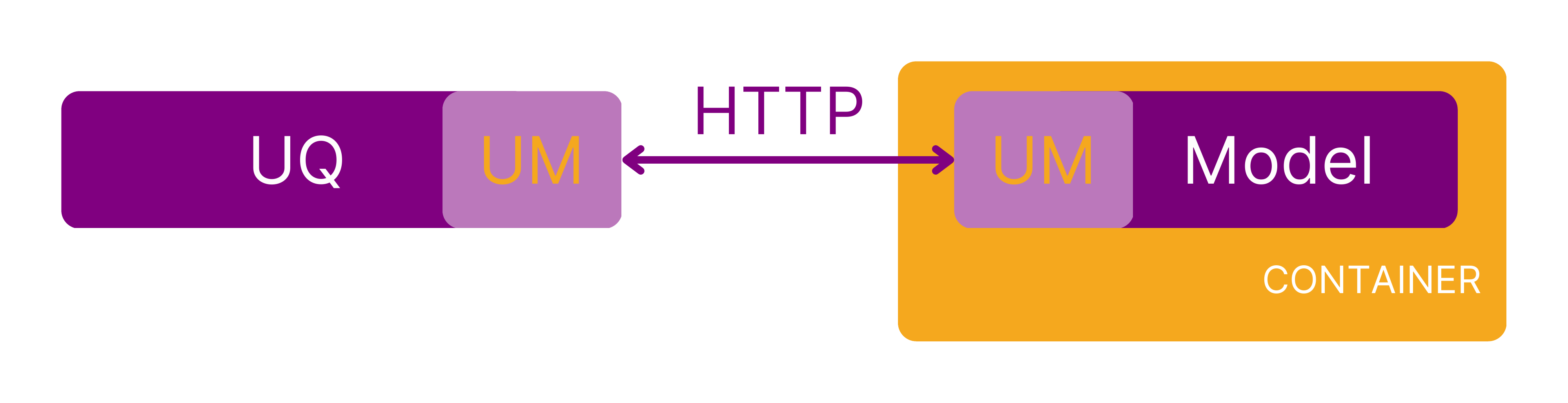}

\caption{\it Monolithic coupling of \gls{UQ} and model software in a single application (left); UM-Bridge providing a universal interface between \gls{UQ} and model applications (right). UM-Bridge integrations (''UM'') handle network communication behind the scenes. Optionally, %UM-Bridge 
models may be containerized.}
\label{fig:microservice}
\end{figure}
%\end{wrapfigure}

The key idea behind UM-Bridge is to implement the above mathematical
interface in an equally universal software interface: We treat
\gls{UQ} algorithm and numerical model as stand-alone applications,
linked only through an \gls{HTTP} based network protocol (see \Cref{fig:microservice}, \cite{UMBridgeSoftwareJOSS}).
In contrast to a monolithic approach, this microservice-inspired architecture enables linking across arbitrary languages and software packages, as well as separating concerns. It should be noted that multilevel, multi-index or multi-fidelity methods (e.g. \cite{MLMC,MLMCMC,Lykkegaard_MLDA,SchwabMLMCMC,MultilevelDILI,lykkegaard_multilevel_2023,teckentrup.etal:MLSC,MIMCMC,piazzola.eal:ferry-paper,sandia:coupled}) such as the one in \Cref{sec:application_mlda} operate on an entire hierarchy of models. Less accurate but cheaper-to-compute models are used to accelerate the \gls{UQ} method, leading to considerable gains in efficiency both in forward and inverse \gls{UQ} problems. While this necessitates multiple models, each individual model still fits in the framework described above.

\subsection{The UM-Bridge protocol}

At its core, the UM-Bridge protocol is built on \gls{HTTP} and JSON for compatibility with many programming languages and easy implementation. All operations a model may provide
are considered optional. For example, models may be implemented with gradients or not, and \gls{UQ} clients can query and adapt
to that. Importantly, this means that the protocol can be extended in the future without breaking compatibility.

UM-Bridge treats model inputs and outputs as lists of vectors, which is inspired the proven internal model interface in MUQ \cite{MUQ}. This enables automatic differentiation with respect to certain components of the input, even when chaining multiple models.

In all the aforementioned integrations of UM-Bridge (see \Cref{tab:lang_support}), the protocol is implemented behind the scenes, making UM-Bridge models available either as simple function calls or fully integrated in the respective software package's model structure.

\subsection{Basic examples}
\label{sec:howto}

Let us illustrate how UM-Bridge works in practice with some simple examples. While intentionally basic, the same approach underlies the applications in \Cref{sec:applications}.

\subsubsection{Clients - Interacting with models}\label{sec:clients}

First, we need a running model server, providing a forward operator $F$. This could be a simulation code  with UM-Bridge support running directly on your system, a custom one as in \Cref{sec:servers}, or a containerized model. For example, we could download and run the containerized tsunami model from the benchmark library (see \Cref{sec:tsunami_model}) via \texttt{docker}:

\begin{lstlisting}[]
docker run -it -p 4242:4242 \
    linusseelinger/model-exahype-tsunami
\end{lstlisting}

Once running, any UM-Bridge client can request evaluations of $F$ from that model server. This could be any supported \gls{UQ} package or a custom code in any supported language (see \Cref{tab:lang_support}). For example, we can call the model from Python via the \texttt{umbridge} Python module:

\begin{lstlisting}[language=Python,breaklines=true,postbreak=\mbox{$\hookrightarrow$\space}]
url = "http://localhost:4242"
model = umbridge.HTTPModel(url, "forward")

print(model([[0.0, 10.0]]))
\end{lstlisting}

All that is needed is the URL of %at which 
the model server %is running,
(here \verb=http://localhost:4242=),
the name of the requested model $F$ (here \verb=forward=)
and a parameter $\theta$ to evaluate $F$ at. The actual evaluation is then a call to the model with a given parameter $\theta=(0,10)$, which returns $F(\theta)$. Optionally, arbitrary model-specific configuration options can be passed, for example to set the discretization level of a multifidelity model:
\begin{lstlisting}[language=Python]
print(model([[0.0, 10.0]], {"level": 0}))
\end{lstlisting}
Similar calls allow access to gradient, Jacobian action and Hessian action, as long as the model implements these operations.

Clients written in other languages follow a similar pattern. For
example, code equivalent to the above would look like this in C++:

\begin{lstlisting}[language=C++]
std::string url = "http://localhost:4242";
umbridge::HTTPModel model(url, "forward");

std::vector<std::vector<double>> outputs
  = model.Evaluate({{0.0, 10.0}});
\end{lstlisting}

\subsubsection{Servers - Defining models}\label{sec:servers}
Defining a model conversely requires implementing the model map
$F$, which could be anything from simple arithmetic all the way to advanced numerical simulations. In addition, model name as well as input and output dimensions are specified. The following minimal example in
Python implements the multiplication of a single input value by two:

\begin{lstlisting}[language=Python]
class TestModel(umbridge.Model):

  def __init__(self):
    super().__init__("forward")

  def get_input_sizes(self, config):
    return [1] # Input dimensions

  def get_output_sizes(self, config):
    return [1] # Output dimensions

  def __call__(self, parameters, config):
    output = parameters[0][0] * 2
    return [[output]]

  def supports_evaluate(self):
    return True

umbridge.serve_models([TestModel()], 4242)
\end{lstlisting}

The code sets \verb=supports_evaluate= to indicate to UM-Bridge that model evaluations are supported. Likewise, gradients, Jacobian- or Hessian-action could be implemented and marked as supported.

The final line instructs UM-Bridge to make the model available to clients, acting as a server. UM-Bridge servers written in other languages follow the same principle.

\subsubsection{Containers - Making models portable}

%Many numerical models consist of a complex software stack and possibly
%associated data sets; installing such software packages often requires
%considerable effort and expert knowledge. However, since UM-Bridge
%models are accessed through a network, they can easily be
%containerized by the package's developers and shared widely.

UM-Bridge models ranging from simple test problems all the way to advanced simulation codes can be containerized for portability, and possibly be published in public repositories.
Such a container can then
be launched on any local machine or remote cluster without
installation effort, and users can readily interact with it as in \Cref{sec:clients}.
%, and the model can be accessed through UM-Bridge
%as if it ran locally within the UQ developer's environment.

Defining such a container comes down to instructions for installing dependencies, compiling if needed, and finally running the desired UM-Bridge server code. For example, a docker container wrapping the example server above can be defined in the following simple Dockerfile:

\begin{lstlisting}
FROM ubuntu:latest

COPY minimal-server.py /

RUN apt update && \
    DEBIAN_FRONTEND="noninteractive" apt install -y python3-pip && \
    pip install umbridge

CMD python3 minimal-server.py
\end{lstlisting}
Once built, \texttt{docker run} can launch an instance of the container as above in \Cref{sec:clients}.

Some legacy tools are built for one-shot simulation runs controlled by an input file, and may be too rigid to be called from a UM-Bridge server code like the one in \Cref{sec:servers}. In such cases, a UM-Bridge server can act
as a small wrapper application that, on every evaluation
request, creates an input file, launches the model software, and then
evaluates the code's output files. Such file based interfaces can be error-prone
when set up individually, but are reproducible and portable when containerized.

\subsection{Scaling up on clusters}
\label{sec:kubernetes}

\begin{figure*}
\centering

\includegraphics[width=0.9\textwidth]{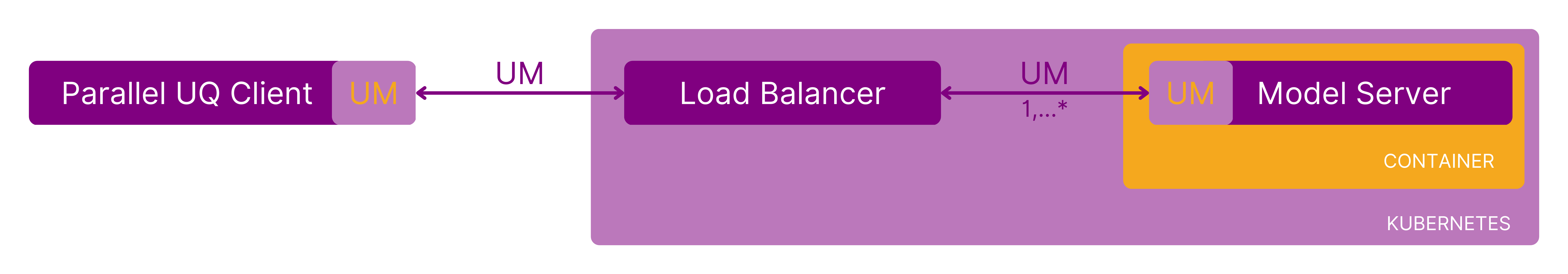}
\caption{\it UM-Bridge provides a general-purpose Kubernetes setup for scaling up \gls{UQ} applications. It runs many parallel instances of containerized models and distributes concurrent requests from a \gls{UQ} client among them. Any UM-Bridge supporting \gls{UQ} or model code readily works with this setup.}
\label{fig:kubernetes_sequential_model}
\end{figure*}

We enable scaling up \gls{UQ} applications by providing universal configurations for cloud clusters and tooling for \gls{HPC} systems. \Cref{fig:kubernetes_sequential_model} illustrates how their key component, a cluster-side load balancer, takes concurrent evaluation requests from a UQ client and distributes them across many independent model instances.

As the client-side interface is entirely identical, a UQ software is completely oblivious to the model being executed on a cluster instead of the local machine. However, the UQ software may now send multiple concurrent evaluation requests to the cluster. Since the model evaluation is typically costly compared to the UQ algorithm itself, a UQ package spawning hundreds of threads on a laptop offloading work to a large cluster is perfectly viable.
Likewise, since every model instance only receives sequential model
evaluations via UM-Bridge, no modification to the model is needed compared to
running on a single machine. In addition, containerized models can be run on cloud
or HPC systems without additional setup cost as long as the HPC
systems supports containers.

%In practice, many UM-Bridge applications can immediately be scaled up
%to clusters using this approach.
Since the UQ software is not involved in distributing evaluations across the cluster, parallelism within many existing UQ packages works out of the box with this setup.
For example, the \gls{sgmk}, MUQ,
QMCPy, and PyMC have all been tested successfully across a number of different models. Some of these applications 
are discussed in \Cref{sec:applications}. 

Note that, due to the flexibility of the interface, more complex architectures are easy to construct. For example, as in \Cref{sec:application_mlda}, a multilevel \gls{UQ} code may at the same time be pointed to a fast surrogate model running on the same workstation and a remote cluster for full-scale model runs, accommodating for heterogeneous resource demand across levels

\subsubsection{UM-Bridge on kubernetes clusters}
\label{sec:umbridge_kubernetes}

One implementation of this architecture is based on kubernetes for cloud systems. The reasoning behind using kubernetes is that
(i) existing model containers can be run without modification,
(ii) the entire setup can be specified in configuration files and is easily reproducible,
and (iii) it can be run on systems ranging from single servers to large-scale clusters and is available on many public cloud systems.

We employ HAProxy as load balancer. HAProxy was originally intended for web services processing large numbers of requests per backend container. However, we set it up such that only a single evaluation request is ever sent to a model server at once, since we assume running multiple numerical model evaluations concurrently on a single model instance would lead to performance degradation.

Kubernetes can be fully controlled through configuration files. The UM-Bridge kubernetes setup can therefore be applied on any kubernetes cluster by cloning the UM-Bridge Git repository and executing
\begin{lstlisting}
kubectl apply -f FILENAME
\end{lstlisting}
on the provided configuration files. This procedure is described in more detail in the UM-Bridge documentation, and takes only a few minutes.

The only changes needed for a custom application are docker image name, number of instances (\texttt{replicas}) and resource requirements in the provided \texttt{model.yaml} configuration. Below is an example of an adapted model configuration, namely the one used in the sparse grids application in \Cref{sec:application_sparse_grids}:

\begin{lstlisting}
apiVersion: apps/v1
kind: Deployment
metadata:
  name: model-deployment
spec:
  replicas: 48
  template:
    metadata:
      labels:
        app: model
    spec:
      containers:
      - name: model
        image: linusseelinger/model-l2-sea:latest
        env:
        - name: OMP_NUM_THREADS
          value: "1"
        resources:
          requests:
            cpu: 1
            memory: 1Gi
          limits:
            cpu: 1
            memory: 1Gi
\end{lstlisting}

In addition, we provide a kubernetes setup that allows for \gls{MPI} parallelism across containers and therefore across compute nodes. It requires the model container to inherit from the \texttt{mpi-operator} base image, but behaves the same in any other regard.

\begin{figure}[ht]
\centering

\begin{tikzpicture}
	\begin{axis}[
		xlabel=Number of model instances,
		ylabel=Run time {[}s{]},
		grid=major,
		xmode=log,
		ymin=0,
        height=4cm,
        width=0.6\linewidth,
    xtick = {2,8,32,128,512},
		xticklabels = {2,8,32,128,512},
	]

% haproxy without debug mode
% c2d-highcpu-112
	\addplot coordinates {
%		(2048,0) % 38 nodes  40960 samples
%		(1024,0) % 20 nodes  20480
		(512,49.429) % 10 nodes   10240
		(256,53.398) % 5 nodes    5120
		(128,50.534) % 3 nodes    2560
		(64,50.058) % 2 nodes     1280
		(32,51.400) % 1 node      640
		(16,47.233) % 1 node
		(8,47.147) % 1 node
		(4,46.930) % 1 node
		(2,46.956) % 1 node
		(1,46.901) % 1 node
	};
% Older run:
%	\addplot coordinates {
%		(210,63.718)
%		(120,65.361)
%		(60,63.303)
%		(30,62.787)
%		(15,62.376)
%		(7,62.272)
%		(3,62.414)
%		(1,62.135)
%	};
	\end{axis}
\end{tikzpicture}
\caption{\it Synthetic weak scaling test of the kubernetes setup on \gls{GKE} for various numbers of parallel model instances, requesting a constant number of evaluations per model instance.}%is requested from the cluster.}
\label{fig:cloud_scalabillity}
\end{figure}
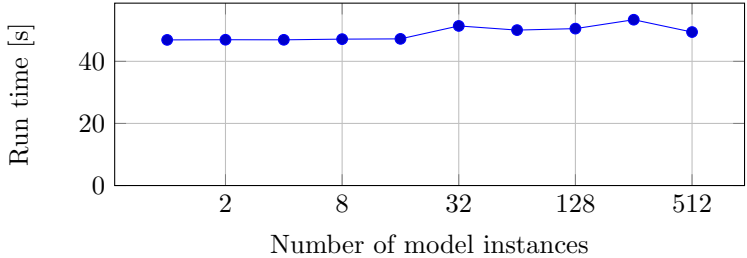
%\end{wrapfigure}
%\end{comment}

\begin{comment}
The results in \Cref{fig:cloud_scalabillity}
show near perfect weak scaling, indicating low parallel overhead. Some
fluctuations are observed for higher process counts, which we believe
mostly results from non-exclusive access to physical nodes in the cloud system.

We did not observe a bottleneck in testing, and assume this setup could be scaled further. In addition, the model itself could be parallelized. For example with a model parallelized across 20 cores, we can expect similar results to the ones presented here fully utilizing a 10,000 core cluster.
\end{comment}

\Cref{sec:application_mlda} demonstrates the UM-Bridge kubernetes setup on 2,800 processor cores running 100 parallel model instances. We further performed synthetic weak scaling tests running up to $N=512$ instances of the L2-Sea model (see \Cref{sec:l2sea_model}),
configured to use 1 processor core each and taking around 2.5s per evaluation. We then requested $20 \cdot N$ model evaluations from the cluster. As \Cref{fig:cloud_scalabillity} shows, we did not observe a significant bottleneck, indicating that this setup could be scaled further. In addition, the model itself could be parallelized. With a model parallelized across 100 cores, we expect similar results to the 512 instance test while fully utilizing a 51,200 core cluster.

\subsubsection{UM-Bridge on HPC systems}

UM-Bridge also includes support for SLURM or PBS based \gls{HPC} systems. In this case, a custom component takes UM-Bridge evaluation requests from the client and translates those into jobs managed by HyperQueue \cite{HyperQueue}. HyperQueue then acts as the load balancer, spawning model instances as \gls{HPC} scheduler jobs and passing evaluation requests to them as needed.

Models may be local installations of the simulation software
or containerized versions. %RS for ease of use. 
While typical \gls{HPC} systems lack Docker support, apptainer \cite{singularity} is increasingly available, where %RS We can therefore achieve 
the same abstraction of parallelization effort as in the kubernetes case is achieved, both 
regarding model and \gls{UQ}.

\section{UQ Benchmark library}
\label{sec:uq-benchmark-library}

An important part of the work we present herein is a sizable library
of UQ benchmark problems.  This collection is a collaboration with more than 15
contributors (coauthors of this paper) from more than 10 international
institutions. %The library contains both generic models and specific benchmarks. 
%Models from the library may be used in multiple benchmarks; for example,
%there are several instances where there is both an inference problem
%and a propagation problem in the benchmarks library using the same
%model.
Each benchmark is implemented as a UM-Bridge-supporting software and
provided as a ready-to-run container. We offer three types: ``Models''
implement a numerical (forward) model of a physical phenomenon. Many
expose a number of configuration options for a high level of control
of the details of the model to be solved.  ``Inference problems'' in turn define a specific inverse \gls{UQ} problem, usually in terms of a Bayesian posterior. ``Propagation benchmarks'' define a very specific forward \gls{UQ} problem to be solved for the model they implement, and are much more restricted in order to ensure comparable and reproducible results.

Each model or benchmark is accompanied by a markdown file that documents how to run the published containers, input and output dimensions, configuration options, and a mathematical description of the map (or a reference to an existing publication).
A documentation website presents the entire library in an
accessible way, drawing from these markdown files behind the
scenes. The Docker containers \cite{docker} for our benchmarks are built automatically in a CI system, ensuring validity through automated testing.

%\Cref{sec:benchmarks} provides detailed descriptions of all current
%benchmarks.

\Cref{sec:models} contains detailed descriptions of all
models, \Cref{sec:inference} describes inference
benchmarks and \Cref{sec:propagation} describes propagation problems. Tables \ref{tbl:models_list}, \ref{tbl:benchmark_list}, and \ref{tbl:propagation_list}
provide brief summaries.

\subsection{Models}

%\todo[inline]{LT: I added the ``cookies problem'' line to Tables \ref{tbl:models_list} and I had to change the flag of the table environment to TBP otherwise with H it would end up after the biblio}
\begin{table}[h] %bp]
    \caption{\it List of models.}%\nopagebreak
    \label{tbl:models_list}
    \centering
    \begin{tabular}{p{0.24\columnwidth}p{0.64\columnwidth}}
    \toprule 
     \multicolumn{2}{c}{\bf Models}\\
     \midrule
         \midrule
    \bf Name & \bf Short description 
    \\ \midrule
       Euler-Bernoulli beam & Deformation of an Euler-Bernoulli beam with a spatially variable stiffness parameter. \\
       \midrule
        L2-Sea & Total resistance estimation of the DTMB 5415 destroyer-type vessel at model scale by potential flow. %& Multidisciplinary coupling with the rigid-body equation of motions; ship performance conditional to operational and geometrical parameters, including multi-fidelity based on grid discretization. \\
        \\ 
        \midrule
        Tsunami  & Propagation of the 2011 Tohoku tsunami modeled by solving the shallow water equations. %& Complex numerical model underlying the tsunami source inference benchmark.
        \\
        \midrule
        Composite material & Elastic deformation of L-shaped composite %RS part 
        with random wrinkles.
        \\
        \midrule
        Tritium desorption & Microscopic transport of tritium through fusion reactor materials using %RS the 
        Foster-McNabb equations.
        \\
        \midrule
        Agent-based disease transmission & Transmission of disease in a heterogenous population using EMOD, a stochastic agent based disease transmission model.
        \\
        \midrule
        Membrane model & Deformation of a membrane for a fixed right-hand side given the stiffness values on an $8\times 8$ grid that makes up the membrane. %& Easily implemented in other software systems.
        \\
        \midrule
       Cookies problem &  %RSA s
       Simplified thermal problem 
       with uncertain conductivity coefficient  
       in 8 circular regions (``%RSthe 
       cookies''), while it is known (and constant) elsewhere (``%RSthe 
       oven''). \\
    \bottomrule
    \end{tabular}
    %\todo[inline]{For tables, put caption at the top (but check with
    %  publisher guidelines first).}
\end{table}

The models in our library range from simple analytic functions to
complex numerical simulations (see \Cref{tbl:models_list}). In contrast to the propagation
benchmarks, models intentionally offer configuration options, allowing
users to explore model variations and to adapt them to their needs. For example, the tsunami model can be set to operate on three different refinement levels. When developing algorithms, operating on fast approximate model levels may be more suitable for quick test runs.

Mathematically, following \Cref{sec:architecture}, we consider a model to be a function $F:\mathbb{R}^n \rightarrow \mathbb{R}^m$ taking a parameter vector $\theta$ to a model output $F(\theta)$. Some models additionally offer derivatives, for example in terms of Hessian actions.

\subsection{Bayesian inference benchmarks}
Inference benchmarks define a (potentially not normalized) probability
density function $\pi:\mathbb{R}^n \rightarrow \mathbb{R}$ whose
properties are to be estimated. They are therefore
restricted to a one-dimensional output. For numerical stability, our
definition of the interface requires implementations to provide the logarithm of~$\pi$. See \Cref{tbl:benchmark_list} for a list of inference problems that are currently part of the library.

Some of the benchmarks define a Bayesian inference problem based on a model and fixed data $y$,
\[ \pi_{\text{posterior}}(\theta) \propto \pi_\text{prior}(\theta) L(y | \theta) , \]
where the likelihood $L(y | \theta)$ relates observed data to model output,
and is itself based on the forward model $F : \mathbb{R}^n
\rightarrow \mathbb{R}^m$ defined by the benchmark. $\pi_\text{prior}$ denotes the prior probability distribution.
Some UQ tools specialized in Bayesian inference may require access
to the individual factors of Bayes' formula above, and
possibly the model output itself. For this reason, our interface
also defines functions to query those components individually.

%\todo[inline]{LT: how come Table \ref{tbl:benchmark_list} benchmarks is not in table environment?}
% You cannot have page breaks in tables, so we have gone for this solution to enable that.

\begin{table}[h!] %bp]
    \caption{\it List of inference benchmarks.}%\nopagebreak
    \label{tbl:benchmark_list}
    \centering
    \begin{tabular}{p{0.24\columnwidth}p{0.64\columnwidth}}
    \toprule 
     \multicolumn{2}{c}{\bf Inference benchmarks}\\
     \midrule
         \midrule
    \bf Name & \bf Short description \\
    \midrule
    Analytic densities  & Infers the PDF of various analytic functions. \\
    \midrule
      Membrane model &
      Infer the PDF of stiffness values of a membrane from measured deformation data.
      %& Represents typical cases for elliptic PDE-based parameter estimation problems.
      \\
      \midrule
      Tritium diffusion posterior & Compute the posterior density of the input parameters given by experimental data.
      \\
      \midrule
      Disease transmission model &
      Agent-based model of disease transmission in an entirely
      susceptible population with correlation between disease
      acquisition and transmission. The aim %RSprovided problem 
      is to infer the disease parameters that result in 40\% of the population infected at the end of the outbreak.
      %& This is a stochastic (e.g., noisy) model.
      \\
      \midrule
      Deconvolution problem & Infer the posterior distribution for a 1D deconvolution problem with Gaussian likelihood and four different prior distributions, configurable via a parameter.
      \\
      \midrule 
      Computed tomography & Compute a posterior distribution for a 2D X-ray CT image reconstruction problem problem, with a Gaussian noise distribution.
      \\
      \midrule
      Inverse heat equation &  Evaluate the posterior distribution for a 1D inverse heat equation with Gaussian likelihood and Karhunen-Lo{\`e}ve parameterization of the uncertain coefficient.
      \\
      \midrule
      Beam inference & Bayesian inverse problem for characterizing the stiffness in an Euler-Bernoulli beam given observations of the beam displacement with a prescribed load.
      \\
      \midrule
      Tsunami Source  & Infer parameters describing the initial displacements leading to the tsunami from the data of two available buoys located near the Japanese coast  
      \\
      \midrule
      Poisson & Estimate a spatially varying diffusion coefficient in an elliptic PDE given limited noisy observations of the PDE solution. 
      \\
      \midrule
      p-Poisson & Estimate a two dimensional flux boundary condition for a nonlinear $p$-Poisson PDE in %RS on the bottom of 
      a three dimensional domain.
      \\
      \bottomrule
    \end{tabular}
\end{table}

\subsection{Uncertainty propagation benchmarks}

We construct benchmark problems like the models above but intentionally restrict them to the minimum necessary options in order to ensure reproducibility. Some benchmarks in the implementation are in fact more restrictive versions of the model codes above.

%\subsubsection{The library of benchmarks and models}\label{sec:library}

%\todo[inline]{The sub-sections above are ordered differently from the
%  tables and the sections in the appendix. Make this consistent.}

\begin{table}[h!]
    \caption{\it List of propagation benchmarks.}
\label{tbl:propagation_list}
    \centering
    \begin{tabular}{p{0.24\columnwidth}p{0.64\columnwidth}}\\%p{0.39\columnwidth}}
    \toprule
     \multicolumn{2}{c}{\bf Propagation benchmarks}\\
     \midrule
    \midrule
    \bf Name & \bf Short description \\%& \bf Distinguishing Feature  \\
    \midrule
    Euler-Bernoulli beam & Modelling the effect of uncertain material parameters on the displacement of an Euler-Bernoulli beam with a prescribed load.\\
    \midrule
    Genz & Multi-dimensional functions for testing quadrature and surrogate methods.\\  %& The input dimensionality, anisotropy and regularity of the functions can be varied to test strength and weaknesses of adaptive methods.\\
    \midrule
      L2-Sea UQ &  Forward UQ of the total resistance in calm water conditional to operational and geometrical uncertain parameters.\\ %& Dimensionality of the problem, fidelity level, and hierarchy can be varied to test UQ methods scalability. \\
    \midrule
      Cookies problem &  Forward UQ of average temperature in a corner of the ``oven'' \\ %& Dimensionality of the problem, fidelity level, and hierarchy can be varied to test UQ methods scalability. \\
      
    \bottomrule
    \end{tabular}
\end{table}

The goal of uncertainty propagation is to determine the effect of
uncertainties in model parameters on model outputs. The
parameter $\theta$ is specified to be a random variable with a distribution
documented in the markdown file. The solution of a propagation
benchmark is then an approximation to the distribution of $F(\theta)$,
or an estimate of some derived quantity like
$\mathbb{E}[G(F(\theta))]$ where $G$ is an ``output functional''. \Cref{tbl:propagation_list} shows the currently available propagation benchmarks.

\subsection{Optimization benchmarks}
UM-Bridge's interface is fundamentally agnostic to what algorithms do
with what the interface provides. This makes it possible to also use
UM-Bridge for other fields like black-box optimization. We have included an initial example
of an optimization benchmark in Table \ref{tbl:optimization_list}
and \Cref{sec:optimisation}.

\begin{table}[h!]
    \caption{\it List of optimization benchmarks.}
\label{tbl:optimization_list}
    \centering
    \begin{tabular}{p{0.24\columnwidth}p{0.64\columnwidth}}%p{0.39\columnwidth}}
    \toprule
     \multicolumn{2}{c}{\bf Optimization benchmarks}\\
         \midrule
    \midrule
    \bf Name & \bf Short description \\ %& \bf Distinguishing Feature  \\
    \midrule
      L2-Sea OPT &  Constraint deterministic global optimization problem for the total resistance reduction in calm water at fixed speed.\\% & Dimensionality of the problem, fidelity level, and hierarchy can be varied to test optimization methods scalability. \\
     \bottomrule
    \end{tabular}
\end{table}

\section{Applications}
\label{sec:applications}
The following sections present some applications of uncertainty
quantification to real-world problems that utilize UM-Bridge and demonstrate at the same time aspects of the work presented in the main
part of the paper. Specifically, \Cref{sec:application_sparse_grids} demonstrates programming language independence by connecting an existing Matlab \gls{UQ} code to a number of model instances written in Fortran. 
\Cref{sec:application_qmc} demonstrates parallel runs of a highly complex numerical solver in a \gls{UQ} application, greatly benefiting from portable models and separation of concerns between model and \gls{UQ} experts.
Finally, \Cref{sec:application_mlda} demonstrates a complex \gls{UQ} application combining fast model approximations running on a workstation and transparently offloaded model runs on a cloud cluster of 2,800 processor cores. Models and \gls{UQ} problems used in \Cref{sec:application_sparse_grids} and \Cref{sec:application_mlda} are available as part of the benchmark library.

\subsection{Accelerating a naval engineering application}
\label{sec:application_sparse_grids}

\subsubsection{Problem description}

% \todo[inline]{LT: this example is the only one with code. Is it intentional? If so, I'd say it explicitely here too. 
% I mean, I know that the purpose of this example is to show language independence, but I guess that in principle
% a reader could be content enough with us saying ``the server is in fortran, the client is in Matlab'' without
% actually showing the code, or viceversa there might be code aspects to show also in the other examples? 
% Unless you really only want to make the point that the code is super easy (but I guess easy/hard depends 
% on the reader) and then once is enough. 
% \\[4pt]
% On the other hand, this snippet is convienent to have when we discuss the forward UQ benchmark so I don't know, 
% let's keep it maybe.
% \\[4pt]
% The snippet to launch the container seems irrelevant here anyway, especially if we 
% add an example earlier on in the Method section. }

In this section we focus on a forward UQ application in the context of 
naval engineering, which we publish as part of the benchmark library 
(see \Cref{sec:l2sea_model} for more details).
The goal is to compute the \gls{pdf} of the resistance $R$ to advancement of a boat, 
i.e., the naval equivalent of the drag force for airplanes.
The boat is advancing in calm water, under uncertain
Froude number $\Frou$ (a dimensionless number proportional to the navigation speed)
and draft $\Dr$ (immersed portion of the hull, directly proportional to the payload).
That is, the set of uncertain inputs is $\theta = (\Frou,\Dr)$, which 
are modeled as follows: 
Froude is a unimodal triangular random variable with support over 
$[\Frou_{\text{a}}, \Frou_{\text{b}}]= [0.25, 0.41]$, i.e.,
\begin{equation}\label{eq:triang-pdf}
\pi_{\Frou}(t) = 
\frac{2}{(\Frou_{\text{b}}-\Frou_{\text{a}})^2} \left(\Frou_{\text{b}}-t\right),    
\end{equation}
while draft is a beta random variable with support over $[\Dr_{\text{a}},\Dr_{\text{b}}]=[-6.776, -5.544]$
and shape parameters $\alpha=10,\beta=10$, i.e.,%$D \sim Beta(D_a,D_b,\alpha,\beta)$, i.e.
\begin{align}
\pi_{\Dr}(t) = & \frac{\Gamma(\alpha+\beta+2)}{\Gamma(\alpha+1)\Gamma(\beta+1)} \times \ldots \label{eq:beta-pdf} \\
& (\Dr_{\text{b}}-\Dr_{\text{a}})^{\alpha+\beta+1}(t-\Dr_{\text{a}})^\alpha(\Dr_{\text{b}}-t)^\beta. \nonumber
\end{align}
%\[
%\pi_{\Dr}(x) = \frac{\Gamma(\alpha+\beta+2)}{\Gamma(\alpha+1)\Gamma(\beta+1)} 
%(\Dr_{\text{max}}-\Dr_{\text{min}})^{\alpha+\beta+1}(x-\Dr_{\text{min}})^\alpha(\Dr_{\text{max}}-x)^\beta
%\]

The computation of $R$ for fixed $\Frou,\Dr$,
-- i.e., the evaluation of the response function $R = R(\theta)$ -- %$R = R(\Frou,\Dr)$ --
is performed by the L2-Sea model \cite{serani:l2sea} which is written
in Fortran and is available as a container from the UM-Bridge benchmark library, 
having being wrapped in an UM-Bridge Python server.
%We couple it via an UM-Bridge wrapper.
%\todo[inline]{LT: I think we need to clarify here that the previous sentence means that s
%ince servers are not available in Fortran we use a python server and embed the call to the
%Fortran executable there}

In order to showcase the ease of parallelizing an existing application by using UM-Bridge, we will show the required command and code snippets in this section. This model can be run locally via the following docker command: 

\begin{lstlisting}
docker run -it -p 4242:4242 linusseelinger/model-l2-sea
\end{lstlisting}

\subsubsection{UQ workflow}

\begin{figure}
    \centering
    \includegraphics[trim={100 60 100 40},clip,width=\linewidth]{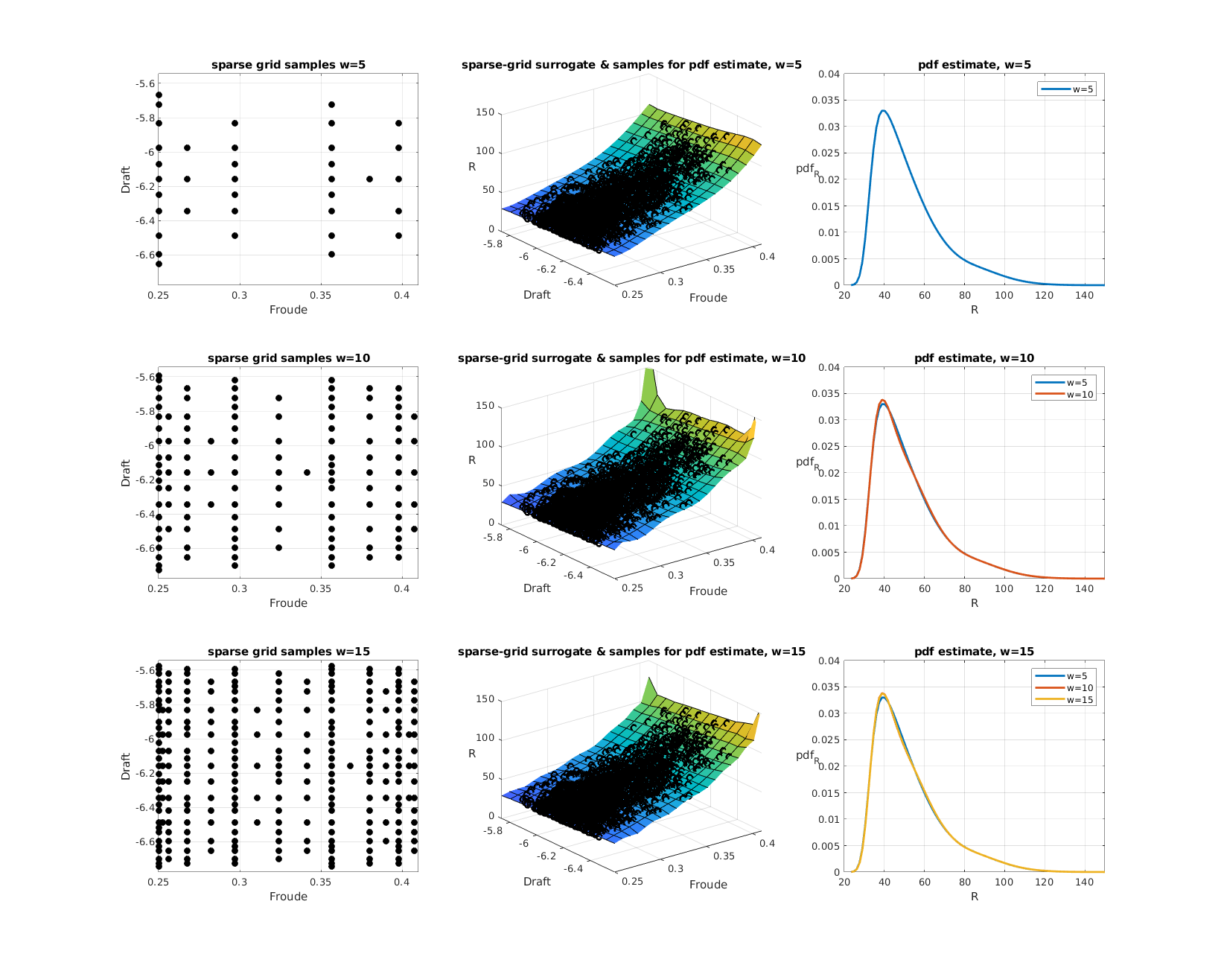}
    \caption{\it Results for the UQ workflow on the L2-Sea model for three 
    sparse-grid levels $w$ (one row for each $w$). $w$ controls the
    number of evaluation points in a sparse grid. Left column: 
    Sparse grids covering the space of possible values of
    $\theta = [\Frou,\Dr]$. Center column: Sparse-grid surrogate models and their
    evaluations at the random sample points used for computing
    the \gls{pdf} of $R$. Right column: Resulting \gls{pdf} of $R$
    for all $w$ up to the current one.}
    \label{fig:results_Lorenzo}
\end{figure}

To compute the probability density function of $R$ we proceed in two steps: 
\begin{enumerate}
    \item We create a surrogate model for the response function $R = R(\theta)$, % $R = R(\Frou,\Dr)$, 
    i.e., an approximation of the actual function based on a limited 
    number of (judiciously chosen) evaluations of $R = R(\theta)$.% $R = R(\Frou,\Dr)$,  
    
    \item We generate a large sample of values $\theta_i=(\Frou_i,\Dr_i)$ %$\Frou,\Dr$ 
    according to their \gls{pdf}.
    We then evaluate the surrogate model for each element of the sample
    (which is considerably cheaper than evaluating the full model $R(\theta)$, % $R(\Frou,\Dr)$, 
    but yields only approximate results), 
    and then use the corresponding values $R_i$ to compute an 
    approximation of the \gls{pdf} of $R$ by a standard kernel density
    method \cite{rosenblatt:kde} (specifically, in this application we use gaussian kernels and automatic selection
    of bandwidth, \cite{bowman1997applied}).
\end{enumerate}

To build the surrogate model, we use the sparse-grids method, and in particular the implementation provided by \gls{sgmk}. This method
requires sampling the space of feasible values of $\theta = (\Frou,\Dr)$ 
with a structured non-Cartesian strategy, depending on the \gls{pdf} of the two uncertain parameters, 
see the left column of \Cref{fig:results_Lorenzo}. The surrogate model is constructed as a sum of certain interpolating polynomials, each based on a subset of samples. The surrogate models are reported in the center column \Cref{fig:results_Lorenzo}. The overshoots in the corners of the domain happen in regions of the parameter space of zero probability; neither full model evaluations nor samples to approximate the \gls{pdf} of $R$ are placed there.

\gls{sgmk} provides ready-to-use functions to generate sparse grids according to several types of 
random variables (uniform, normal, exponential, beta, gamma, triangular);  
once the grid is created, it is a simple matter of looping through its points
and calling the L2-Sea solver for each one using the point coordinates as values for the inputs $\theta$. %$\Frou,\Dr$. 
In \gls{sgmk} this is as easy as the following snippet:

\begin{lstlisting}[]
% uri of L2-Sea solver
uri    = 'http://104.199.68.148'; 
model  = HTTPModel(uri,'benchmark_UQ');

% setting up L2-Sea configuration
fid = 3;
config = struct('fidelity',fid);

% wrap UM-Bridge call - theta must be a row vector
R = @(theta) model.evaluate(theta(:)',config);

% define nodes for Froude and Draft
Fr_a=0.25;   Fr_b=0.41;
knots_Fr = @(n) knots_triangular_leja(n,Fr_a,Fr_b);
lev2knots_Fr = @lev2knots_lin;
  
T_a=-6.776; T_b=-5.544; alpha=10; beta=10;
knots_T = @(n) knots_beta_leja(n,alpha,beta,T_a,T_b,'sym_line','on_file');
lev2knots_T = @lev2knots_2step;

% build sparse grid
N=2; w=5;
S = create_sparse_grid(N,w,{knots_Fr,knots_T},{lev2knots_Fr,lev2knots_T});
Sr = reduce_sparse_grid(S);

% call L2-Sea on each point (supports parallel evaluations, some omissis for brevity)
R_values = evaluate_on_sparse_grid(f,Sr,...);
\end{lstlisting}

% \begin{lstlisting}[]
% % uri of L2-Sea solver
% uri    = 'http://104.199.68.148'; 
% model  = HTTPModel(uri,'forward');

% % setting up L2-Sea configuration
% fid = 3;
% config = struct('fidelity',fid,'sinkoff','y','trimoff','y');

% % L2-Sea actually takes 16 inputs but we use only the first two and set the others to zero. Result must be row
% inputs = @(theta) [theta(:)' zeros(1,14)];

% % wrap UM-Bridge call 
% R = @(theta) model.evaluate(inputs(theta),config);

% % define nodes for Froude and Draft
% Fr_a=0.25;   Fr_b=0.41;
% knots_Fr = @(n) knots_triangular_leja(n,Fr_a,Fr_b);
% lev2knots_Fr = @lev2knots_lin;
  
% T_a=-6.776; T_b=-5.544; alpha=10; beta=10;
% knots_T = @(n) knots_beta_leja(n,alpha,beta,T_a,T_b,'sym_line','on_file');
% lev2knots_T = @lev2knots_2step;

% % build sparse grid
% N=2; w=5;
% S = create_sparse_grid(N,w,{knots_Fr,knots_T},{lev2knots_Fr,lev2knots_T});
% Sr = reduce_sparse_grid(S);

% % call L2-Sea on each point (supports parallel evaluations, some omissis for brevity)
% R_values = evaluate_on_sparse_grid(f,Sr,...);
% \end{lstlisting}

Once we have obtained a characterization of the surrogate model in
the form of \lstinline{R_values}, we generate a random sample of values of $\theta_i=(\Frou_i,\Dr_i)$
according to their \gls{pdf} by rejection sampling \cite{casella2004generalized}, and  
evaluate the surrogate model on such values with a straightforward call
to a \gls{sgmk} function: 
\begin{lstlisting}
surr_evals=interpolate_on_sparse_grid(S,Sr,R_values,random_sample);
\end{lstlisting}
These evaluations are shown as black dots in the center column of \Cref{fig:results_Lorenzo}.
These surrogate model evaluations are used as input to the \texttt{ksdensity} estimator provided by Matlab to compute an approximation of the \gls{pdf} of $R$.
% \textcolor{red}{LT: I have simplified the snippet here, but we could actually just remove it}:
%\begin{lstlisting}
%[ksd_pdf,ksd_points]=ksdensity(surrogate_evals,'support','positive','Bandwidth',0.1);
%\end{lstlisting}
%\begin{lstlisting}
%ksdensity(surr_evals,'support','positive');
%\end{lstlisting}

Results are reported in the right column of \Cref{fig:results_Lorenzo}, and as expected, 
the estimated \gls{pdf} stabilizes as we add more points to the 
sparse grid and the surrogate model becomes more reliable. The three rows are
obtained repeating the snippets above three times, increasing the so-called 
\textit{sparse grid level} \lstinline{w}, which is 
an integer value  that controls how many points are to 
be used in the sparse grid procedure, specifically $w=5,10,15$, corresponding
to $36,121,256$ points. Note that the three sparse grids produced are 
nested, i.e., they are a sub-set of one another, so that in total only $256$ calls to L2-Sea are needed.
 \gls{sgmk} is able to take advantage of this and only evaluate at the 
\textit{new} sparse-grid points in each grid level.%, i.e., it will call L2-Sea $36, 85, 135$ times.

\subsubsection{Computational aspects}
In order to allow parallel model evaluations, 48 instances of the
model were run on \texttt{c2d-high-cpu} nodes of \gls{GKE} using the
kubernetes configuration in \Cref{sec:kubernetes}. For the test case
presented here, one evaluation of the L2-Sea model takes about 30-35s
on a single physical CPU core. We disable \gls{SMT} on \gls{GKE} to
ensure stability in L2-Sea run times.

On the \gls{UQ} side, \gls{sgmk} was run on a regular laptop, connecting to the cluster. Due to UM-Bridge, the \gls{sgmk} Matlab code can directly be coupled to the Fortran model code. Further, no modification to \gls{sgmk} is necessary in order to issue parallel evaluation requests: The call \lstinline{R_values = evaluate_on_sparse_grid(f,Sr,...)} in \gls{sgmk}
is internally already using Matlab's \lstinline{parfor} in order to loop over parameters to be evaluated. By opening a parallel session in Matlab with $48$ workers 
(via the \texttt{parpool} command), the very same code executes the requests in parallel, and the cluster transparently distributes requests across model instances.

Using this approach, we obtained a total run time of around $290$
seconds instead of the sequentially expected
$30 \times 256 = 7680$ seconds, corresponding to a speedup of more than $26$.
Three factors affect the speedup: The evaluations are split in three batches, none of the batches having cardinality equal to a multiple of the $48$ parallel model instances; the CPU-time for evaluating the L2-Sea model at different values of $\theta_i$ is not identical; and finally a minor overhead in the \gls{UQ} method itself.

Apart from the limited number of evaluations needed for the specific application, the main limit in scalability we observe here is Matlab's memory consumption: approximately 500 MB allocated per worker, regardless of the actual workload. This could be improved by sending batches of requests per worker, but not without modifying the \gls{sgmk} code.

%\bigskip

%\textcolor{red}{say that for the next applications you don't show the code}

\subsection{Investigating material defects in composite aero-structures}
\label{sec:application_qmc}

\subsubsection{Problem description}

\begin{figure*}[t!]
    \centering
    \includegraphics[width=0.78\linewidth]{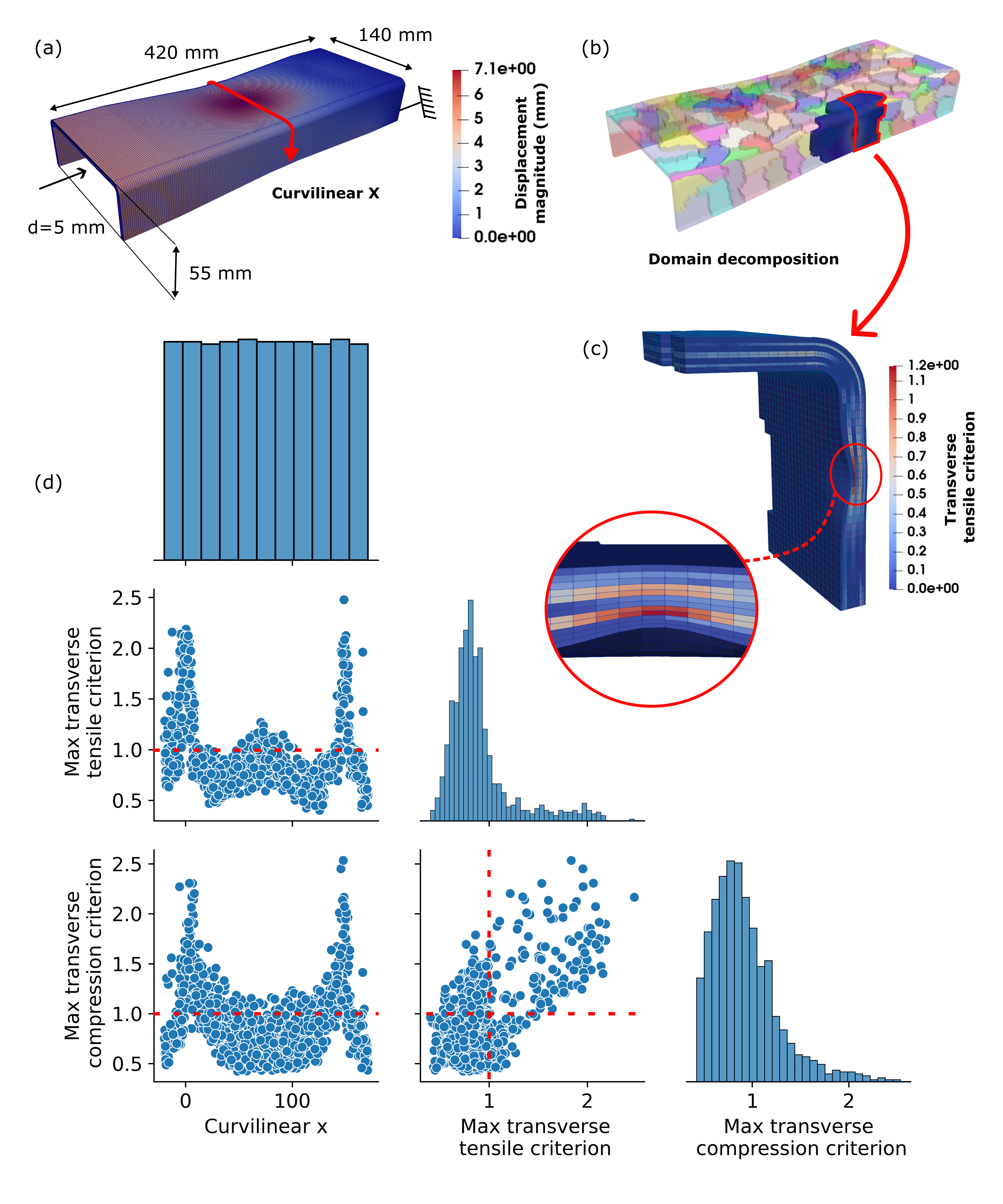}
    \caption{\it 
Offline-Online framework and results: Geometric description at macro-scale and displacement field of the pristine (offline) model (a). The domain decomposition and the targeted subdomains for a set of defect parameters (b) show the fraction of the domain that needs to be recomputed online. The update of the solution for a wrinkle defect (c) shows the local variation of the transverse tensile failure criterion at the defect location. The pairwise relationships of the defect location along the curvilinear axis X (model input) (depicted on (a)) and two damage criteria (model output): the transverse compressive and tensile failure criteria are plotted as a scatter plots (off-diagonal) and histograms (diagonal) in (d).
}
\label{fig:offline-online}
\end{figure*}

In this section we examine the effect of random material defects on structural aerospace components made out of composite laminated material. These may span several metres in length, but are still governed by a mechanical response to processes on smaller, sub-millimetre length scales. Such a multi-scale material displays an intricate link between meso-scale (ply scale) and macro-scale behaviour (global geometric features) that cannot be captured with a model describing only the macro-scale response. This link between material scales is even stronger as abnormalities or defects, such as wrinkles (out-of-plane fibre waviness)~\Cref{fig:offline-online}(c), may occur during the fabrication process. To simulate such objects, we use a multi-scale spectral generalised finite element method (MS-GFEM) \cite{Benezech:2023}. The approach constructs a coarse approximation space (i.e., a reduced order model) of the full-scale underlying composite problem by solving local generalised eigenproblems in an A-harmonic subspace. 

By operating on subdomains, this MS-GFEM approach allows for reusing the solutions of local problems in places not affected by localized changes (such as a localized defect in an aerospace part). To exploit this structure, we use an offline-online framework in which the MS-GFEM is applied to a pristine model offline (see \Cref{fig:offline-online}(a)), and only the eigenproblems in subdomains intersecting local defects are recomputed online to update the approximation space (see \Cref{fig:offline-online}(b)). Online recomputations can be performed by a single processor.

For this application, we simulate the behaviour of an aerospace component (namely a C-shaped laminated composite spar, see~\Cref{fig:offline-online}) under compression. To model the wrinkling defect, a geometric transformation is applied to the grid as depicted on~\Cref{fig:offline-online}(c). To assess the severity of the defect, the we use two damage criteria, representing the ability of the ply to fail in the direction transverse to the fibre orientation in tension and compression (see \Cref{fig:offline-online}(d)). The full-scale model consists of 1.4 million degrees of freedom. The MS-GFEM approximation reduces it to $12,819$, which corresponds to a model order reduction factor of around $106$. The method is implemented in C++ and part of dune-composites, a \gls{DUNE} \cite{DUNE} module.

\subsubsection{UQ workflow}

We define a forward \gls{UQ} problem by assuming a random defect parameter with a uniform distribution $\theta \sim \mathcal{U}([-20,171] \times [150,270] \times [0,90] \times [3.5,4.5])$. The first two components are the position of the defect along the curvilinear axes of the width and length of the part, the third is the orientation of the wrinkle, and the fourth is the size of the wrinkle. Position and size values are given in mm and the angle in degrees.

To solve the \gls{UQ} problem, we employ the Quasi-Monte Carlo method implemented in QMCPy \cite{QMCPy}, drawing 1024 samples from QMCPy's \texttt{DigitalNetB2} generator.
We observed that the main influence on the defect is the position along X (maxima located at the corners of the C section) for both criterion, as shown in~\Cref{fig:offline-online}(d).
We consider the material to be capable of initiating failure mechanisms under load when any criterion reaches $1$ (the threshold is represented as a dashed red line in the figure). We observed a different distribution for both criterion with a probability of failure of $20.41 \%$ for the tensile criterion and $32.42\%$ for the compressive criterion. This is due to the complex interaction between the global geometric feature and the local material behaviour (the laminated stacking sequence is shown on the right-hand side of \Cref{fig:l2sea_results_section}).
This interaction is not trivial and cannot be anticipated by looking only at the pristine output fields. This methodology is ideal and necessary to conduct an efficient parametric study of defects in large composite parts.

\subsubsection{Computational aspects}

We conducted the numerical experiment by running the model container on a 12-core workstation, which was outfitted with a simple k3s installation in order to support the kubernetes setup of \Cref{sec:kubernetes}. QMCPy was run on the same system, with QMCPy's UM-Bridge integration set to 12 parallel evaluations. QMCPy's UM-Bridge integration then issues parallel requests through Python's \texttt{multiprocessing} framework, which the kubernetes setup in turn processes.

Online model evaluations on a single core average at around 57 minutes of run time, with most runs between 30 and 90 minutes. The main reason for this variability is that the number of subdomains that need to be recomputed varies depending on defect location. We observe a total run time of 81.2 hours for 1024 \gls{QMC} samples on the 12-core workstation.

The preparatory offline run computing the low-dimensional basis on the entire domain was conducted on the Hamilton HPC Service of Durham University, UK. On 128 processor cores it took 8min04s. Compared to computing such a full MS-GFEM solution each time, the online method contributes a speedup of roughly 18 in addition to the \gls{UQ} method's parallelization.

Containerization turned out crucial in this application: The MS-GFEM method is complex to set up and multiple instances running on the same system might lead to conflicts, which is alleviated by containerization. In addition, sharing model containers between collaborators greatly accelerated development.

\subsection{Multilevel Delayed Acceptance for tsunami source inversion}
\label{sec:application_mlda}

\subsubsection{Problem description}

We solve a Bayesian inference problem based on a tsunami model, which we make available in the benchmark library (see \Cref{sec:inference:tsunami}). It models the propagation of the 2011 Tohoku tsunami by solving the
shallow water equations with wetting and drying. 
For the numerical solution of the PDE, we apply an ADER-DG method
implemented in the ExaHyPE framework \cite{Reinarz2020,
Rannabauer2018}.
Details on the model and its discretization can be found
in \cite{Seelinger2021}, and include a model with a smoothed bathymetry data incorporating wetting and drying through the use of a finite volume subcell limiter with polynomial order $2$ and a total of $1.7 \cdot 10^5$ degrees of freedom ($2187$ spatial); and
a model using fully resolved bathymetry data and a limited ADER DG scheme of polynomial degree $2$ with a total of $1.7\cdot 10^7$ degrees of freedom ($6561$ spatial).
Bathymetry data was obtained from
GEBCO \url{https://www.gebco.net/data_and_products/gridded_bathymetry_data/}.

\subsubsection{UQ workflow}

The aim is to obtain information about the parameters describing the initial
displacements in bathymetry leading to the tsunami from the data of
two DART buoys located near the Japanese coast. (The data for DART buoys 21418 and 21419 was obtained from NDBC \url{https://www.ndbc.noaa.gov/}.)
The posterior distribution of model inputs, i.e. the source location
of the tsunami, was sampled with the Multilevel Delayed Acceptance
(MLDA) MCMC algorithm \cite{Lykkegaard_MLDA, lykkegaard_multilevel_2023} using the open-source Python package tinyDA \cite{tinyDA}. MLDA is a scalable MCMC algorithm, which can broadly be understood as a hybrid of Delayed Acceptance (DA) MCMC \cite{christen_markov_2005} and Multilevel MCMC \cite{dodwell_multilevel_2015}. The MLDA algorithm works by recursively applying DA to a model hierarchy of arbitrary depth, see \cite{lykkegaard_multilevel_2023}.

%\begin{figure}[htbp]
%    \centering
%    \includegraphics[width=\linewidth]{mlda-concept.pdf}
%    \caption{\it The MLDA algorithm. Each level $\ell>0$ above the coarsest is sampled by recursively employing Delayed Acceptance (DA) MCMC. On the coarsest level ($\ell=0$), any Metropolis-Hasting algorithm can be used.}
%    \todo[inline]{I don't think this figure is useful for the current paper -- this paper isn't about MLDA, but about UM-Bridge, after all. But if you need to include it: Perhaps center the second figure under the first, the same for the third, and put the fourth under, rather than beside, the third?}
%    \label{fig:mlda}
%\end{figure}

We used a three-level model hierarchy consisting of the fully resolved model on the finest level, the smoothed model on the intermediate level and a Gaussian Process emulator on the coarsest level. The \gls{GP} emulator was trained on $1024$ low-discrepancy samples of the model response from the smoothed model. The \gls{GP} emulator employed a constant mean function, a Matérn-$\frac{5}{2}$ covariance function with Automatic Relevance Determination (ARD) and a noise-free Gaussian likelihood. The \gls{GP} hyperparameters were optimized using Type-II Maximum Likelihood Estimation \cite{rasmussen_gaussian_2006}. 

\subsubsection{Computational aspects}

\begin{figure}
    \centering
    \includegraphics[width=.5\linewidth]{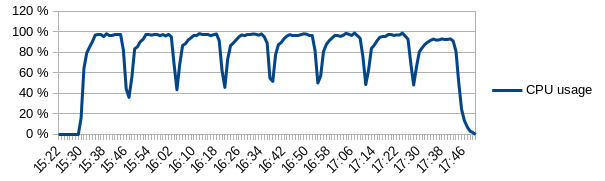}
    \caption{\it CPU usage on \gls{GKE} kubernetes cluster during MLDA
    run for the tsunami model. Intermittent drops are caused by runs of the intermediate model, which require fewer resources.}
    \label{fig:mlda_cpu_usage_k8s}
\end{figure}

The model was run using the kubernetes setup of \Cref{sec:kubernetes} on \gls{GKE} with 100 \texttt{c2d-highcpu-56} nodes running 100 instances of the model container. In total, $2800$ physical CPU cores were used. Each model instance was run on one \texttt{c2d-highcpu-56} node exploiting the Intel TBB parallelization capabilities of ExaHyPE \cite{Reinarz2020}.

Parallelization of the \gls{UQ} method was achieved through $100$ independent MLDA samplers. tinyDA allows running multiple samplers by internally making use of Ray \cite{liaw2018tune}, a Python library for distributed computing. The \gls{UQ} software was run on a workstation, connecting to the cluster via tinyDA's UM-Bridge integration. Gaussian process evaluations were conducted directly on the workstation due to their low computational cost, while model evaluations were offloaded to the cluster.

Overall run time was 2h20min, with very effective use of the cluster as indicated by its CPU utilization in \Cref{fig:mlda_cpu_usage_k8s}. For parameter $(0,0)$, the smoothed model takes 1m1s to evaluate, while the fully resolved model takes 15m4s. We computed a total of 800 evaluations of the fully resolved model and 1400 of the smoothed model. Disregarding the additional inexpensive \gls{GP} evaluations, we obtain a parallel speedup of 96.38, giving a close to perfect speedup for 100 MLDA chains.

\section{Discussion} % How does it fit into scientific landscape, what is the impact

UM-Bridge fills a major gap in the \gls{UQ} software ecosystem: There
is currently no common platform linking arbitrary \gls{UQ} algorithms and model software, let alone general-purpose cloud and \gls{HPC} support or language-independent benchmarks.

\begin{table*}[t]
\centering
\caption{\it Exemplar list of \gls{UQ} software, describing which type of connection between \gls{UQ} and model is supported (multiplicity): one particular \gls{UQ} package to one model (one-to-one), one particular \gls{UQ} package to $M$ different models (one-to-many), or $N$ UQ codes to $M$ models (many-to many). Further, the table describes whether the model can be implemented in arbitrary languages (model language agnostic), how the interface is realised, e.g. via reading/writing files or in software (technology) and which forms of \gls{HPC} support the package provides (scalability).}
\label{tbl:existpack}
\begin{tabular}{p{3.6cm}p{2.3cm}p{2cm}p{2.8cm}p{3cm}}
\toprule
\bf Package & \multicolumn{3}{c}{\bf Link} &  \bf Scalability\\
\midrule
        & \bf Multiplicity & \bf Model language agnostic & \bf Technology & \\
        \midrule\midrule
hIPPYlib-MUQ \cite{kim2022hippylibmuq} & one-to-one & No & C++, Python & Single node \\
MUQ \cite{MUQ}    & one-to-many & No & C++, Python & HPC (MPI) \\
PyMC \cite{pyMC3}, QMCPy \cite{QMCPy} & one-to-many & No & Python & Single node \\
Dakota \cite{dakota} & one-to-many & Yes & C++, Files & HPC (SLURM) \\
lagun \cite{lagun}  & one-to-many & Yes & Files & HPC (SLURM) \\
UQLab \cite{UQLab}  & one-to-many & Yes & Matlab, files & HPC (SLURM)\\
UM-Bridge \cite{UMBridgeSoftwareJOSS}  & many-to-many & Yes & C++, Julia, Matlab, Python, R & HPC (SLURM, PBS), Kubernetes \\
\bottomrule
\hline
\end{tabular}
%\todo[inline]{Complete the caption text. RS: This table isn't referenced in the text. Add a reference to it somewhere. Also, can you make sure this float doesn't create such a bad layout for this page, wasting lots of space.}
\end{table*}

\begin{comment}
In practice, most publications that couple UQ methods and complex models
do so via input and output files: The \gls{UQ} component generates a
set of parameters, passes it to the modelling component via some kind
of input file, executes the model, and later processes the
outcomes. This approach precludes the use of more complex UQ
approaches that perhaps require more information from the simulator;
it also prevents systematic comparisons between methods which are often
only demonstrated on trivial test cases (say a superposition of
Gaussians) and a single complex model that happens to be available to
the authors of the paper, but is
otherwise inaccessible to the broader community.
\end{comment}

In practice, most publications that couple advanced UQ methods and complex models
do so in one of two ways: Either they directly embed the model in the
UQ code, which requires compatible programming languages and
entails unwieldy software stacks. Or, the \gls{UQ} component
passes parameters to the modelling component via some kind
of input file, executes the model software, and then processes the
outcomes.

Both approaches do not lend themselves to containerized models, and the parallelization infrastructure for \gls{HPC} support remains up to each individual \gls{UQ} code.
Since the interfaces are incompatible across UQ packages,
they lead to user lock-in: It is impractical for models to
implement interfaces to multiple UQ packages and it is often
difficult to translate data and data structures between programming
languages.
All of this further precludes systematic comparisons between UQ methods, which are often
only demonstrated on trivial test cases (say a superposition of
Gaussians) and a single complex model that is
inaccessible to the broader community.

UM-Bridge addresses those problems by providing a uniform and
language-independent set of interfaces. As a consequence, it is able
to couple $N$ UQ codes and $M$ model applications using only $N+M$
implementations of the coupled interface, rather than the $N\cdot M$
modifications otherwise necessary. In addition, UM-Bridge provides
generic parallelization platforms that serve \emph{any} \gls{UQ} code and \emph{any} model, and
it addresses the issue of different languages and complex, perhaps conflicting dependencies, build systems, or parallelization schemes. \Cref{tbl:existpack} gives an overview of how commonly used \gls{UQ} packages
realize the interface between \gls{UQ} and model and how that compares to UM-Bridge.
%\todo[inline]{LT: I think \Cref{tbl:existpack} and the sentence %introducing it 
%are a bit critical. The way the sentence is written now seems to suggest that
%the software listed here exhaustively list commonly used UQ packages, and
%the fact that the first column of the table is the list of packages further
%stresses that point. What if we twist it a bit and change the order of the
%columns such that the first one is the type of multiplicity (one-to-one, one-to-many, many-to-many) and then in the second column becomes ``\textit{examples}
%of such multiplicity''? Then we change the text and the caption accordingly. 
%By the way, one example we might want to add is the software Tasmanian, which 
%uses libensamble to interface with external solvers}
%AR: Clarified that this is not an exhaustive list, but rather representative examples of the types of connections packages tend to have

In building UM-Bridge, we were inspired by projects
in other domains
that aim at coupling with modeling software. For example, the
Functional Mockup Interface (FMI) \cite{blochwitz_functional_2011},
prominent in commercial codes like ANSYS and Matlab, standardizes this
process by encapsulating models as Functional Mockup Units (FMUs),
promoting interoperability between different simulation tools.
%Users typically export their models as FMUs, containing information about structure, parameters, and inputs. 
As another example, preCICE
(Precise Code Interaction Coupling Environment) \cite{preCICE}
facilitates multi-physics simulations by coupling grid interfaces
between different specialized simulation codes through the preCICE interface.

In the end, a UQ project's success hinges on whether its benefits can be
demonstrated using practical examples of relevance to the
community. To this end, we have built the UM-Bridge benchmarks library
that provides a ready made set of benchmarks that can be used to test the performance of new methods, and  which can be used as a reference for comparison with existing methods. In other areas such as numerical linear algebra \cite{Dongarra1988}, geophysics \cite{Harris2009,Harris2018,Xing2016}, HPC \cite{Fischer2020}, model order reduction \cite{morWiki}, etc., such benchmark collections are widely used. To the best of our knowledge no such collection currently exists for \gls{UQ}.

Beyond its initial focus on \gls{UQ}, the protocol can adapt to the needs of other fields through extensions. Ongoing work, e.g.,  adds an optional ``fast path'' through shared memory whenever client and server run on the same machine, enabling transfer of large amounts of data like entire random fields.

\paragraph{Limitations}
Algorithms requiring information beyond the current UM-Bridge interface can not be integrated. Also, for large data transfers or extremely fast models, the network overhead may become significant. Both limitations are somewhat mitigated by support for future protocol extensions. At the time of writing, smooth handling of model code crashes is planned but not yet supported.

%\todo[inline]{LT: what about the fact that at the moment commercial software like Abaqus and Ansys cannot be encapsulated in a container AFAIK?}
% LS: Don't think that's a limitation on our end, containers are useful but not required (e.g. the HPC setup even scales models up without containerization)

\begin{comment}
\todo[inline]{The following section reads like marketing-speech. Name
  the ``more and more UQ packages'' -- otherwise, it's all
  vaporware. Similarly, is industry connection really happening? If
  yes, provide a concrete example. I think the last paragraph is the
  only one that provides actual value.}
  LS: Condensed it down a lot. Bit tricky to mention specific packages that have not yet completed support, since we don't have a guaranteed commitment from them. Same for industry, we mainly have completed applications at digiLab, and OpenGoSim is not completed yet.
\end{comment}

\section{Outlook}

We aim to establish UM-Bridge across the \gls{UQ} community, serving as a catalyst for highly efficient \gls{UQ} methods applied to challenging problems, and acting as a link between \gls{UQ} and applications as well as between academia and industry. With growing computational resources and the advance of science, we expect growing complexity in \gls{UQ} applications, making the need for separation of concerns and advanced \gls{UQ} tools even more prominent.

\section{Acknowledgements}

D. Aristoff gratefully acknowledges support from the National Science Foundation via awards DMS-1818726 
and DMS-2111277.

W.~Bangerth was partially
supported by the National Science Foundation under award OAC-1835673
as part of the Cyberinfrastructure for Sustained Scientific Innovation (CSSI)
program; by award DMS-1821210; 
and by award EAR-1925595.

J. Bénézech research was supported by the UK Engineering and Physical Sciences Research Council (EPSRC) through the Programme Grant EP/S017038/1 “Certification of Design: Reshaping the Testing Pyramid” and made use of the Hamilton HPC Service of Durham University. The code is developed and integrated in the Distributed and Unified Numerics Environment (DUNE) https://www.dune-project.org/.

J. D. Jakeman was supported by the US Department of Energy’s Office of Advanced Scientific Computing Research program.
Sandia National Laboratories is a multi-mission laboratory managed and operated by National Technology \& Engineering Solutions of Sandia, LLC (NTESS), a wholly owned subsidiary of Honeywell International Inc., for the U.S. Department of Energy’s National Nuclear Security Administration (DOE/NNSA) under contract DE-NA0003525. This written work is authored by an employee of NTESS. The employee, not NTESS, owns the right, title and interest in and to the written work and is responsible for its contents. Any subjective views or opinions that might be expressed in the written work do not necessarily represent the views of the U.S. Government. The publisher acknowledges that the U.S. Government retains a non-exclusive, paid-up, irrevocable, world-wide license to publish or reproduce the published form of this written work or allow others to do so, for U.S. Government purposes. The DOE will provide public access to results of federally sponsored research in accordance with the DOE Public Access Plan.

A. Reinarz and L. Seelinger gratefully acknowledge support by Cristian Mezzanotte and the Google Cloud Platform research credits program. 

The work of N. A. B. Riis, J. S. Jørgensen and A.M.A. Alghamdi was supported by The Villum Foundation (grant no. 25893).

R. Scheichl and L. Seelinger gratefully acknowledge support by the state of Baden-Württemberg through bwHPC, as well as the German Research Foundation (DFG) through grant INST 35/1597-1 FUGG, as well as its Excellence Strategy EXC 2181/1 - 390900948 (the Heidelberg STRUCTURES Excellence Cluster).

L. Tamellini has been partially supported by the project 202222PACR
"Numerical approximation of uncertainty quantification problems for PDEs by multi-fidelity methods (UQ-FLY)", 
funded by European Union—NextGenerationEU.
Lorenzo Tamellini and Massimiliano Martinelli have been partially supported 
by the ICSC—Centro Nazionale di Ricerca in High Performance Computing, Big Data, and Quantum Computing, 
funded by European Union—NextGenerationEU.
% \textcolor{red}{need to add EU logos here}

M. Diez, R. Pellegrini, and A. Serani are partially supported by the Office of Naval Research through NICOP grant N62909-21-1-2042, administered by Woei-Min Lin, Elena McCarthy, and Salahuddin Ahmed of the Office of Naval Research and Office of Naval Research Global, and their work has been conducted in collaboration with the NATO task group AVT-331 on ``Goal-driven, multi-fidelity approaches for military vehicle system-level design''.

%\newpage

%\bibliographystyle{elsarticle-num} 
%\bibliography{paper_abbreviated}

%\printbibliography
%\end{refsection}

%%%%%%%%%%%%%%%%%% The following includes the appendix after the main document. Used for arxiv version.
%\begin{comment}
%\begin{refsection}
%\newrefcontext[labelprefix=S] % Add prefix to reference labels in order to differentiate from main part references

%\clearpage %enforce newpage, even after bibliography

\bibliographystyle{elsarticle-num} 
\bibliography{paper_abbreviated}

\appendix

%\section{Appendix: Applications}
%\label{sec:applications}
%\input{applications}

%\newpage
\section{Appendix: Benchmark library}
\label{sec:benchmarks}
This appendix documents all models and benchmark problems currently
part of the UM-Bridge benchmark library, using the three categories
defined in Section~\ref{sec:uq-benchmark-library}.

\FloatBarrier

\subsection{Models}\label{sec:models}
In this section we describe the models contained in the benchmark library, these 
implement a numerical (forward) model of a physical phenomenon. Each model is accompanied by a markdown file that documents how to run the published container, input and output dimensions, configuration options, and a mathematical description of the map (or a reference to an existing publication).

\subsubsection{Euler-Bernoulli beam}
This model describes the deformation of a beam with a spatially
variable stiffness parameter. Let $u(x)$ denote the vertical
deflection of the beam and $f(x)$ denote the vertical force acting on
the beam at point $x$ (positive for upwards, negative for
downwards). We assume that the displacement can be well approximated
using Euler-Bernoulli beam theory and thus satisfies the fourth-order \gls{PDE}
\begin{equation}
    \frac{\partial^2}{\partial x^2}\left[ r E(x) \frac{\partial^2 u}{\partial x^2}\right] = f(x),
\end{equation}
where  $E(x)$ is an effective stiffness and $r$ is the beam radius.
For a beam of length $L$, the cantilever boundary conditions take the form
$$u(x=0) = 0,\quad \left.\frac{\partial u}{\partial x}\right|_{x=0} = 0$$
and
$$\left.\frac{\partial^2 u}{\partial x^2}\right|_{x=L} = 0, \quad  \left.\frac{\partial^3 u}{\partial x^3}\right|_{x=L} = 0.$$

This \gls{PDE} is solved with a finite difference method and the beam
stiffness is defined at each of the (typically $N=31$) nodes in the
discretization. The displacement of the beam is also returned at these
$N$ points. The beam radius is set to $r=0.1$ and the value of $f(x)$ is fixed.

This model takes in $E(x)$ at $N$ finite difference nodes and returns the value of $u(x)$ at those nodes.   

\subsubsection{L2-Sea model}
\label{sec:l2sea_model}
This model describes the calm-water resistance $R$ of a destroyer-type vessel by potential flow. % , \textcolor{}{} see also Section~\ref{sec:application_sparse_grids}. 
Specifically, the vessel under investigation is the DTMB 5415 (at model scale), which is a widely used benchmark for towing tank experiments \cite{olivieri2001-TechRep}, CFD studies \cite{sadat2015cfd}, and hull-form optimization \cite{serani2016-AOR}, considering both deterministic \cite{grigoropoulos2017mission} and stochastic \cite{serani2021hull} formulations.

A potential flow solver is used to evaluate the hydrodynamic loads, based on the Laplace equation
\begin{equation}
    \nabla^2\phi = 0
\end{equation}
where $\phi$ is the velocity scalar potential, from which the velocity
is computed through $\mathbf{u}=\nabla\phi$. $\phi$ is computed numerically through the Dawson linearization \cite{dawson1977-NSH} of the potential flow equations, using the boundary element method \cite{landrini1996steady} (see Fig. \ref{fig:l2sea_results_section}). Finally, the total resistance $R$ is estimated as the sum of the wave and the frictional resistance: The wave resistance component is estimated by integrating the pressure distribution over the hull surface, obtained using Bernoulli's theorem
\begin{equation}
    \frac{p}{\rho} + \frac{\left(\nabla\phi\right)^2}{2}-gz = const;
\end{equation}
the frictional resistance component is estimated using a flat-plate approximation based on the local Reynolds number \cite{schlichting2000-BLT}.

The steady equilibrium involving two degrees of freedom (sinkage and
trim) is computed iteratively through the coupling between the hydrodynamic loads and the rigid-body equation of motion. 
%The model can exploit multiple grid discretization levels; further details can be found in \cite{serani:l2sea}. 

The model takes as inputs 16 parameters, i.e., $\theta \in \mathbb{R}^{16}$: 
the Froude number $\Frou$, the ship draft $\Dr$ 
(we have already introduced these quantities in Appendix \ref{sec:application_sparse_grids}), 
and $N=14$ shape modification parameters 
$\mathbf{\thetashape} = [\thetashape_1,\ldots,\thetashape_{14}]$ that change the shape of the hull of the vessel,
i.e., $\theta = (\Frou,\Dr,\mathbf{s})$.
More specifically, the shape of the hull is described by a function $\mathbf{g}$, depending on 
the Cartesian coordinates $\boldsymbol{\xi}$ and on the shape parameters $\mathbf{\thetashape}$ as follows:
\begin{equation}\label{eq:reducedspace}
    \mathbf{g}(\boldsymbol{\xi},\mathbf{\thetashape})=\mathbf{g}_0(\boldsymbol{\xi}) 
    + \boldsymbol{\gamma}(\boldsymbol{\xi},\mathbf{\thetashape})
\end{equation}
where $\mathbf{g}_0$ is the original geometry and $\boldsymbol{\gamma}$ is a shape modification vector obtained by a physics-informed design-space dimensionality reduction \cite{serani2019stochastic}
\begin{equation*}
    {\boldsymbol{\gamma}}(\boldsymbol{\xi},\mathbf{\thetashape}) = \sum_{k=1}^N \thetashape_k \boldsymbol{\psi}_k(\boldsymbol{\xi})
    \label{e:exp_gamma}
\end{equation*}
with $\boldsymbol{\psi}$ a set of orthonormal functions.
Note that the shape parameters and the associated shape modifications are organized in a hierarchical order, 
meaning that the first parameters produce larger design modifications than the last ones \cite{serani2021hull}.

%\textcolor{red}{LT: removed figures that had appeared in a previous paper} 

%(see \Cref{sec:optimisation:l2sea} for more details on their definition)}. 
We now discuss bounds for the parameters collected in $\theta$. Concerning $\Frou$,
although in principle it can be any number between 0 (excluded) and the maximum velocity
of the ship (corresponding to $\Frou=0.41$, in practice the mesh is suitable for values in the range $[0.25, 0.41]$.
$\Dr$ must be negative (otherwise the ship would be above water), but not too much (or the ship would be completely under water);
for meshing reasons again, the suitable range for $\Dr$ is $[-6.776, -5.544]$. 
Finally, the shape modification parameters $\thetashape_i$ must be chosen in the range [-1,1].

Moreover, three configuration parameters can be set to control the behavior for this model: 
\textit{fidelity} is an integer value that controls the mesh refinement, ranging from to 1 (highest resolution, default) to 7 (lowest resolution), with a refinement ratio of 2$^{0.25}$, see \cite{serani:l2sea} for more details;
% The multifidelity levels are defined by the computational grid size. Specifically, the benchmark is defined with seven grid (fidelity) levels with a refinement ratio of 2$^{0.25}$, e.g., see the finest (G1) and coarsest (G7) grids and associated numerical solutions in Fig. \ref{fig_sea:meshes} and \ref{fig_sea:GridStudy}
\textit{sinkoff} is a character that can be set to \texttt{'y'} (default) / \texttt{'n'} to respectively disable/enable sinkage secondary movements (up/down translation on the vertical axis); similarly, \textit{trimoff} can be set to \texttt{'y'} (default) / \texttt{'n'} to respectively disable/enable trim secondary movement (up/down rotation about its transverse axis, i.e. bow to stern). 

The model returns the total resistance $R$ as well as $4$ geometrical quantities 
(related to the beam, draft, and sonar dome dimensions) whose value changes when playing with the
shape parameters $\mathbf{s}$; we postpone their definition to \Cref{sec:optimisation:l2sea}.

\subsubsection{Tsunami model}
\label{sec:tsunami_model}
This model describes the propagation of the 2011 Tohoku tsunami by
solving the shallow water equations, see also Section~\ref{sec:application_mlda}. For the numerical solution of the PDE, we apply an ADER-DG method implemented in the ExaHyPE framework \cite{ExaHyPE}.

This benchmark creates a sequence of three models:
\begin{enumerate}
    \item In the first model the bathymetry is approximated only by a depth average over the entire domain and the discretisation is a DG discretisation of order $2$.
    \item In the second model the bathymetry is smoothed using a Gaussian filter and the DG discretisation is supplemented with a finite volume subcell limiter allowing for wetting and drying.
    \item In the third model the full bathymetry data is used and the same DG discretisation is used as in the second model.
\end{enumerate}

The model takes two inputs giving the location of the displacement initiating the tsunami. The model outputs four values giving the maximal wave heights and time at two buoys located off the coast.

%\todo[inline]{It would be useful to include a description of what the inputs and outputs of this model are.}

%\begin{figure}
%\centering
%   \includegraphics[width=0.3\textwidth]{Images/l0-bath.jpg}
%   \includegraphics[width=0.3\textwidth]{Images/l1-bath.jpg}
%   \includegraphics[width=0.3\textwidth]{Images/l2-bath.jpg}
%    \caption{\it The three-level tsunami test case, each point represents an accepted sample at level $l=0, 1, 2$ The dashed lines show the expected value $\mathbb{E}(Q_0)$ or $\mathbb{E}(Q_0) + \sum_l \mathbb{E}[Q_l - Q_{l-1}]$ with the point $(0,0)$ in red for reference.}
%  \label{fig:exahype_tsunami}
%\end{figure}

\subsubsection{Composite material}

This benchmark implements the 3D anisotropic linear elasticity equations for a composite part with randomised wrinkle. The maximum deflection of the part is estimated by embedding a wrinkle into a high fidelity \gls{FE} simulation using the high performance toolbox dune-composites \cite{Reinarz2018}. A simplified model for a 3D bending test was used. The curved composite parts were modelled with shortened limbs of length 10 mm. A unit moment was applied to the end of one limb using a multi-point constraint, with homogeneous Dirichlet conditions applied at the end of the opposite limb. This gives the same stress field towards the apex of the curved section as a full 3D bending test. The analysis assumes standard anisotropic 3D linear elasticity and further details on the numerical model and discretisation can be found in \cite{Reinarz2018}.

The wrinkle defect is defined by a deformation field $W:\Omega \rightarrow \mathbb R^3$ mapping a composite component from a pristine state to the defected state. 

The wrinkles are defined by the wrinkle functions
$$
  W(x,\xi) =  g_1(x_1)g_3(x_3)\sum_{i=1}^{N_w} a_i f_i(x_1,\lambda),
$$
where $g_i(x_i)$ are decay functions, $f_i(x_1,\lambda)$ are the first
$N_w$ \gls{KL} modes parameterized by the length scale $\lambda$ and
$a_i$ the amplitudes.
The wrinkles are prismatic in $x_2$, i.e., the wrinkle function  is assumed to have no $x_2$ dependency. For more details on the wrinkle representation see \cite{Sandhu2018}.

The amplitude modes and the length scale can be taken as random variables, so that the stochastic vector is defined by $\boldsymbol \theta = [a_1,a_2,\ldots,a_{N_w},\lambda]^T$.

%\todo[inline]{Should it have been $\theta$ instead of $\xi$ to denote
%the stochastic input?}

The model outputs a single scalar value, the maximum deflection of the composite part under an applied pressure of $50$ Atmospheres.

%\todo[inline]{This section defines what the inputs are, but not what
%the outputs are.}

\subsubsection{Tritium desorption}
\label{sec:tritium_model}

In this model, we consider macroscopic tritium transport processes through fusion materials using the Foster-McNabb equations, as described in \cite{hodille_macroscopic_2015, delaporte-mathurin_finite_2019} using the Achlys \cite{stephen_dixon_2021_6412090} software package, which is built on top of the \gls{FE} library MOOSE \cite{lindsay2022moose}. Specifically, the following equations are solved:
\begin{align*}
    \frac{\partial C_{m}}{\partial t} &= \nabla  \cdot \left( D \left(T \right) \nabla  C_{m} \right) - \sum_i \frac{\partial C_{t,i}}{\partial t} + S_\text{ext},\\
    \frac{\partial C_{t,i}}{\partial t} &= \nu_m \left(T\right) C_m \left(n_i - C_{t,i} \right) - \nu_i\left(T\right) C_{t,i}, \\
    \rho_m C_p \frac{\partial T}{\partial t} &= \nabla \cdot \left(k \nabla T \right).
\end{align*}
Here, $C_{m}$ represents the mobile concentration of \gls{HI}, $C_{t,i}$ represents the trapped concentration of \gls{HI} at the $i$th trap and $T$ is the temperature of the material.
Furthermore, the evolution of the extrinsic trap density $n_3$ is modelled as
%\begin{equation*}
%    \frac{dn_{3}}{dt} = (1 - r) \phi \left[ \left(1-\frac{n_3}{n_{3a,max}}\right)\eta_a \: f(x) + \left(1-\frac{n_3}{n_{3b,max}}\right)\eta_b \: \theta(x) \right]
%\end{equation*}
described in \cite{delaporte-mathurin_finite_2019}, taking into account additional trapping sites that are created as the material is damaged during implantation.

The simulation encompasses three distinct phases, namely an \textit{implantation} phase, where the material is exposed to a particle source, a \textit{resting} phase, and a \textit{desorption} phase, where the material is heated and the desorption rate is computed.

The input parameters of the forward model consist of the detrapping energy of the \gls{HI} traps, $E_1$, $E_2$, and $E_3$, and the densities of the intrinsic traps, $n_1$ and $n_2$. All other model parameters are kept fixed. The output of the model is the desorption rate in atomic fraction as a function of the temperature $T \in [300, 800]$ during the desorption phase, discretized on a grid of $500$ interpolaton points. This model supports the evaluation of the forward map.

\subsubsection{Agent-based disease transmission model}
\label{sec:emod_abm_model}

This model simulates the transmission of disease in a heterogenous population using EMOD \cite{bershteyn18}, a stochastic agent based disease transmission model.

Our simulation consists of 100,000 individuals who are susceptible to a disease that is introduced into the population with a probabilistic rate of 1 infection per day for the first 5 days. This disease has a latent period that follows a Gaussian distribution with mean 3 days and standard deviation of 0.8 days. The infectious period $P$ is assumed to also follow a Gaussian distribution with a mean of 8 days and a standard deviation of 0.8. The infectivity of the disease ($I_t$; how likely it is for an infectious individual to infect another) is assumed to follow a log-normal distribution and is determined by the simulation's $R_0$ and its variance:
$ \mu_{I_t} = \log\left(\frac{{R_0}}{\mu_{P}}\right) - 0.5\sigma_{I_t}^2 $
where 
$ \sigma_{I_t} = \log\left(\frac{\sigma_{R_0}^2}{2R_0^2} + 0.5\right). $

%One of the challenges associated with this model is that it is stochastic.  Even for the same parameters the total number of infections will differ between simulations. Furthermore, it is not guaranteed that there will always be an outbreak: sometimes, by chance, there are not enough initial infections to generate a large outbreak.  You can reduce this challenge by increasing the number of initial infections via the `daily\_import\_pressures' configuration parameter (e.g., from 1 infections per day to 10 infections per day for the first 5 days).

The likelihood of an individual to acquire and then transmit the disease is correlated. There is no waning immunity. The benchmark is fitting to an attack fraction of 0.40 with a standard deviation of 0.025. There are bounds on $R_0 > 0$; the variance of $R_0 > 0$; and the correlation between acquisition and transmission must lie between 0 and 1 (inclusive).

The input parameters are the reproductive number, its logarithmic standard deviation, and correlation between acquisition and transmission. Optional configuration parameters are a flag to fix the simulation random seed, the number of infected individuals introduced into the simulation for the first 5 days, and the epsilon parameter utilized by scipy.optimize.approx\_fprime (used to estimate the gradient). The simulation terminates when there are no longer any infected individuals (minimum run time of 50 days) and returns daily timeseries `New Infections', `Infected', `Infectious Population', `Susceptible Population', `Symptomatic Population', `Recovered Population', and `Exposed Population'.

% \todo[inline]{We need to state what the inputs and outputs are.}

\subsubsection{The membrane model}
\label{sec:models:membrane}

The ``membrane model'' describes the vertical deformation of a membrane for a fixed right hand side, as a function of the membrane's stiffness properties. This forward model is described in great detail in \cite{AristoffBangerth} as a building block for a Bayesian inverse problem.

More specifically, the model considers the vertical deflection $u(\mathbf x)$ of a membrane given the spatially variable stiffness coefficient $a(\mathbf x)$ for a known force $f(\mathbf x)=10$. The model uses a specific, prescribed discretization of the following partial differential equation to obtain $u(\mathbf x)$:
\begin{align*}
    -\nabla \cdot [a(\mathbf x) \nabla u(\mathbf x)] = f(\mathbf x),
\end{align*}
augmented by zero boundary conditions. The forward model considers as inputs $64$ values that define $a(\mathbf x)$ on an $8\times 8$ subdivision of the domain, computes the discretized deformation $u(\mathbf x)$, and
then computes an output vector of size $169$ by evaluating the deflection $u(\mathbf x)$ on a $13\times 13$ grid of points.

The detailed description in \cite{AristoffBangerth} allows the
implementation of this model in a number of programming
languages. Indeed, \cite{AristoffBangerth} provides links to a C++
implementation based on the \textsc{deal.II} library \cite{dealII94}
that is used for the containers described herein, as well as Matlab and Python implementations.

%This model is the basis for the Bayesian inverse problem benchmark described in Section \ref{sec:inverse:membrane}.

\subsubsection{The cookies problem}
\label{sec:models:cookies}

\begin{figure}
    \centering
    \includegraphics[width = .5\linewidth]{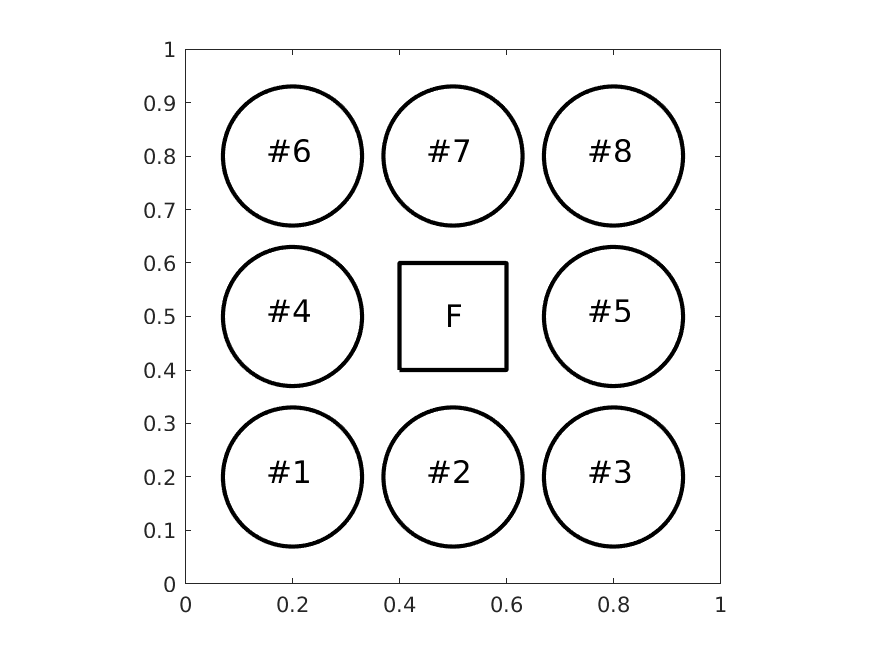}
    \caption{\it Computational domain for the cookies problem, \cref{sec:models:cookies}.}
    \label{fig:cookies-geometry}
\end{figure}

This model implements the so-called ``cookies problem'' \cite{back.nobile.eal:comparison,kressner.tobler:cookies,ballani.gras:tensor}, i.e., a simplified thermal equation in which the conductivity coefficient is uncertain in 8 circular subdomains (``the cookies''), whereas it is known (and constant) in the remaining of the domain (``the oven''), see \cref{fig:cookies-geometry}. More specifically, the temperature solves the stationary thermal equation
\begin{align*}
    -\nabla \cdot [a(\mathbf{x},\mathbf{\theta}) \nabla u(\mathbf{x}, \mathbf{\theta})] = f(\mathbf{x}),
\end{align*}
with homogeneous Dirichlet boundary conditions and forcing term defined as
\[
f(\mathbf{x}) = 
\begin{cases} 
100 &   \text{if } \,  \mathbf{x} \in F \\
0   &   \text{otherwise} 
\end{cases}
\]
where $F$ is the square $[0.4, 0.6]^2$ in the center of the domain, again see \cref{fig:cookies-geometry}. 
The 8 subdomains with uncertain diffusion coefficient (the cookies) are circles with radius 0.13 and center coordinates reported in \cref{tab:cookies_centers}.
\begin{table}[tb]
    \centering
        \begin{tabular}{ccccccccc}%{c|c|c|c|c|c|c|c|c}
        Cookie & 1   & 2   & 3   & 4   & 5   & 6   & 7   & 8  \\
        \hline
        $x$      & 0.2 & 0.5 & 0.8 & 0.2 & 0.8 & 0.2 & 0.5 & 0.8 \\
        $y$      & 0.2 & 0.2 & 0.2 & 0.5 & 0.5 & 0.8 & 0.8 & 0.8 \\
        \end{tabular}
        \caption{\it Centers of subdomains for the cookies problem, \cref{sec:models:cookies}.}
    \label{tab:cookies_centers}
\end{table}
The uncertain diffusion coefficient is defined as
\begin{equation}\label{eq:cookies-diffusion-coeff}
a = 1 + \sum_{n=1}^8 \theta_n \chi_n(\mathbf{x})
\end{equation}
where $\theta_n>-1$ and 
\[
\chi_n(\mathbf{x}) = 
\begin{cases} 
1   &   \text{inside the n-th cookie} \\ 
0   &   \text{otherwise} 
\end{cases}
\]
The output of the model is the integral of the solution over $F$, i.e. 
\begin{equation}\label{eq:cookies-qoi}
    \Psi = \int_F u(\mathbf{x}) d \mathbf{x}.
\end{equation}
The PDE is solved by an isogeometric method with splines of degree $p$ (tunable by the user, default $p=4$) and maximum continuity $p-1$ \cite{BeiraoDaVeiga:2014},
based on a quadrilateral mesh with $N_{el}\times N_{el}$ elements, with $N_{el}=100 \times k$ where $k$ is an integer value (``fidelity'') that 
can be set by the user (default $k=2$). The solver is provided by the C++ library IGATools \cite{pauletti2015igatools}. Through the config option,
the user can also set the number of CPU threads to be allocated to solving the PDE (default $1$). 
% This model only provides evaluation of the forward map.

\subsection{Inference benchmarks}\label{sec:inference}
In this section we describe the inference benchmarks contained in the benchmark library, these 
implement a specific inverse \gls{UQ} problem, usually in terms of a Bayesian posterior. 

\subsubsection{Analytic functions}
\begin{figure}
  \centering
     \includegraphics[width=0.24\linewidth]{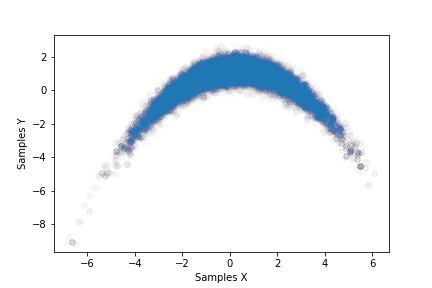}
     \includegraphics[width=0.24\linewidth]{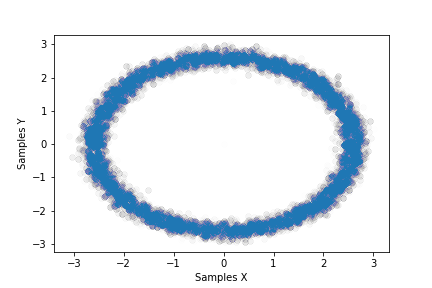}
    \includegraphics[width=0.24\linewidth]{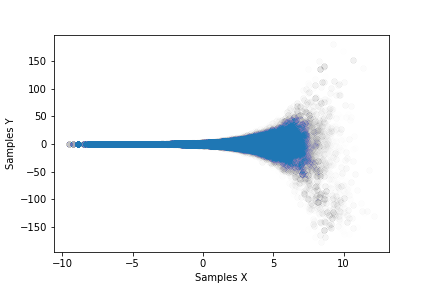}
    \includegraphics[width=0.24\linewidth]{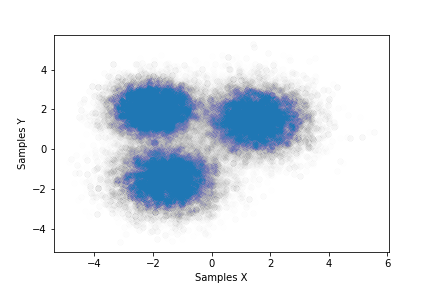}
      \caption{\it Samples drawn from the distribution defined through the \gls{pdf} $\pi$ for the analytic banana, donut, funnel and Gaussian mixture respectively.}%, and their corresponding contour.}
\label{fig:analytic}
\end{figure}

%\todo[inline]{Mikkel: Can each of the analytic functions maybe be put into paragraphs, rather than subsections.}
The benchmarks in this series of four benchmarks consist of an analytically defined \gls{pdf} $\pi : \mathbb{R}^2 \rightarrow \mathbb{R}$ resembling the shape of a banana, donut, funnel, and a Gaussian mixture, see \cref{fig:analytic}. They are based on transformed normal distributions; the variance may be adjusted for all benchmarks.

\paragraph{Analytic Banana Benchmark}
We begin with a normally distributed random variable
$$Z \sim \mathcal{N}\left(\begin{pmatrix} 0 \\ 4 \end{pmatrix}, \text{scale} \begin{pmatrix} 1.0 & 0.5\\ 0.5 & 1.0 \end{pmatrix}\right),$$
and denote its \gls{pdf} by $f_Z$.

In order to reshape the normal distribution, define a transformation $T : \mathbb{R}^2 \rightarrow \mathbb{R}^2$
$$ T(\theta) := \begin{pmatrix} \theta_1 / a \\ a \theta_2 + a b (\theta_1^2 + a^2) \end{pmatrix}. $$

Finally, the benchmark \gls{pdf} is defined as
$$ \pi(\theta) := f_Z(T(\theta)). $$
\begin{table}[t!]
  \centering
  \begin{tabular}{cccp{3cm}}
  \toprule
  Name & Type & Default & Description \\
  \midrule
  a & double & 2.0 & Transformation parameter \\
  b & double & 0.2 & Transformation parameter \\
  scale & double & 1.0 & Scaling factor applied to the underlying normal distribution's variance \\
  \bottomrule
  \end{tabular}
\caption{\it Configuration parameters the for Analytic Banana benchmark.}
\label{tbl:config_analytic}
\end{table}

The parameters $a$, $b$, and `scale' that appear in the formulas above
are configurable, see Table \ref{tbl:config_analytic}. This benchmark
supports only evaluation of the forward map, but not its derivatives.

\paragraph{Analytic Donut benchmark}

%\begin{table}[h!]
%  \centering
%  \begin{tabular}{cccp{3cm}}
%  \toprule
%  Config & Type & Default & Description \\
%  \midrule
%  - & - & - & - \\
%  \bottomrule
%  \end{tabular}
%\caption{\it Configuration parameters for the Analytic Funnel benchmark.}
%\end{table}

%\begin{figure}
%  \centering
%     \includegraphics[width=0.49\textwidth]{Images/analytic-donut-samples.png}
     %\includegraphics[width=0.49\textwidth]{Images/analytic-donut-contour.png}
%      \caption{\it Samples drawn from the distribution defined through the \gls{pdf} $\pi$.}%, and their corresponding contour.}
%\label{fig:analytic-donut}
%\end{figure}

The \gls{pdf} $\pi$ is defined as
$$ \pi(\theta) := - \frac{(\| \theta \| - r)^2}{\sigma^2}, $$
where $r = 2.6$ and $\sigma^2 = 0.033$.
This benchmark supports the `Evaluate', `Gradient', and `ApplyJacobian' operations.

\paragraph{Analytic Funnel Benchmark}

%\begin{table}[h!]
%  \centering
%  \begin{tabular}{cccp{3cm}}
%  \toprule
%  Config & Type & Default & Description \\
%  \midrule
%  - & - & - & - \\
%  \bottomrule
%  \end{tabular}
%\caption{\it Configuration parameters for the Analytic Funnel benchmark.}
%\end{table}

%\begin{figure}
%  \centering
%     \includegraphics[width=0.49\textwidth]{Images/analytic-funnel-samples.png}
     %\includegraphics[width=0.49\textwidth]{Images/analytic-funnel-contour.png}
%      \caption{\it Samples drawn from the distribution defined through the \gls{pdf} $\pi$.}%, and their corresponding contour.}
%\label{fig:analytic-funnel}
%\end{figure}

First, define a helper function
$$ f(\theta,m,s) := - \frac12 \log(2 \pi) - \log(s) - \frac12 ((\theta-m)/s)^2, $$
where here $\pi=3.141\ldots$.
The logarithm of the output \gls{pdf} is then defined as
$$ \log(\pi(\theta)) := f(\theta_1, 0, 3) + f\left(\theta_2, 0, \exp\left(\frac12 \theta_1\right)\right). $$
This benchmark supports the `Evaluate', `Gradient', and `ApplyJacobian' operations.

\paragraph{Analytic Gaussian Mixture Benchmark}
%\begin{table}[h!]
%  \centering
%  \begin{tabular}{cccp{3cm}}
%  \toprule
%  Config & Type & Default & Description \\
%  \midrule
%  - & - & - & - \\
%  \bottomrule
%  \end{tabular}
%\caption{\it Configuration parameters for the Analytic Funnel benchmark.}
%\end{table}

%\begin{figure}
%  \centering
%     \includegraphics[width=0.49\textwidth]{Images/analytic-gaussian-mixture-samples.png}
     %\includegraphics[width=0.49\textwidth]{Images/analytic-gaussian-mixture-contour.png}
%      \caption{\it Samples drawn from the distribution defined through the \gls{pdf} $\pi$.}%, and their corresponding contour.}
%\label{fig:analytic-gaussian-mixture}
%\end{figure}

Let
\begin{align*}
X_1 &\sim \mathcal{N}\left(\begin{pmatrix} -1.5 \\ -1.5 \end{pmatrix}, 0.8 I\right),\\
X_2 &\sim \mathcal{N}\left(\begin{pmatrix} 1.5 \\ 1.5 \end{pmatrix}, 0.8 I\right),\\
X_3 &\sim \mathcal{N}\left(\begin{pmatrix} -2 \\ 2 \end{pmatrix}, 0.5 I\right).
\end{align*}
Denote by $f_{X_1}, f_{X_2}, f_{X_3}$ the corresponding \gls{pdf}s. The \gls{pdf} $\pi$ is then defined as
$$ \pi(\theta) := \sum_{i=1}^3 f_{X_i}(\theta). $$
This benchmark supports the `Evaluate', `Gradient', and `ApplyJacobian' operations.

\subsubsection{Membrane inverse benchmark}
\label{sec:inverse:membrane}

%Whereas Section~\ref{sec:models:membrane} describes the forward problem of computing the vertical deformation of a membrane given its stiffness properties, the \textit{membrane inverse benchmark} does the opposite: Given $169$ measurement values, along with a probability distribution describing how accurately we know them, the goal is to estimate the statistical distribution of the $64$ stiffness values that may have produced these measurements, subject to a prior probability distribution for these coefficients.

This benchmark for Bayesian inversion is, like the forward model in Section~\ref{sec:models:membrane}, described in substantial detail in \cite{AristoffBangerth}. In short, whereas the forward model maps the $64$ coefficients $\theta\in\mathbb R^{64}$ that describe $a(\mathbf x)=a^\theta(\mathbf x)$ to a vector of measurements of $u(\mathbf x)$ at $139$ distinct locations, denoted by $z^\theta\in \mathbb R^{139}$, the inverse problem consists of inferring $\theta$ from a specific set of measurements $\hat z$. To this end, we define the probability that a given set of coefficients $\theta$ is correct as
\begin{align}
    \pi(\theta|\hat z) = L(\hat z|\theta) \pi_\text{pr}(\theta),
\end{align}
where the likelihood $L$ and prior probability are defined as
\begin{align}
      L(z|\theta)
  &=
  \exp\left(-\frac{\|z-z^\theta\|^2}{2\sigma^2}\right)\\
  &=
  \prod_{k=0}^{168} \exp\left(-\frac{(z_k-z^\theta_k)^2}{2\sigma^2}\right),
  \\
    \pi_\text{pr}(\theta) &= \prod_{i=0}^{63} \exp\left(-\frac{(\ln(\theta_i)-\ln(1))^2}{2\sigma_\text{pr}^2}\right),
\end{align}
with $\sigma=0.05, \sigma_\text{pr}=2$, and the values of $\hat z$ as tabulated in \cite{AristoffBangerth}.
The benchmark only supports evaluation of the forward map.

The description of the benchmark in \cite{AristoffBangerth} also contains very detailed statistics for mean and covariance matrix of the posterior distribution for these $64$ values, along with estimates for the accuracy for these mean values. As a consequence, any inference scheme using the implementation of this benchmark provided through the containers described in Sections~\ref{sec:architecture} can compare their results with those provided in \cite{AristoffBangerth}; this includes the workload-accuracy trade-off of inference algorithms for this benchmark.
\subsubsection{Bayesian calibration of tritium desorption}
\label{sec:tritium_benchmark}
This benchmark performs Bayesian model calibration of the tritium desorption model outlined in Section \ref{sec:tritium_model}.
The benchmark defines a probability density $\pi:\mathbb
R^5\to \mathbb R$ where the inputs $\theta$ are $E_1,E_2,E_3$ (the detrapping
energies used in the model description), and $n_1,n_2$ (the density of
intrinsic traps).

\begin{table}[t!]
\centering
\begin{tabular}{lr}
\toprule
Parameter & Prior distribution \\
\midrule
$E_1$ & $\mathcal U(0.7, 1.0)$ \\
$E_2$ & $\mathcal U(0.9, 1.3)$ \\
$E_3$ & $\mathcal U(1.1, 1.75)$ \\
$n_1$ & $\mathcal U(5 \cdot 10^{-4}, 5 \cdot 10^{-3})$ \\
$n_2$ & $\mathcal U(10^{-4}, 10^{-3})$ \\
\bottomrule
\end{tabular}
\caption{\it Prior distributions for the Tritium desorption model.}
\label{tbl:tritium_prior}
\end{table}

\begin{figure}
\centering
   \includegraphics[width=.48\linewidth]{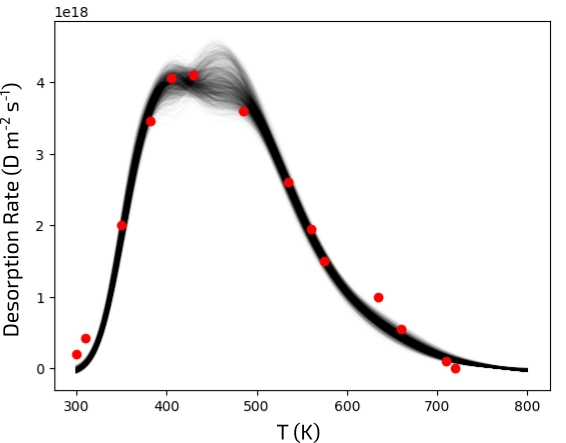}
    \caption{\it Posterior samples ($N=1000$) of the Tritium desortion rate (black curves) and the experimental data from \cite{ogorodnikova_deuterium_2003} (red dots). Both are plotted against the temperature $T$.}
  \label{fig:tritium_benchmark}
\end{figure}

The prior distributions over model parameters are all assumed to be uniform as shown in Table \ref{tbl:tritium_prior}.
This prior distribution is then conditioned on the experimental data
of \cite{ogorodnikova_deuterium_2003}, where the desorption rate
of \gls{HI} was measured for a tungsten lattice, to yield the
posterior distribution $\pi(\theta | d)$. More formally, the
(unnormalized) posterior distribution of parameters $\theta$ given
data $d = \{\phi_i\}_{i=0}^{N_d-1}$ with $N_d = 14$ measurements of
the desorption rate $\phi$ is computed using Bayes Theorem $\pi(\theta
| d) \propto  \mathcal L(d | \theta) \pi_\text{pr}(\theta)$. The measurement errors $\varepsilon$ are assumed to follow a Gaussian distribution, so that
\begin{equation*}
    \mathcal L(d | \theta) \propto \exp \left(-\frac{1}{2\sigma_\varepsilon^2} \sum_{i=0}^{13}(\mathcal F_i(\theta) - d_i)^2 \right),
\end{equation*}
with $\sigma_\varepsilon = 10^{17}$.
%The posterior distribution was sampled using the $\mathrm{DREAM}_{\mathrm{(ZS)}}$ proposal algorithm \cite{ter_braak_differential_2008, vrugt_accelerating_2009}.
Figure \ref{fig:tritium_benchmark} shows $N = 1000$ posterior samples of the Tritium desorption curve along with the experimental data from \cite{ogorodnikova_deuterium_2003}.

\subsubsection{Agent-based disease transmission benchmark}

This benchmark uses EMOD \cite{bershteyn18}, a stochastic agent-based disease transmission model, to look at how the reproductive number, $R_0$, and the correlation between an individual's acquisition and transmission correlation can affect the ultimate attack fraction and timing of the outbreak peak. This correlation can greatly affect the size and timing of the outbreak for a fixed reproductive number.
The model underlying this benchmark is described in Section \ref{sec:emod_abm_model}.

The benchmark computes a probability distribution $\pi: {\mathbb
R}^3 \to {\mathbb R}$ from three inputs that correspond to $R_0$, the
variance of $R_0$, and the correlation between acquisition and transmission. The benchmark is fitting to an attack fraction of 0.40 with a standard deviation of 0.05 and an outbreak peak at 175 days from the start of the simulation with standard deviation of 10 days.
The benchmark has additional configuration parameters, described in Table \ref{tbl:config_emod}. The benchmark only supports evaluation of the forward map.

\begin{table}[t!]
\caption{\it Configuration parameters for the disease model benchmark.}
\label{tbl:config_emod}
  \centering
  \begin{tabular}{p{1.5cm}p{1cm}p{1cm}p{3cm}}
  \toprule
  Config & Type & Default & Description \\
  \midrule
  \footnotesize refresh\textunderscore  seed & \footnotesize int (bool) & \footnotesize 1 & \footnotesize Change random number seed \\
  \footnotesize daily\textunderscore import\allowbreak\textunderscore pressure & \footnotesize double & \footnotesize 1.0 & \footnotesize Average daily importations for the first 5 simulated days \\
  \footnotesize log\textunderscore  level & \footnotesize string & \footnotesize ERROR & \footnotesize Level of logging \\
  \footnotesize epsilon & \footnotesize list & \footnotesize \begin{tabular}{c}[0.001,\\~0.001,\\~0.001]\end{tabular} & \footnotesize Increment used in scipy optimize \textrm{approx\textunderscore fprime} to estimate the gradient \\
  \bottomrule
  \end{tabular}
\end{table}

One of the challenges of this benchmark is that the computation
of $\pi(\theta)$ is not deterministic because it evaluates a
stochastic model whose outputs (even for a fixed $\theta$) depends on
random numbers: Even for the same parameters the total number of
infections will differ between simulations. Furthermore, it is not
guaranteed that there will always be an outbreak: sometimes there are
not enough initial infections to cause an outbreak. This challenge can
be decreased by increasing the number of initial infections via the \texttt{daily\_import\_pressures} configuration parameter (e.g., from 1 infection per day to 10 infections per day for the first 5 days).

% \todo[inline]{I'm not entirely sure I understand this paragraph. Are
% you saying that the computation of $\pi(\theta)$ is not deterministic?
% If so, perhaps replace the first sentence of the preceding paragraph
% by ``One of the challenges of this benchmark is that the computation
% of $\pi(\theta)$ is not deterministic because it evaluates a
% stochastic model whose outputs (even for a fixed $\theta$) depends on
% random numbers: Even...''}

\subsubsection{1D deconvolution benchmark}\label{sec:deconv}

This benchmark is based on a 1D deconvolution test problem from the
library
CUQIpy\footnote{\url{https://cuqi-dtu.github.io/CUQIpy/}} \cite{riis2023cuqipy,alghamdi2023cuqipy}. It
defines a posterior distribution $\pi: {\mathbb R}^{128} \to {\mathbb R}$ with a Gaussian likelihood and four different choices of prior distributions with configurable parameters.
Supported features of this benchmark are: `Evaluate' and `Gradient'.

The 1D periodic deconvolution problem is defined by the inverse problem
$$
\mathbf{b} = \mathbf{A}\mathbf{x} + \mathbf{e},
$$
where $\mathbf{b}$ is an $m$-dimensional random vector representing
the observed data, $\mathbf{A}$ is an $m\times n$ matrix representing
the convolution operator, $\mathbf{x}$ is an $n$-dimensional random
vector representing the unknown signal we would like to recover, and $\mathbf{e}$ is an $m$-dimensional random vector representing the noise. 
The posterior distribution over $\mathbf{x}$ given $\mathbf{b}$ is
defined as
$$
\pi(\mathbf{x}\mid \mathbf{b}) \propto L(\mathbf{b}\mid \mathbf{x})\pi_\text{pr}(\mathbf{x}), 
$$
where $L(\mathbf{b}|\mathbf{x})$ is a likelihood function and $\pi_\text{pr}(\mathbf{x})$ is a prior distribution. 
The noise is assumed to be Gaussian with a known noise level, and thus the likelihood is defined via
$$
\mathbf{b} \mid \mathbf{x} \sim \mathcal{N}(\mathbf{A}\mathbf{x}, \sigma^2\mathbf{I}_m),
$$
where $\mathbf{I}_m$ is the $m\times m$ identity matrix and $\sigma=0.01$.
The prior can be configured by choosing from the following assumptions about $\mathbf{x}$:

\begin{itemize}
    \item `Gaussian': Gaussian (normal) distribution: \\$x_i \sim \mathcal{N}(0, \delta)$,
    \item `GMRF' Gaussian Markov Random Field: \\$x_i-x_{i-1} \sim \mathcal{N}(0, \delta)$,
    \item `LMRF' Laplace Markov Random Field: \\$x_i-x_{i-1} \sim \mathcal{L}(0, \delta)$,
    \item `CMRF' Cauchy Markov Random Field: \\$x_i-x_{i-1} \sim \mathcal{C}(0, \delta)$,
\end{itemize}
where $\mathcal{L}$ is the Laplace distribution and $\mathcal{C}$ is
the Cauchy distribution. The parameter $\delta$ is the prior parameter
and is configurable (as an additional parameter that can be provided
via UM-Bridge; its default is $\delta=0.01$.

\begin{figure}[t!]
  \centering
     \includegraphics[width=\columnwidth]{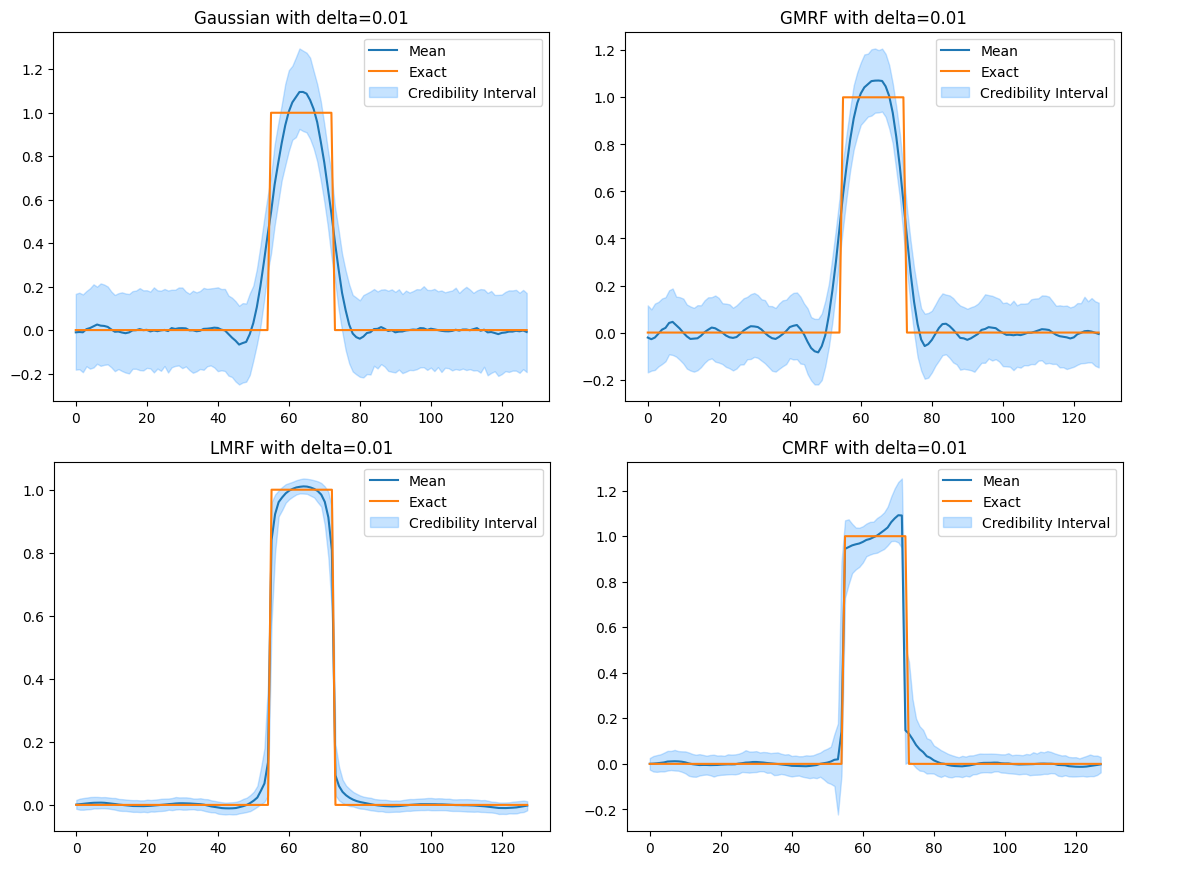}
      \caption{\it Sample mean and $95\%$ credibility intervals of the posteriors defined for the deconvolution 1D benchmark.}
\label{fig:Deconvolution1D-samples}
\end{figure}
In \cref{fig:Deconvolution1D-samples} we show examples of posterior samples obtained for the four different choices of prior. The posterior mean and $95\%$ credibility intervals are shown. The samples are obtained using the built-in automatic sampler selection in CUQIpy \cite{riis2023cuqipy,alghamdi2023cuqipy}.

%The choice of prior is specified by providing the name to the HTTP model. In this case \texttt{Deconvolution1D\_Gaussian}, \texttt{Deconvolution1D\_GMRF}, \texttt{Deconvolution1D\_CMRF}, and \texttt{Deconvolution1D\_LMRF}, respectively.

%In addition to the HTTP models for the posterior, there is also an HTTP model for the exact solution to the problem. This model is called \texttt{Deconvolution1D\_ExactSolution} and returns exact phantom used to generate the synthetic data when called.

\subsubsection{Computed tomography benchmark}

This benchmark focuses on 2D image reconstruction in X-ray computed
tomography (CT) \cite{CTbook}. In CT, X-rays are passed through an
object of interest and projection images are recorded at a number of orientations. Materials of different density absorb X-rays by different amounts quantified by their linear attenuation coefficients. A linear forward model describes how the 2D linear attenuation image of an object results in a set of projections measured, typically known as a sinogram.

\begin{figure}[tb]
    \centering
    \includegraphics[width=0.49\linewidth]{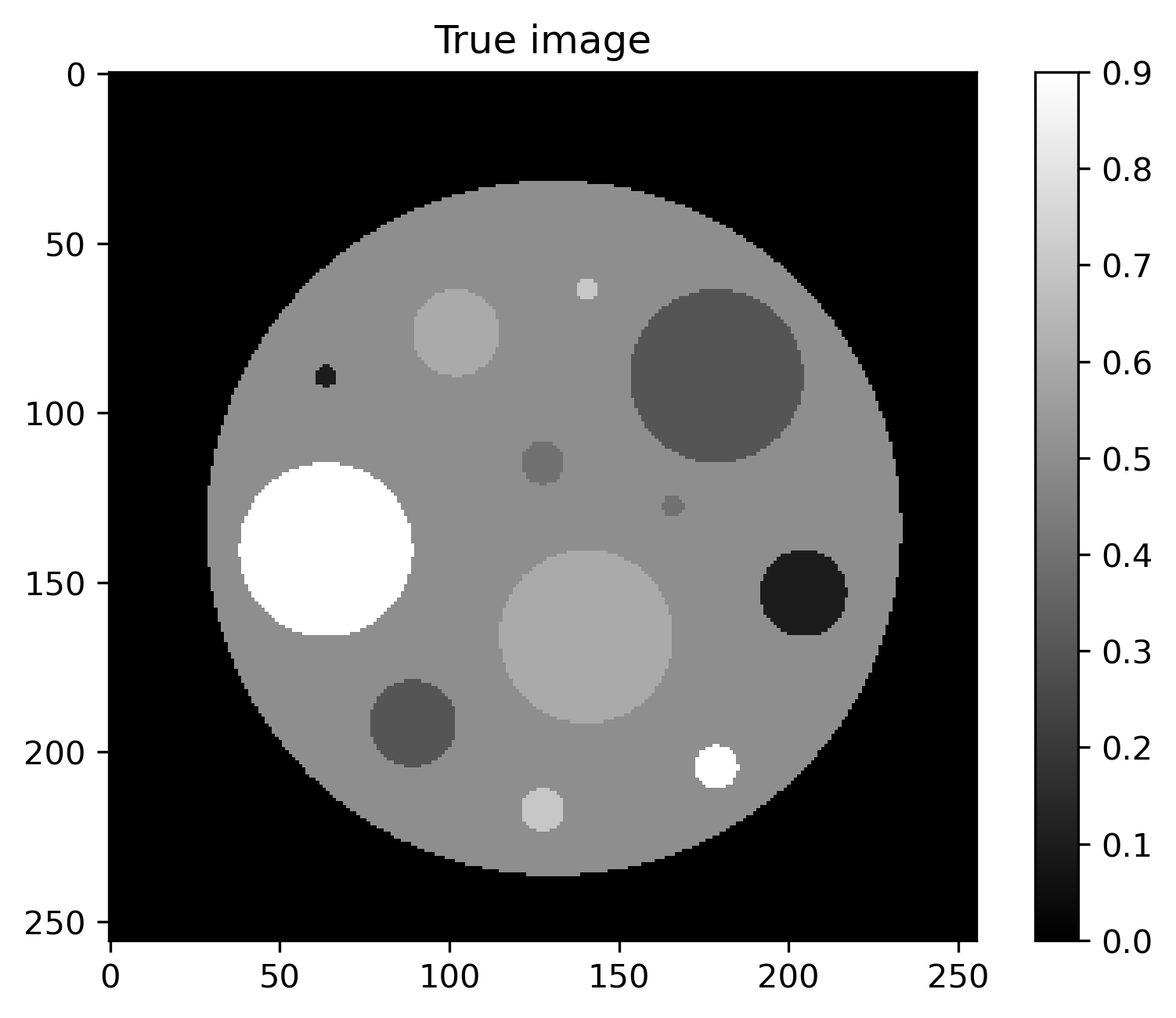}
    \includegraphics[width=0.49\linewidth]{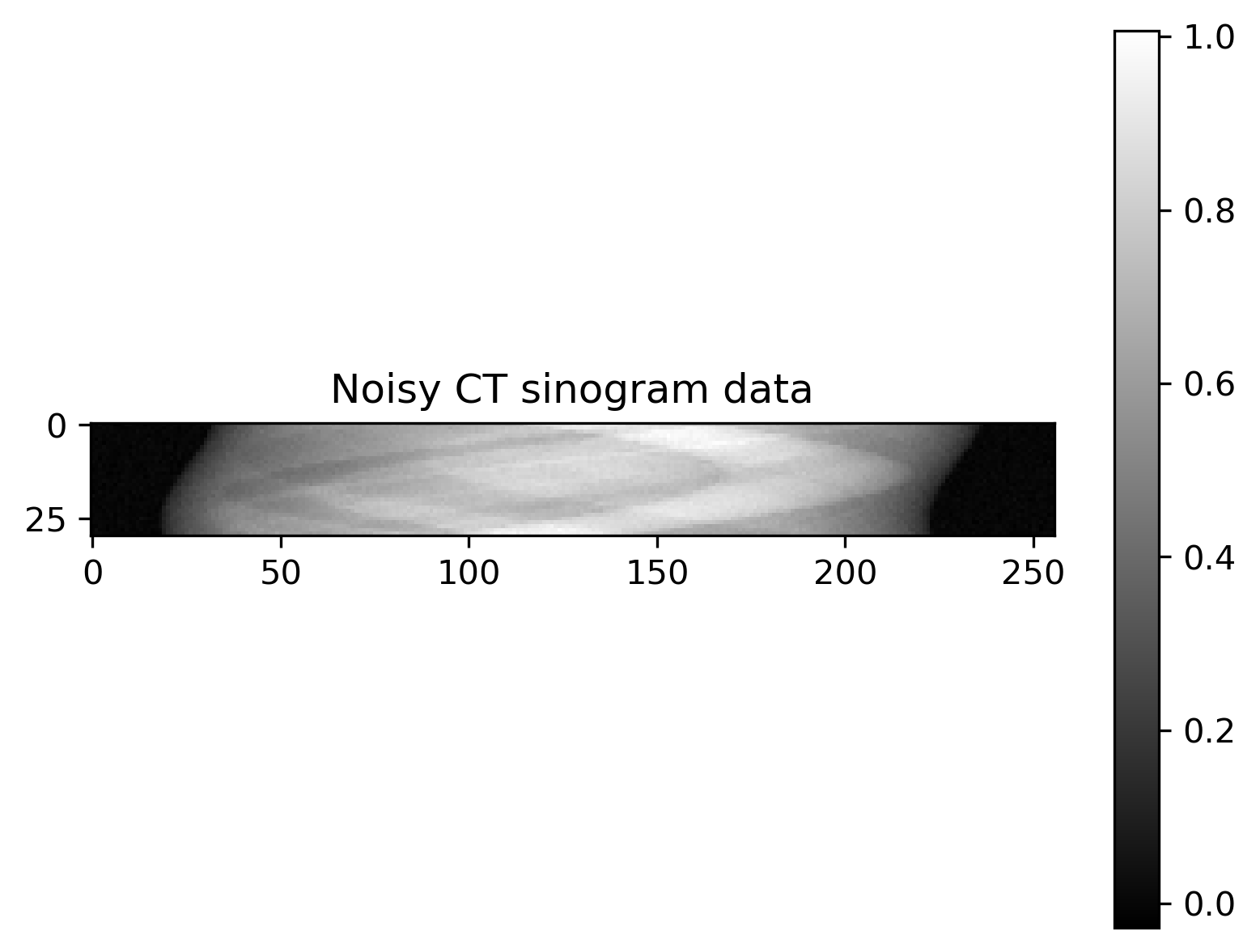}
    \caption{\it Left: Exact solution image. %Top right: noise-free
    %sinogram data. 
    Right: noisy CT sinogram data.} %Bottom right:
%    Noise. }
%    \todo[inline]{First, these are not our pictures (I assume), and so
%    they would need proper attribution. But moreover, I don't think it
%    is useful that we provide these images, they only take up space;
%    let's just provide a link instead.}
%\todo[inline]{J. S. J\o{}rgensen: The pictures for the three CUQIpy benchmarks (1D deconvolution, CT and 1D heat) are generated by the benchmark codes included in UM-bridge contributed by the CUQIpy developers and co-authors of this article. For example for the CT case: \url{https://github.com/UM-Bridge/benchmarks/tree/main/benchmarks/cuqi-ct} As such, I don't think any attribution for pictures is needed, aside from the existing citation of the underlying library CUQIpy with which the benchmarks are implemented. It is up to the main authors to decide whether to include these pictures. We find the pictures do a good job of conveying the nature and content of the benchmark, similarly to how pictures are used for other benchmarks.  We agree the CT figure contains quite some whitespace and would be happy to rearrange to reduce this, if desired. }
    \label{fig:ct_image_data}
\end{figure}

The inverse problem of CT takes the same form as in the previous section,  %\eqref{eq:deconvinverseproblem},
%\begin{align}
%\mathbf{b} = \mathbf{A}\mathbf{x} + \mathbf{e},    
%\end{align}
with $\mathbf{b}$ now an $m$-dimensional random vector representing the observed (vectorized) CT sinogram data, $\mathbf{A}$ is an $m\times n$ matrix representing the CT projection operator or ``system matrix'', $\mathbf{x}$ is an $n$-dimensional random vector representing the unknown (vectorized) square image, and $\mathbf{e}$ is an $m$-dimensional random vector representing the noise. The underlying true image and the %clean and
noisy CT sinogram data %as well as the noise image 
are shown in Fig.~\ref{fig:ct_image_data}.

The matrix $\mathbf{A}$ is generated by using the MATLAB package AIR Tools II\footnote{\url{https://github.com/jakobsj/AIRToolsII}} \cite{Hansen2018} . %Specifically, the following Matlab code was used to gener  ate the forward model matrix:
%\begin{lstlisting}[language=Matlab]
%[A, b, x] = paralleltomo(256, 0:6:174, N);
%save A256_30.mat A
%\end{lstlisting}
The forward model matrix corresponds to an image of $256\times256$ pixels, hence $n=256^2$, scanned in a parallel-beam geometry with 30 projections taking over 180 degrees with angular steps of 6 degrees, each projection consisting of 256 X-ray lines, i.e., $m = 30\cdot 256=7680$. See the above links for more details.

This benchmark follows the same structure and provides the same features as the 1D Deconvolution benchmark in \cref{sec:deconv}, providing a posterior distribution of $\mathbf{x}$ given $\mathbf{b}$. It also employs %benchmark uses the library 
CUQIpy\footnote{\url{https://cuqi-dtu.github.io/CUQIpy/}} \cite{riis2023cuqipy,alghamdi2023cuqipy}  to specify the linear forward model representing CT and uses the same Gaussian noise model. %CUQIpy offers support for GPU-accelerated CT forward models using the Core Imaging Library (CIL) \cite{Jorgensen2021,Papoutsellis2021} (\url{https://ccpi.ac.uk/cil/}) through the CUQIpy-CIL plugin. For simplicity, the present benchmark is based on a matrix constructed with the Matlab library AIR Tools II \cite{Hansen2018} (\url{https://github.com/jakobsj/AIRToolsII}). It defines a posterior distribution for a 2D X-ray CT image reconstruction problem problem, with a Gaussian noise distribution ie likelihood and four different choices of 2D prior distributions with configurable parameters.

%The input for this benchmark is a $256\times 256$ pixel image. %The output is a
%probability distribution $\pi((\mathbf{b}\mid \mathbf{x})$. This
%benchmark supports the `Evaluate' and `Gradient' operations.

% This benchmark defines a posterior distribution over $\mathbf{x}$ given $\mathbf{b}$ as
% \begin{align}
% \pi(\mathbf{x}\mid \mathbf{b}) \propto L(\mathbf{b}\mid \mathbf{x})\pi_\text{pr}(\mathbf{x}),
% \end{align}
% where $L(\mathbf{b}|\mathbf{x})$ is a likelihood function and $\pi_\text{pr}(\mathbf{x})$ is a prior distribution. 
% The noise is assumed to be Gaussian with a known noise level, and so the likelihood is defined via
% \begin{align}
% \mathbf{b} \mid \mathbf{x} \sim \mathcal{N}(\mathbf{A}\mathbf{x}, \sigma^2\mathbf{I}_m),
% \end{align}
% where $\mathbf{I}_m$ is the $m\times m$ identity matrix and
% $\sigma=0.01$.
% \todo[inline]{This paragraph is a verbatim copy from the previous
% section. Let's just refer to that instead of taking up all this space.}

The same prior configurations for $\mathbf{x}$ as in the 1D deconvolution benchmark are available, but in 2D versions where distributions stated above are applied to both vertical and horizontal finite differences, $x_{i,j}-x_{i-1,j}$ and $x_{i,j}-x_{i,j-1}$ for $x_{i,j}$ for $i,j = 1, \dots, 256$.

% The prior can be configured by choosing of the following assumptions about $\mathbf{x}$, where we here consider x a square image with pixels $x_{i,j}$ for $i,j = 1, \dots, 256$:
% \begin{itemize}
%     \item `Gaussian': Gaussian (normal) distribution: $x_{i,j} \sim \mathcal{N}(0, \delta)$,
%     \item `GMRF' Gaussian Markov Random Field: \\$x_{i,j}-x_{i-1,j} \sim \mathcal{N}(0, \delta)$, $x_{i,j}-x_{i,j-1} \sim \mathcal{N}(0, \delta)$,
%     \item `LMRF' Laplace Markov Random Field: \\$x_{i,j}-x_{i-1,j} \sim \mathcal{L}(0, \delta)$, $x_{i,j}-x_{i,j-1} \sim \mathcal{L}(0, \delta)$,
%     \item `CMRF' Cauchy Markov Random Field: \\$x_{i,j}-x_{i-1,j} \sim \mathcal{C}(0, \delta)$, $x_{i,j}-x_{i,j-1} \sim \mathcal{C}(0, \delta)$,
% \end{itemize}
% where $\mathcal{L}$ is the Laplace distribution and $\mathcal{C}$ is the Cauchy distribution. The parameter $\delta$ is the prior parameter and is configurable (see above).

%\todo[inline]{The previous paragraph is a verbatim copy of one in the
%previous section, except that we are now in 2d. I'm sure this whole
%paragraph could be said in two sentences with substantially less text.}

%\todo[inline]{J. S. J\o{}rgensen: Indeed, this section can be shortened substantially and refer to the previous section and just state changes.}

\begin{figure}[tb]
    \centering %0.243
\includegraphics[width=0.24\linewidth,clip,trim=1.5cm 0.85cm 2.4cm 0.75cm]{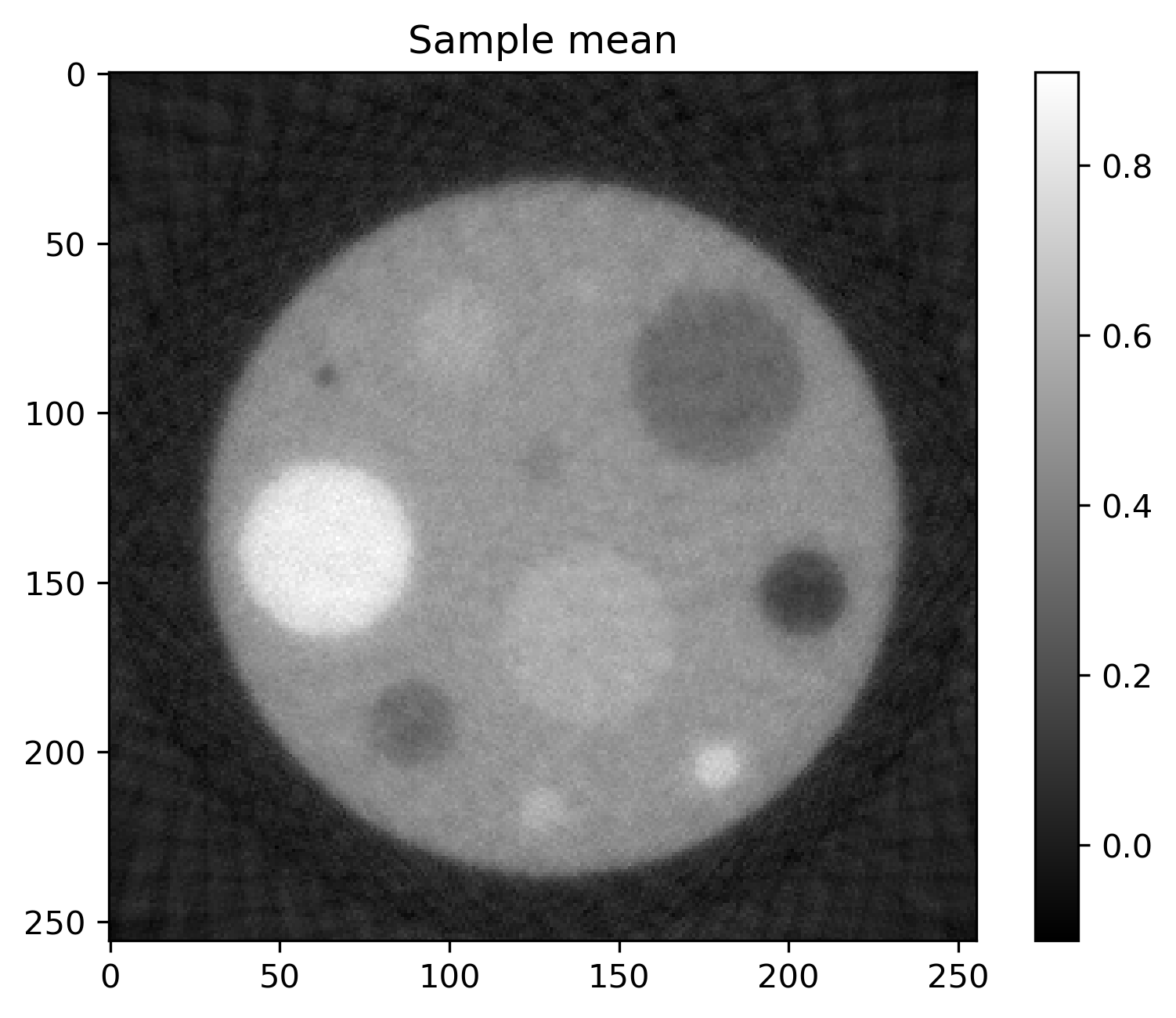}
\includegraphics[width=0.24\linewidth,clip,trim=1.5cm 0.85cm 2.4cm 0.75cm]{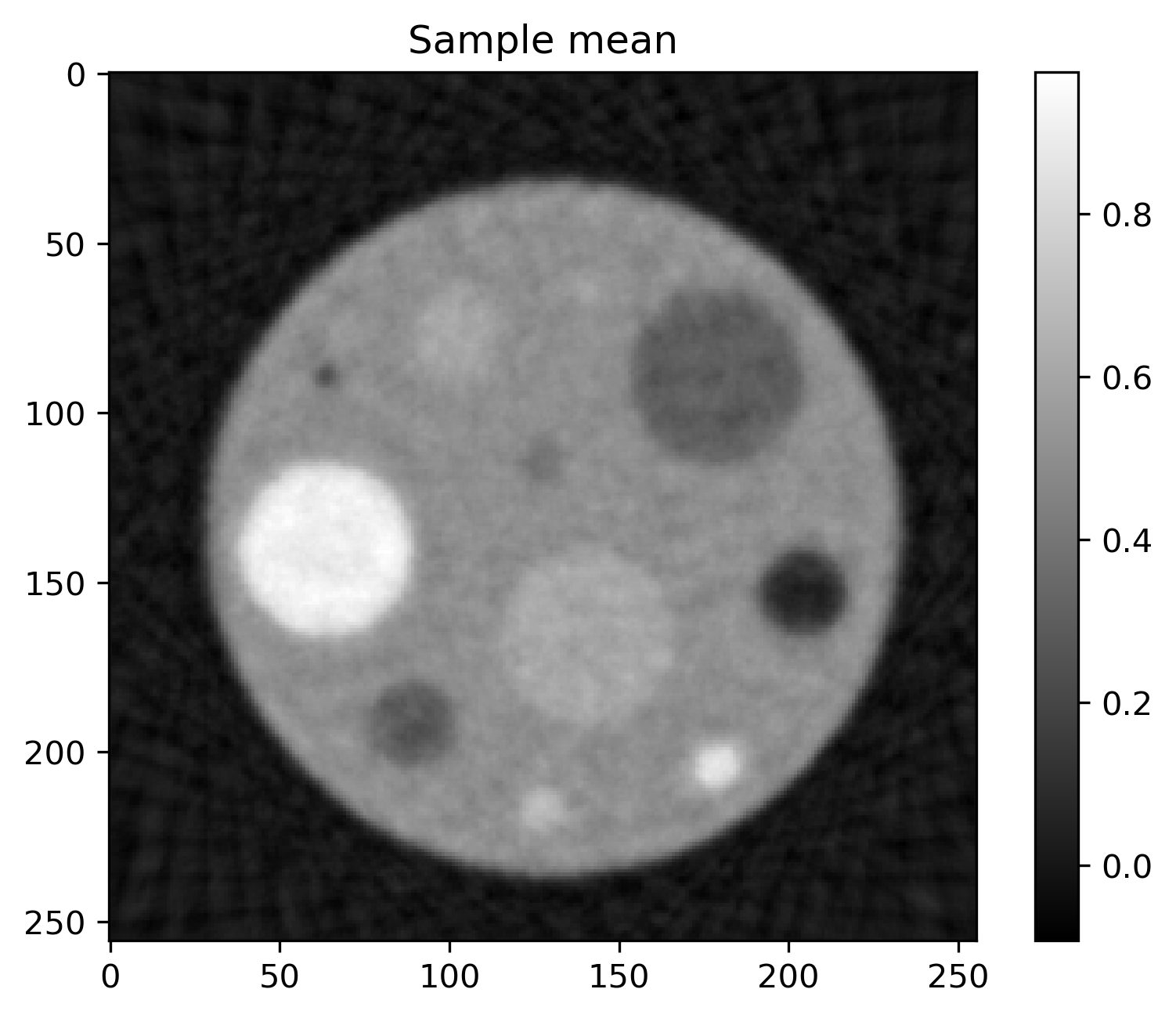}
\includegraphics[width=0.24\linewidth,clip,trim=1.5cm 0.85cm 2.44cm 0.75cm]{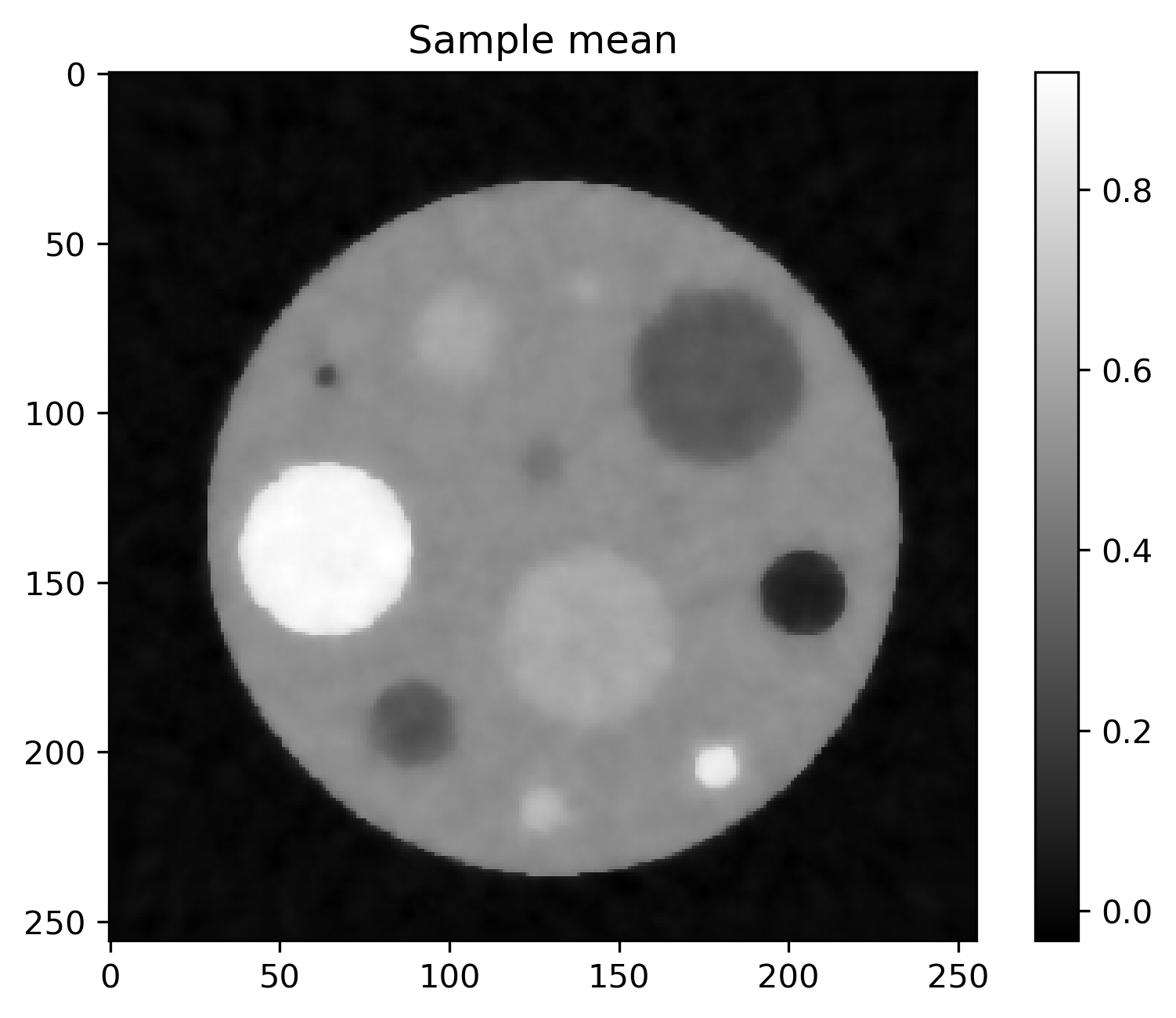}
\includegraphics[width=0.24\linewidth,clip,trim=1.5cm 0.85cm 2.4cm 0.75cm]{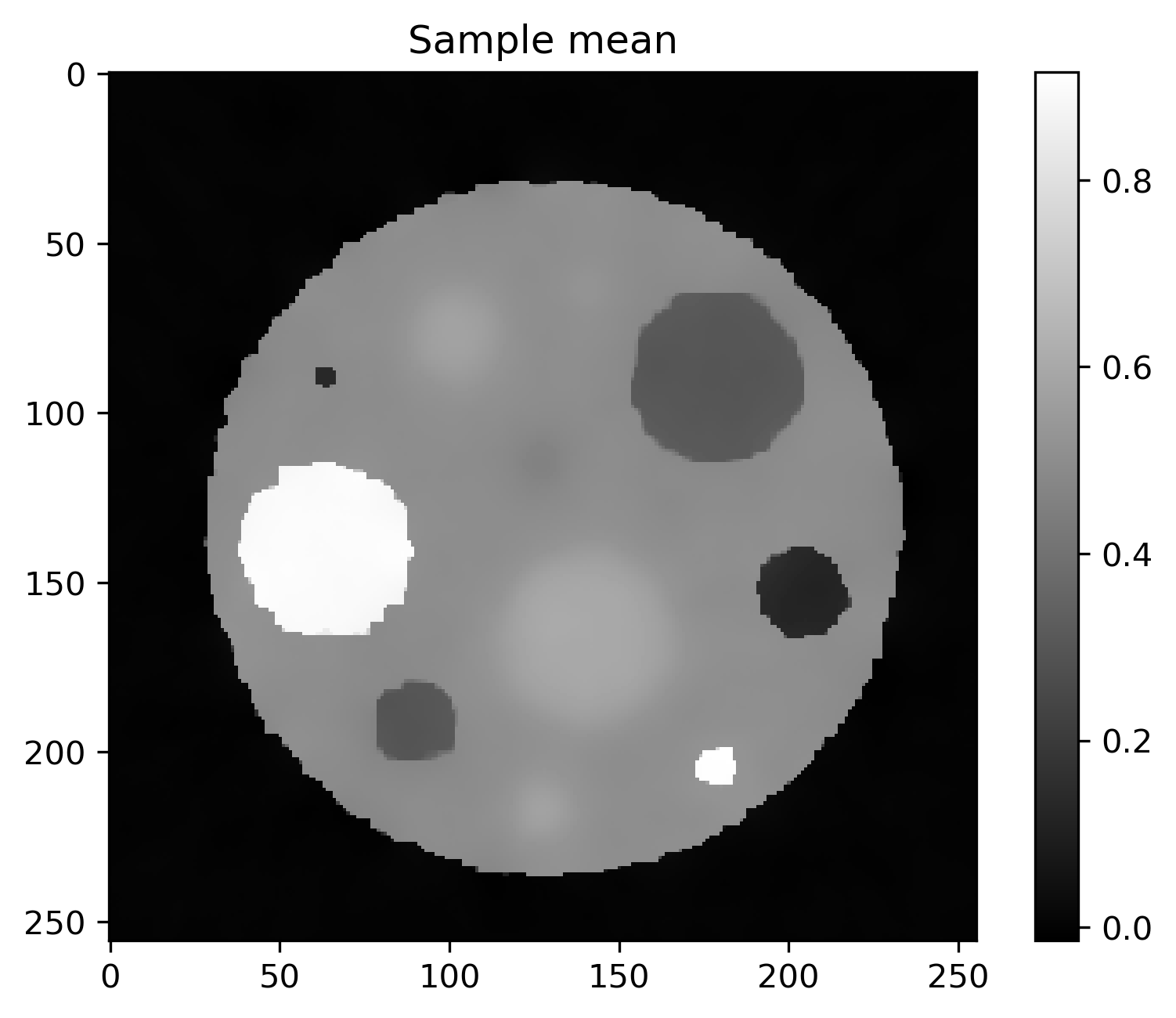}\\
\includegraphics[width=0.24\linewidth,clip,trim=1.3cm 0.85cm 2.75cm 0.75cm]{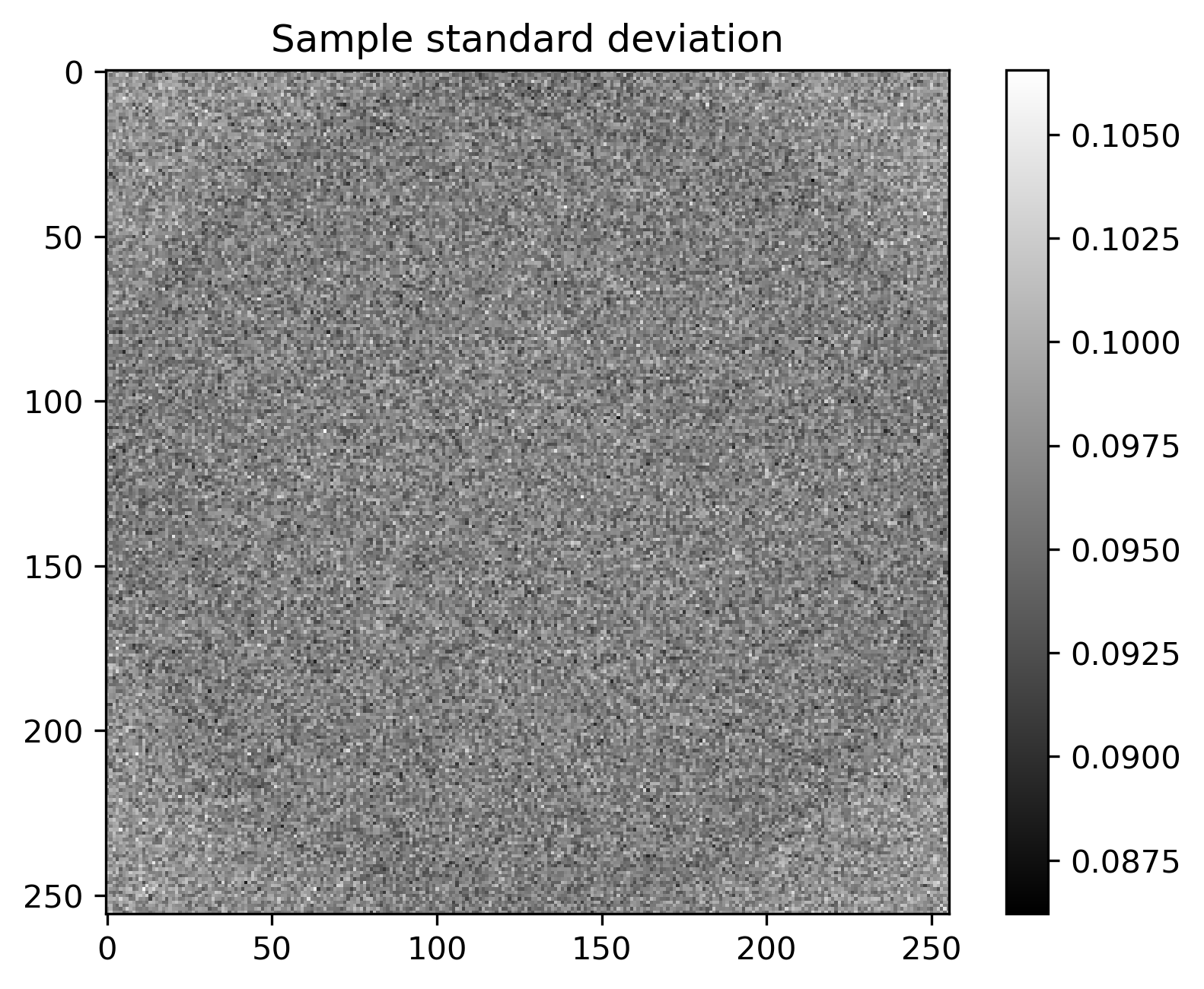}
\includegraphics[width=0.24\linewidth,clip,trim=1.3cm 0.85cm 2.55cm 0.75cm]{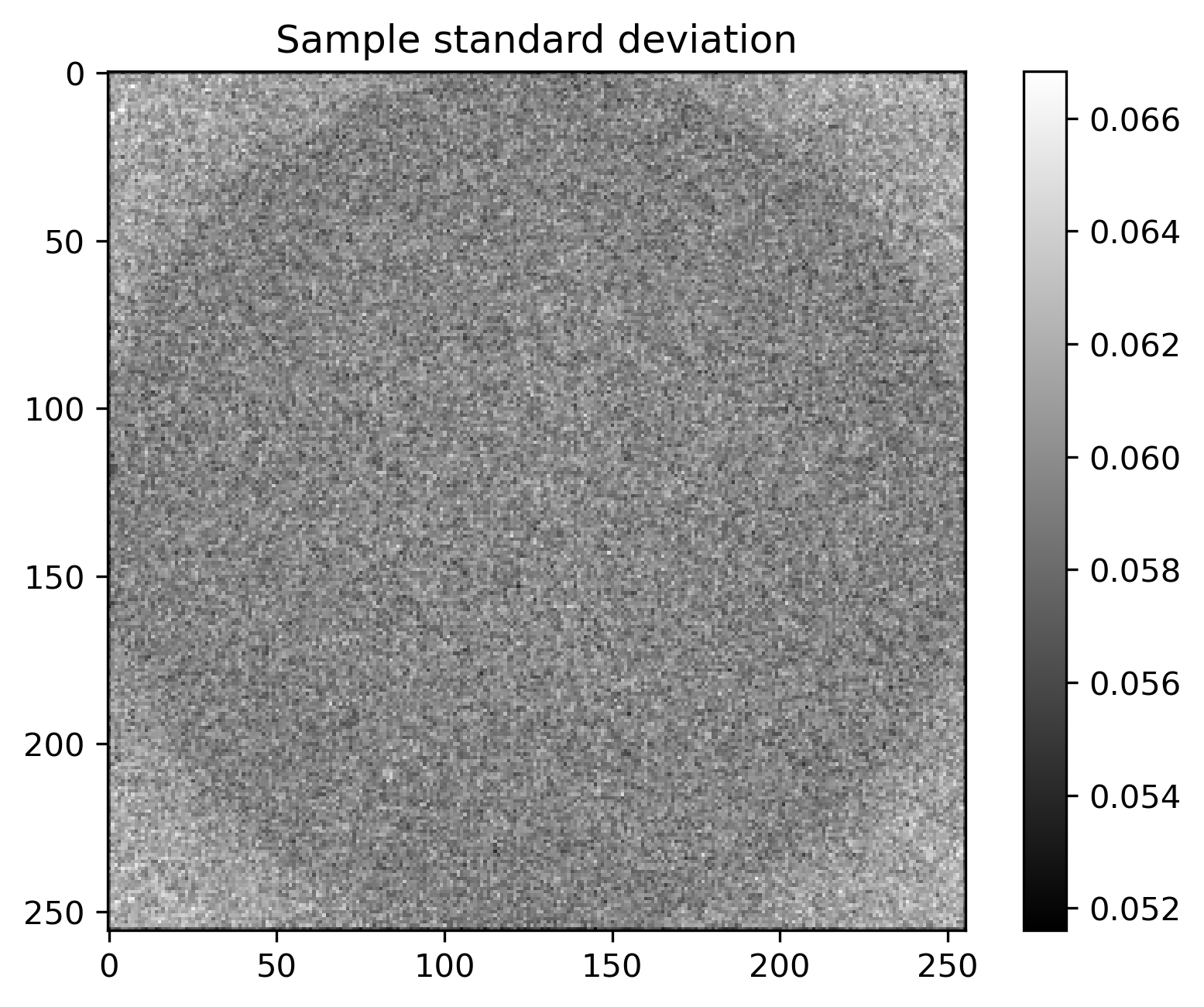}
\includegraphics[width=0.24\linewidth,clip,trim=1.3cm 0.85cm 2.35cm 0.75cm]{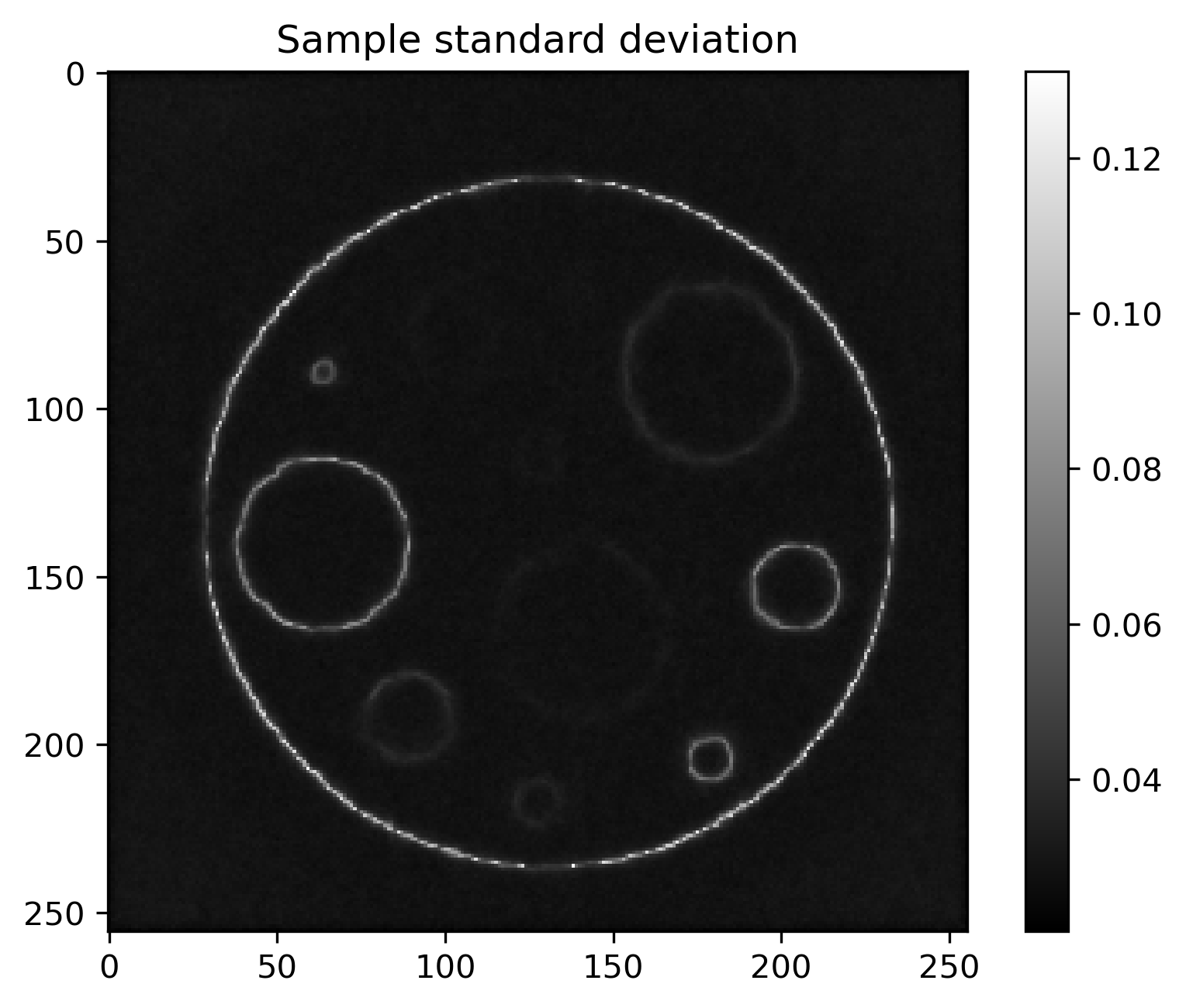}
\includegraphics[width=0.24\linewidth,clip,trim=1.3cm 0.85cm 2.35cm 0.75cm]{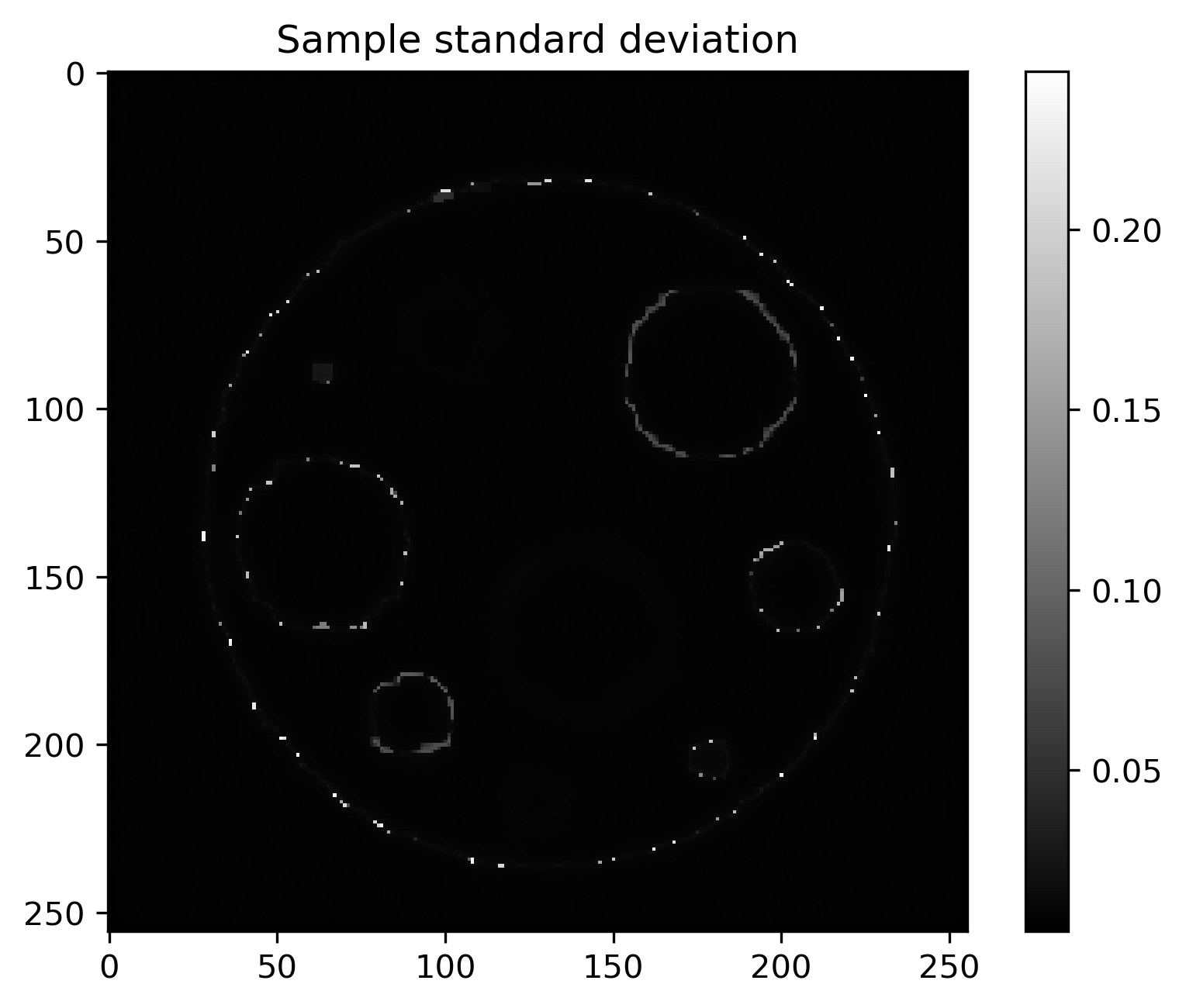}
    \caption{\it Top: Posterior mean images for Gaussian, GMRF,
    LMRF, CMRF priors, with $\delta=0.01$, except LMRF where
    $\delta=0.1$. Bottom: Posterior standard deviation images for
    the same cases.}
    \label{fig:ct-results}
\end{figure}

In \cref{fig:ct-results} we show examples of posterior samples obtained for the four different choices of prior. The posterior mean and standard deviation images are shown. MCMC samples are obtained using the built-in automatic sampler selection provided by CUQIpy. %\cite{riis2023cuqipy,alghamdi2023cuqipy}.

%The choice of prior is specified by providing the name to the HTTP model. In this case \texttt{CT\_Gaussian}, \texttt{CT\_GMRF}, \texttt{CT\_LMRF}, and \texttt{CT\_CMRF}, respectively. 

%In addition to the HTTP models for the posterior, there is also an HTTP model for the exact solution to the problem. This model is called \texttt{CT\_ExactSolution} and returns the exact phantom used to generate the synthetic data when called.

\subsubsection{One dimensional heat Bayesian inverse problem benchmark}

This benchmark is built using the library
CUQIpy \cite{riis2023cuqipy,alghamdi2023cuqipy}.
It defines a posterior distribution of a Bayesian inverse problem
governed by a 1D heat equation, with a Gaussian likelihood
and \gls{KL} parameterization of the uncertain
parameters. Specifically, the inputs $\mathbf x\in {\mathbb R}^{20}$ are
20 KL coefficients, and the output is the posterior probability
$\pi(\mathbf x \mid \mathbf b)$ where $\mathbf b$ are given data.
Posteriors
for two cases are available: In the first case, the data is available
everywhere in the domain (the interval $[0,1]$) with a noise level of $0.1\%$; in the other case, the data is available only on the left half of the domain and with a noise level of $5\%$. For more details about this Bayesian inverse problem, see \cite{alghamdi2023cuqipy}. 
The two cases are accessible via UM-Bridge via the
names \texttt{Heat1DSmallNoise} and \texttt{Heat1DLargeNoise},
respectively. These benchmarks only support evaluation of the forward map.

The underlying inverse problem of this benchmark is to infer an initial temperature profile $g(\xi)$ at time $\tau=0$ on the unit interval $\xi \in [0,1]$ from measurements of temperature $u(\xi, \tau)$ at time $\tau=\tau^\text{max}$.

%In this example, the governing partial differential equation that can be solved to find the temperature $u(\xi, \tau)$ given the initial temperature profile $g(\xi)$ is

%\begin{align}
%    \frac{\partial u(\xi,\tau)}{\partial t} -  \frac{\partial^2 u(\xi,\tau)}%{\partial \xi^2}   & = 0, \quad \xi\in[0,1], \\&\quad 0\le \tau \le %\tau^\mathrm{max}= 0.01, \nonumber\\
%    u(0,\tau)= u(1,\tau)&= 0,\nonumber\\
%    u(\xi,0)&= g(\xi), 
%\end{align}
%assuming zero source term and a constant diffusion coefficient of value 1. We discretize the system using first order finite difference method in space on a regular grid of $n=100$ nodes, and forward Euler in time. We denote by $\mathbf{g}$ and $\mathbf{u}$ the discretization of $g$ and $u$.

We parameterize the discretized initial condition $\mathbf{g}$ using a truncated \gls{KL} expansion to impose some regularity and spatial correlation, and reduce the dimension of the discretized unknown parameter from $n$ to $n_\text{KL}$, where $n_\text{KL} \ll n$ and $n_\text{KL}=20$ is the number of the truncated KL expansion coefficients. For a given (discretized) expansion basis $\mathbf{a}_1, ..., \mathbf{a}_n$, the parameterization of $\mathbf{g}$ in terms of the KL expansion coefficients  $x_1, ..., x_n$  can be written as  \cite{alghamdi2023cuqipy}
$$
\mathbf{g} =  \sum_{i=1}^{n} x_i \mathbf{a}_i \approx \sum_{i=1}^{n_\text{KL}} x_i \mathbf{a}_i. 
$$
In the Bayesian setting, we consider  $\mathbf{x}=[x_1, ..., x_{n_\text{KL}}]^\mathsf{T}$  a random vector representing the unknown truncated \gls{KL} expansion coefficients. Consequently, the corresponding $\mathbf{g}$ is a random vectors as well. We define the inverse problem as 
$$
\mathbf{b} = \mathbf{A}(\mathbf{x}) + \mathbf{e},
$$
where $\mathbf{b}$ is an $m$-dimensional random vector representing the observed data, measured temperature profile at time $\tau^\text{max}= 0.01$ in this case, and $\mathbf{e}$ is an $m$-dimensional random vector representing the noise. $\mathbf{A}$ is the forward operator that maps $\mathbf{x}$ to the exact data $\mathbf{A}(\mathbf{x})$. Applying $\mathbf{A}$ involves solving the heat \gls{PDE} above for a given realization of $\mathbf{x}$ and extracting the observations from the space-time solution $\mathbf{u}$.  

This benchmark defines a posterior distribution over $\mathbf{x}$ given $\mathbf{b}$ as
$$
\pi(\mathbf{x}\mid \mathbf{b}) \propto L(\mathbf{b}\mid \mathbf{x})\pi_\text{pr}(\mathbf{x}),
$$
where $L(\mathbf{b}|\mathbf{x})$ is a likelihood function and $\pi_\text{pr}(\mathbf{x})$ is a prior distribution. The noise is assumed to be Gaussian with a known noise level, and so the likelihood is defined via
$$
\mathbf{b} \mid \mathbf{x} \sim \mathcal{N}(\mathbf{A}(\mathbf{x}), \sigma^2\mathbf{I}_m),
$$
where $\mathbf{I}_m$ is the $m\times m$ identity matrix and $\sigma$ defines the noise level. The prior $\pi_\text{pr}(\mathbf{x})$ on the \gls{KL} coefficients is assumed to be a standard multivariate Gaussian. 

\begin{figure}[t!]
  \centering
     \includegraphics[width=0.95\columnwidth]{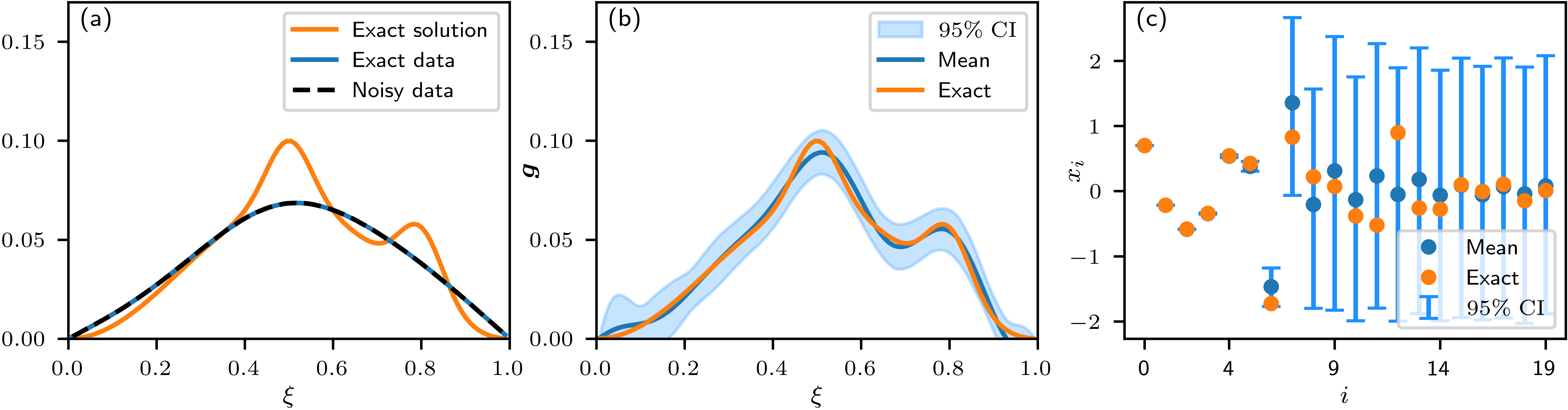}
      \caption{\it The small noise case of the 1d inverse heat
     benchmark. (a) Noisy data, exact data (solution at $\tau=\tau^\text{max}$), and exact solution (solution at $\tau=0$). Note that for the small noise case, noisy and exact data are visually indistinguishable.  (b)
     Samples' $95\%$ credible interval computed after mapping the
     samples of KL coefficients $\mathbf{x}$ to samples of the corresponding function $\mathbf{g}$. (c) \gls{KL} coefficients samples' $95\%$ credible intervals. }
\label{fig:heat1D-small-noise}
\end{figure}

Figure~\ref{fig:heat1D-small-noise}\emph{a} shows the exact solution (the true initial condition), the exact data $\mathbf{A}(\mathbf{x})$, and a synthetic noisy data $\mathbf{b}^\text{obs}$ for the small noise case. We generate $50,000$ samples of the posterior using the component-wise Metropolis–Hastings (CWMH) sampler \cite{riis2023cuqipy, alghamdi2023cuqipy}. We show in Figure~\ref{fig:heat1D-small-noise}\emph{b} the samples' $95\%$ credible interval computed after mapping the \gls{KL} coefficients $\mathbf{x}$ samples to the corresponding function $\mathbf{g}$ samples; and we show the KL coefficients $\mathbf{x}$ samples $95\%$ credible intervals in Figure~\ref{fig:heat1D-small-noise}\emph{c}. %Figure~\ref{fig:heat1D-large-noise}, similarly, shows the set up and the solution of the large noise case.

%\begin{figure}[h!]
%  \centering
%     \includegraphics[width=0.95\columnwidth]{Images/heat1D/fig_large_noise.png}
%      \caption{The large noise case. (a) Noisy data, exact data, and exact solution (b) Samples $95\%$ credible interval computed after mapping the KL coefficients $\mathbf{x}$ samples to the corresponding function $\mathbf{g}$ samples. (c) KL coefficients samples $95\%$ credible intervals. }
%\label{fig:heat1D-large-noise}
%\end{figure}

%In addition to the two HTTP models for the posterior, there is also an HTTP model for the exact solution of the problem. This model is called \texttt{Heat1DExactSolution} and returns the exact initial heat profile used to generate the synthetic data. The map from the coefficients $\mathbf{x}$ to the discretized function $\mathbf{g}$ is provided via the HTTP model \texttt{KLExpansionCoefficient2Function} and the projection of $\mathbf{g}$ on the coefficient space $\mathbf{x}$ is provided by the HTTP model \texttt{KLExpansionFunction2Coefficient}. 

%Using [CUQIpy](https://cuqi-dtu.github.io/CUQIpy/), this benchmark is defined in the files `heat1D_problem.py`, `data_script.py`, and `server.py` provided here.

\subsubsection{Beam inference}
\label{sec:inverse:ebbeam}

\begin{figure}
  \centering
     \includegraphics[width=.5\linewidth]{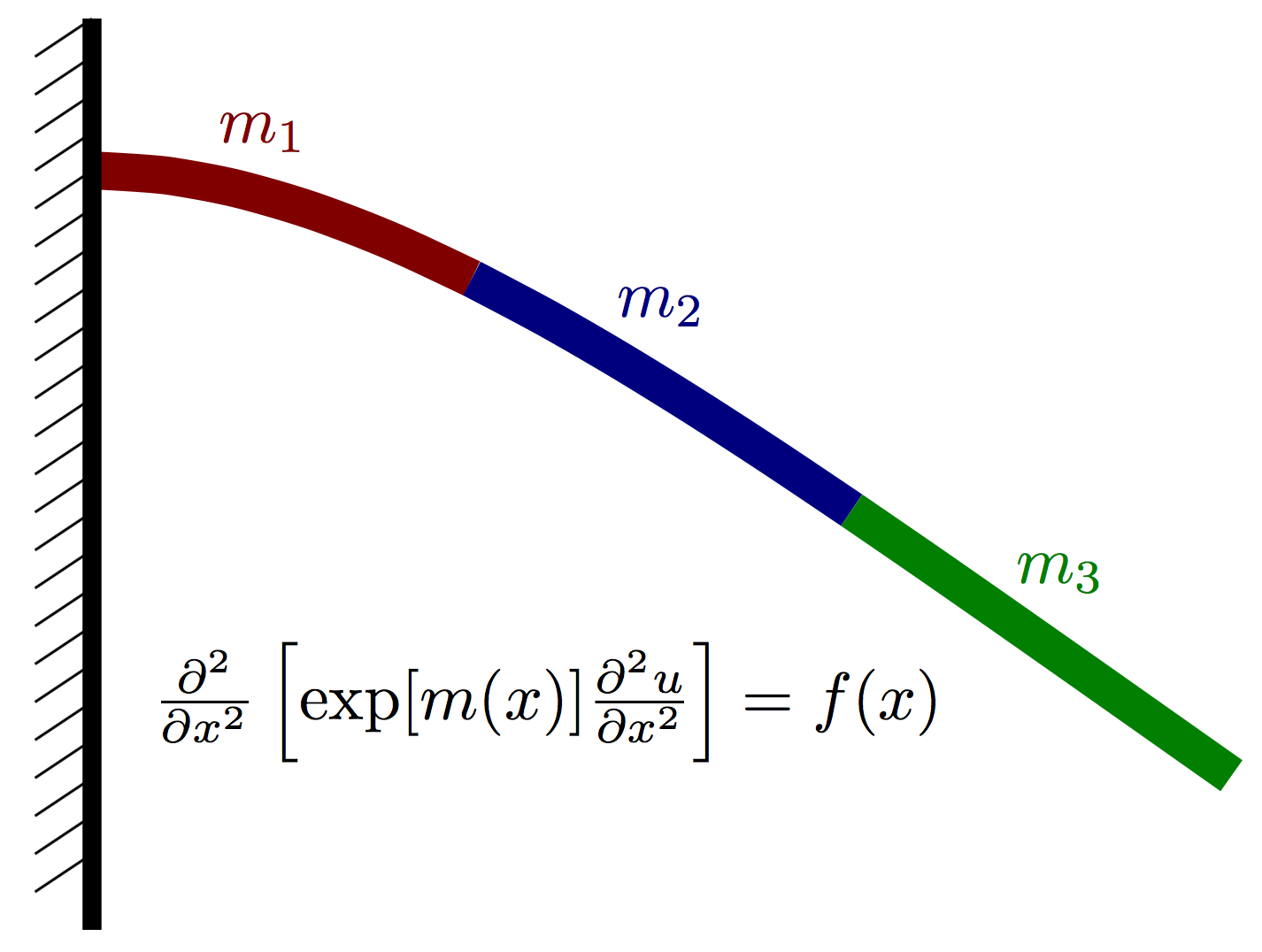}
      \caption{\it Schematic of the Euler-Bernoulli beam with the material properties $m_1, m_2$ and $m_3$.}
\label{fig:beam}
\end{figure}

This benchmark is a Bayesian inverse problem for characterizing the stiffness of an Euler-Bernoulli beam given observations of the beam displacement under a prescribed load. The Young’s modulus is piecewise constant over three regions as shown in Figure \ref{fig:beam}. Synthetic data is used. A realization of a Gaussian process is used to define the \textit{true} material properties. Additive noise is added to the output of the model to create the synthetic data.
The benchmark is then defined by a probability density function
$\pi:{\mathbb R}^3 \to {\mathbb R}$ where the three inputs correspond
to Young's modulus on the three regions.
The benchmark supports the `Evaluate', `Gradient', `ApplyJacobian',
and `ApplyHessian' operations. The latter three are evaluated using finite differences.

The output probability is defined as usual as the product of a
likelihood and a prior:
\begin{equation*}
    \pi(m \mid y) = L(y \mid m) \pi_\text{pr}(m).
\end{equation*}
In order to define the likelihood,
let $N_x$ denote the number of finite difference nodes used to discretize the Euler-Bernoulli PDE above.  For this problem, we will have observations of the solution $u(x)$ at $N_y$ of the finite difference nodes.  Let $\hat{u}\in\mathbb{R}^{N_x}$ denote a vector containing the finite difference solution and let $y\in\mathbb{R}^{N_y}$ denote the observable random variable, which is the solution $u$ at $N_y$ nodes plus some noise $\epsilon$, i.e.
$$y = B\hat{u} + \epsilon,$$
where $\epsilon \sim N(0, \sigma_y)$ and $B$ is the finite difference mass matrix.  The solution vector is given by
$$\hat{u} = [K(m)]^{-1}\hat{f},$$
where $K(m)$ is the stiffness matrix of the problem.
%\todo[inline]{We should state what the $B$ matrix is. Does it just
%select individual nodes from $u$?}

Combining this with the definition of $y$, we have the complete forward model
$$y = B[K(m)]^{-1} \hat{f} + \epsilon$$.
The likelihood function then takes the form
$$L(y | m) = N\left(\, B [K(m)]^{-1} f,\,\,\Sigma_y \,\right).$$

For the prior $\pi_\text{pr}(m)$, we assume each value is an independent normal random variable
\begin{equation*}
    m_i \sim N(\mu_i, \sigma_i^2).
\end{equation*}

\subsubsection{Tsunami source inference}
\label{sec:inference:tsunami}

The aim of this benchmark is to obtain the parameters describing the
initial displacements leading to the Tohoku tsunami from the data of
two available buoys located near the Japanese coast. It is based on
the model described in Section~\ref{sec:tsunami_model}, which also describes the different levels available.
The likelihood of a set of parameters given the simulation results is
computed using a weighted average of the maximal wave height and the time at which it is reached. The likelihood is given by a normal distribution with mean given by maximum waveheight and the time at which it is reached for the the two DART buoys 21418 and 21419 (This data can be obtained from NDBC). The covariance matrix shown in Table \ref{tbl:tsunami_covar} depends on the level, but not the probe point.
%\todo[inline]{We need to define ``level''.}

\begin{table}[t!]
\caption{\it Mean and variances for the Tsunami test case.}
\label{tbl:tsunami_covar}
\centering
\begin{tabular}{cccc}
\toprule
$\mu$ & $\Sigma$ ($l=0$) & $\Sigma$ ($l=1$) & $\Sigma$ ($l=2$) \\
\midrule
1.85232 & 0.15 & 0.1 & 0.1\\
0.6368  & 0.15 & 0.1 & 0.1\\
30.23   & 2.5 & 1.5 & 0.75\\
87.98  & 2.5  & 1.5 & 0.75 \\ 
\bottomrule
\end{tabular}
%\todo[inline]{I don't understand the table. It says ``covariance'',
%but that should be a matrix. Are these the diagonal elements? In that
%case, they are simply the ``variances''.}
\end{table}

The prior cuts off all parameters which would lead to an initial displacement which is too close to the domain boundary. Some parameters may lead to unstable models, e.g., a parameter which initialises the tsunami on dry land; in this case we treat the parameter as unphysical and assign an almost zero likelihood. %Due to constant parameter dimension across levels, we only need to choose a proposal density for the coarsest level. We choose Adaptive Metropolis \cite{AMMCMC, AMMCMC2} provided by MUQ. As initial prior we set $\mathcal{N}(0, 10 I)$.
The description of this benchmark in \cite{Seelinger2021} also contains details on a solution using parallel MLMCMC implemented in the MUQ library \cite{MUQ}.

%\todo[inline]{In the previous paragraph, I don't understand what
%``proposal density'' means. Presumably, we define the benchmark in
%terms of a probability density -- where does a ``proposal density''
%enter here? Perhaps the second half of the paragraph above is not
%about the benchmark definition, but about how the benchmark is solved
%-- but that is besides the point here, and should either be omitted or
%needs to be its own paragraph that makes clear what the text is about.}
% Thanks for catching that, indeed it is the solution method and does not belong here

%\begin{table}[h!]
%  \centering
%  \begin{tabular}{ccp{3cm}}
%  \toprule
%  Mapping & Dimensions & Description \\
%  \midrule
%  input  & [2] & x and y coordinates of a proposed tsunami origin. \\
%  output & [4] & Arrival time and maximum water height at two buoy points. \\
%  \bottomrule
%  \end{tabular}
%\caption{\it Properties of the Tsunami benchmark.}
%\todo[inline]{I inlined the information from this table into the text,
%  as I have for all other benchmarks (because I can do it in one
%  sentence, rather than needing 1/4 of a column for it). So I propose
%  to remove this table as well, though I am confused how the output
%  has dimension 4. Should it not simply be a probability? I presume
%  that the dimension 4 refers to the forward model, which is then
%  merged into a probability?}
%\label{tbl:tsunami_prop}
%\end{table}

The inputs of this benchmark are the $x$ and $y$ coordinates of a
proposed tsunami origin; the output is a probability given the 
arrival time and maximum water height at two buoy points of the model compared to measured data provided by NDBC.
The benchmark only supports evaluation of the forward map.

\subsubsection{Coefficient field inversion in a two-dimensional Poisson linear PDE}

This is an implementation of the Bayesian inverse problem presented
in \cite{kim2022hippylibmuq}. The inverse problem estimates a
spatially varying diffusion coefficient in an elliptic PDE given
limited noisy observations of the PDE solution -- the benchmark is
thus related to the one discussed in
Section~\ref{sec:inverse:membrane}, but uses a different
parameterization. The coefficient in the equation is discretized with finite elements and efficient derivative information is available through the adjoint techniques implemented in hIPPYlib and hIPPYlib-MUQ.

%\begin{table}[h!]
%  \centering
%  \begin{tabular}{ccp{3cm}}
%  \toprule
%  Mapping & Dimensions & Description \\
%  \midrule
%  input  & [$256^2$] & Signal $\mathbf{x} \in \mathbb{R}^{256\times 256}$ \\
%  output & [1] & Log PDF $\pi(\mathbf{b}\mid \mathbf{x})$ \\
%  \bottomrule
%  \end{tabular}
%\caption{\it Properties of the CT benchmark. - NEEDS UPDATE - Noemi and Umberto}
%\end{table}
%Supported features of this benchmark are: Evaluate and Gradient.

\begin{figure}[tb]
  \centering
    \hspace{-0.3in}
    \includegraphics[width=0.49\columnwidth]{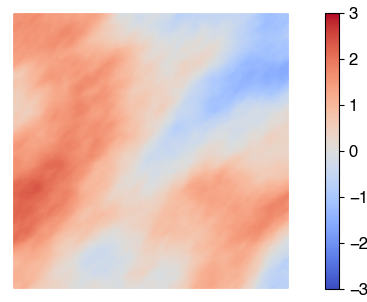}
    \includegraphics[width=0.49\columnwidth]{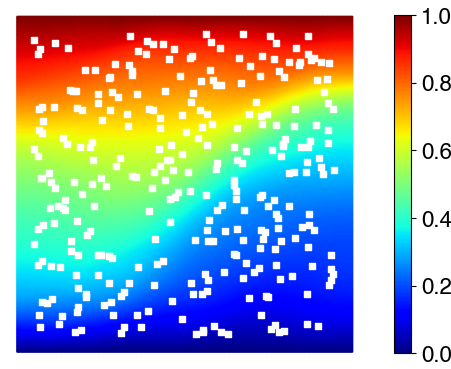}
  \caption{\it True parameter field (left) and the corresponding state field
  (right) for the Poisson problem. The locations of the observation points are
  marked as white squares in
  the right figure.}
  \label{fig:ex1_mtrue_observation}
\end{figure}

%\begin{figure}[tb]
%  \centering
%    \hspace{-0.3in}
%    \includegraphics[width=0.44\columnwidth]{Images/hippylib-figures/ex1_prior_mean.png}
%    \includegraphics[width=0.44\columnwidth]{Images/hippylib-figures/ex1_prior_sample2.png}\\
%    \includegraphics[width=0.44\columnwidth]{Images/hippylib-figures/ex1_prior_sample1.png}
%    \includegraphics[width=0.53\columnwidth]{Images/hippylib-figures/ex1_prior_sample0_colorbar.png}
%  \caption{Prior mean (leftmost) and three sample fields drawn from the prior
%  distribution for the Poisson problem.}
%  \label{fig:ex1_prior_samples}
%\end{figure}

%\begin{figure}[tb]
%  \centering
%    \hspace{-0.3in}
%    \includegraphics[width=0.44\columnwidth]{Images/hippylib-figures/ex1_la_mean.png}
%    \includegraphics[width=0.44\columnwidth]{Images/hippylib-figures/ex1_la_sample2.png}\\
%    \includegraphics[width=0.44\columnwidth]{Images/hippylib-figures/ex1_la_sample1.png}
%    \includegraphics[width=0.53\columnwidth]{Images/hippylib-figures/ex1_la_sample0.png}
%\caption{The MAP point (leftmost) and three sample fields drawn from the
%  Laplace approximation of the posterior distribution for the Poisson problem.}
%  \label{fig:ex1_la_samples}
%\end{figure}

Specifically, the benchmark is based on the following model:
$$
\begin{aligned}
-\nabla\cdot\left[\exp(m) \nabla u\right] &= 0  && \text{ in } \Omega, \\
u & = g && \text{ on } \partial\Omega_D, \\
\exp(m) \nabla u \cdot \hat{n} & = 0  && \text{ on } \partial\Omega_N.
\end{aligned}
$$
Here, $\Omega$ is the domain, $u(x)$ is the solution,
$g$ are prescribed Dirichlet conditions on the boundary
$\partial\Omega_D\subseteq \partial \Omega$, and the last equation
represents homogeneous Neumann conditions on the remainder of the
boundary $\partial \Omega_N
= \partial \Omega \setminus \partial \Omega_D$.   The field $m(x)$ is
discretized with linear finite elements on a uniform triangular mesh
of the unit square $\Omega = [0,1]^2$ with $32$ cells in each
coordinate direction, and is thus described by 1,089 values that form
the input to the problem.  The state is discretized with quadratic elements.

The prior distribution on the field $m(x)$ is a Gaussian process defined through a stochastic differential equation. %In Figure~\ref{fig:ex1_prior_samples} we show the prior mean and three sample fields drawn from the prior distribution for the Poisson problem. See Section 5.1.2 of \cite{kim2022hippylibmuq} for details. 
The likelihood is formed from observations of $u(x)$ at 300 points $x^{i}$ drawn uniformly over $[0.05,0.95]^2$, as shown in Figure~\ref{fig:ex1_mtrue_observation}. These observations are obtained by solving the forward problem on the finest mesh with the true parameter field shown on the left in Figure~\ref{fig:ex1_mtrue_observation} and then adding a random Gaussian noise with a relative standard deviation of $0.5\%$ to the resulting state.
Model, prior, likelihood, and sampling of the posterior, as well as convergence diagnostics are implemented with the hIPPYlib-MUQ\footnote{\url{https://hippylib.github.io/muq-hippylib/}} package, which enables efficient calculation of gradients, Jacobian actions, and Hessian actions using adjoint and tangent linear techniques.

Based on these definitions, the benchmarks then returns
the Laplace approximation of the posterior 
probability that maps the inputs $m \in{\mathbb R}^{1089}$ to $\mathbb
R$. The benchmark supports the `Evaluate', `Gradient', `ApplyJacobian', and
`ApplyHessian' operations.

\subsubsection{Coefficient field inversion in a Robin boundary condition for a three-dimensional $p$-Poisson nonlinear PDE}

\begin{figure}[tb]
  \centering
    \hspace{-0.3in}
    \includegraphics[width=0.49\columnwidth]{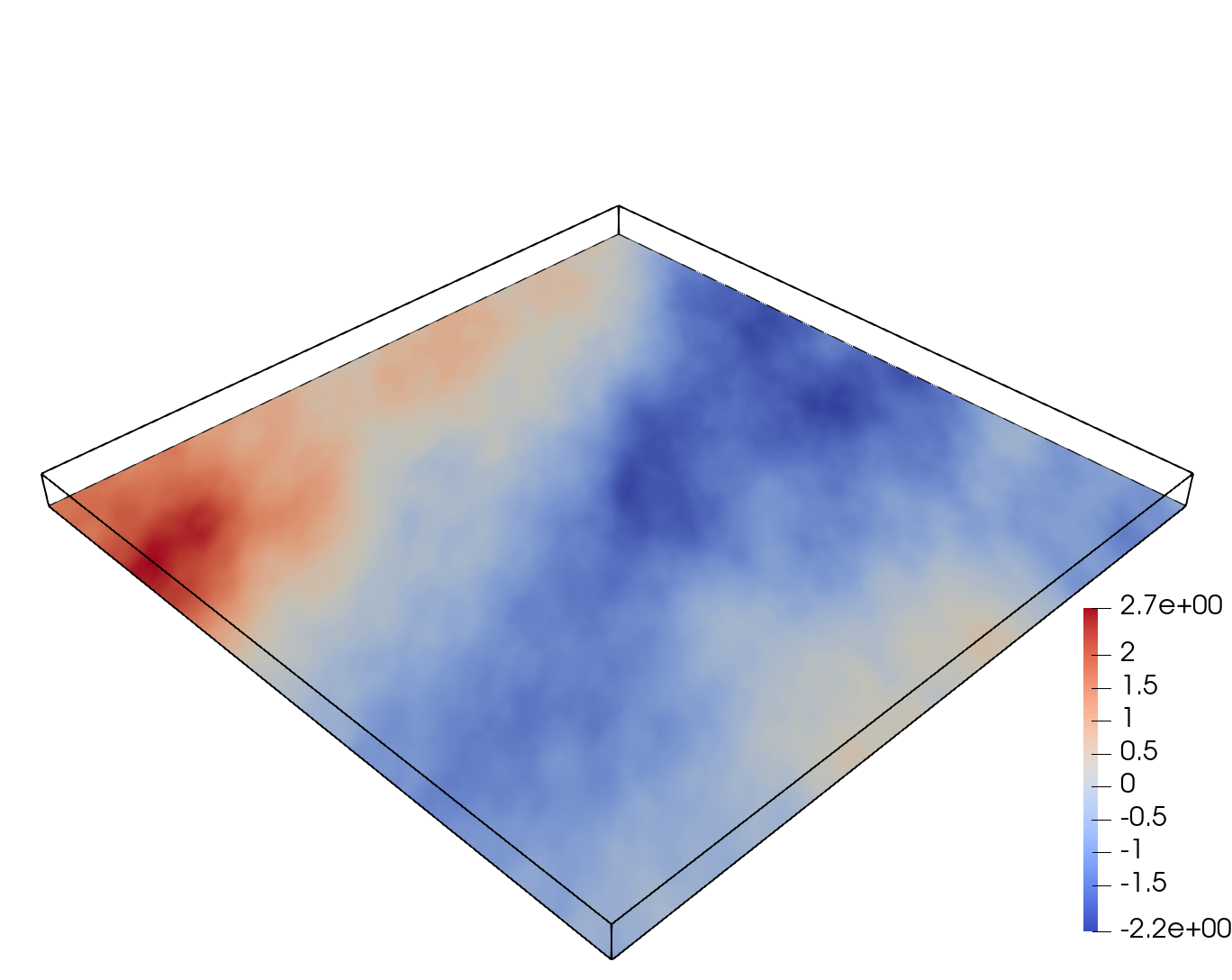}
    \includegraphics[width=0.49\columnwidth]{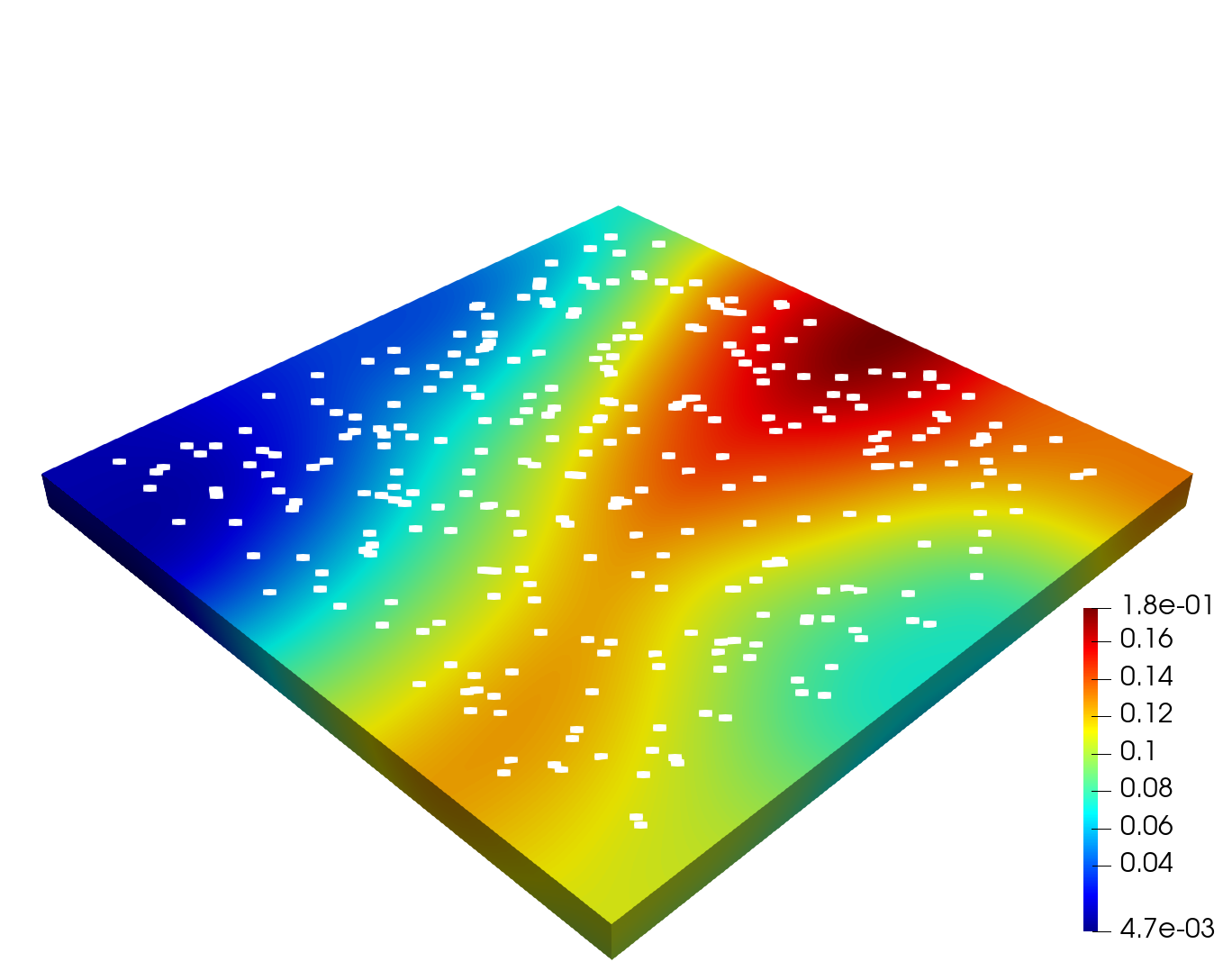}
  \caption{\it Left: True parameter field on the bottom surface.
    Right: Corresponding state field and
    $l=300$ observation points (white square marks) on the top surface. 
    %Right:
    %The MAP point on the bottom surface.
    }
    \label{fig:ex2_mtrue_observation}
\end{figure}

This benchmark focuses on a three-dimensional nonlinear Bayesian
inverse problem presented in~\cite{kim2022hippylibmuq}. The inverse
problem estimates a two-dimensional flux boundary condition on the
bottom of a thin three-dimensional domain with nonlinear $p$-Poisson PDE.  Observations of the PDE solution at the top of the domain are used, as shown in Figure~\ref{fig:ex2_mtrue_observation}.

%\begin{table}[h!]
%  \centering
%  \begin{tabular}{ccp{3cm}}
%  \toprule
%  Mapping & Dimensions & Description \\
%  \midrule
%  input  & [$256^2$] & Signal $\mathbf{x} \in \mathbb{R}^{256\times 256}$ \\
%  output & [1] & Log PDF $\pi(\mathbf{b}\mid \mathbf{x})$ \\
%  \bottomrule
%  \end{tabular}
%\caption{\it Properties of the CT benchmark. - NEEDS UPDATE - Noemi and Umberto. {\color{red}JSJ: This table is duplicated from the CT section, to remove or to modify here to match the contents of the section?}}
%\end{table}
%Supported features of this benchmark are: Evaluate and Gradient.

This benchmark then defines a Bayesian posterior density over a spatially-variable boundary condition $m(x)$ given noisy observations of the solution of the nonlinear partial differential equation
\begin{equation*}
   \begin{aligned}
-\nabla\cdot\left[|\nabla u|_\epsilon^{p-2} \nabla u\right] &= f  && \text{ in } \Omega, \\
u & = g && \text{ on } \partial\Omega_D, \\
|\nabla u|_\epsilon^{p-2} \nabla u \cdot \hat{n} & = m  && \text{ on } \partial\Omega_N,
\end{aligned} 
\end{equation*}
where $\Omega$ is the domain, $u(x)$ is the solution and $g$ are prescribed Dirichlet conditions on the boundary $\Omega_D\subseteq \partial \Omega$.
In this benchmark we set $p=3$ and $f=0$.  The domain, illustrated in the figure above, is $\Omega = [0,1]^2\times [0,0.05]$.

The output of the benchmark is a probability that maps a discretized
version of $m(x)$ at 233,389 nodes to the product of a prior and a likelihood.
The prior distribution on the boundary condition $m(x)$ is a Gaussian
process defined through a stochastic differential equation; see \cite[Section 5.1.2]{kim2022hippylibmuq} for details.
The likelihood is formed from observations of $u(x)$ at 300 points $x^{i}$ drawn uniformly over the top surface of the domain.  Gaussian noise with a standard deviation of $0.005$ is assumed.
Like the benchmark of the previous section,
model, prior, likelihood, and sampling of the posterior, as well as convergence diagnostics are implemented with the hIPPYlib-MUQ package.

This benchmark supports the `Evaluate', `Gradient', `ApplyJacobian', and
`ApplyHessian' operations.
%Briefly discuss results and refer to the TOMS paper.

\subsection{Propagation benchmarks}\label{sec:propagation}
In this section we describe the propagation benchmarks contained in the benchmark library, these 
implement a specific forward \gls{UQ} problem.

\subsubsection{Euler-Bernoulli beam}

%\todo[inline]{This section could be simplified by referring to the
%     inverse benchmark. But if we switch B2 and B3, the reference is
%     backward, so I will not touch this section for the moment.}
This is an uncertainty propagation problem modeling the effect of uncertain material parameters on the displacement of an Euler-Bernoulli beam with a prescribed load. Refer to \Cref{sec:inverse:ebbeam} for details. %The Young's modulus is piecewise constant over three regions as shown below, and the three constants are assumed to each be uncertain with a uniform distribution.

\begin{table}[t!]
\caption{\it Properties of the Euler-Bernoulli beam propagation benchmark.}
 \label{tbl:eb_prprop}
  \centering
  \begin{tabular}{ccp{3cm}}
  \toprule
  Mapping & Dimensions & Description \\
  \midrule
  input  & [3] & The value of the beam stiffness in each lumped region.\\
  output & [31] & The resulting beam displacement at 31 equidistant grid nodes. \\
  \bottomrule
  \end{tabular}
\end{table}

The inputs and outputs of this model for this propagation benchmark can be found in
Table \ref{tbl:eb_prprop}. Supported features of this benchmark are
the  `Evaluate', `Gradient', `ApplyJacobian', and `ApplyHessian' operations. The latter three are evaluated using finite differences.

\begin{figure}
  \centering
    \hspace{-0.3in}
    \includegraphics[width=.5\linewidth]{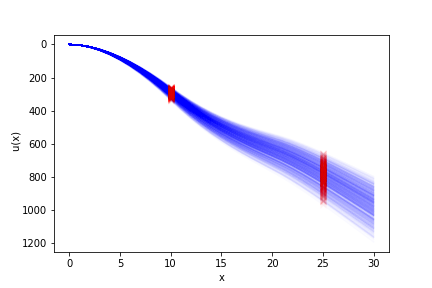}
  \caption{\it Samples of the Euler-Bernoulli beam model output
    $F(M)$ with the quantity of interest (solution at nodes $10$ and $25$) marked with crosses.}
  \label{fig:eb_propagation}
\end{figure}

For the prior, we assume the material parameter to be a uniformly distributed random variable
$$ M \sim U_{[1, 1.05]^3}. $$
The goal is to identify the distribution of the resulting quantity of interest, the deflection of the beam at finite difference node $10$ and $25$ of the discretisation. Samples from the model output $\hat u$ are illustrated in \cref{fig:eb_propagation}.
%\todo[inline]{What are the $Q$, $F$, $M$ symbols? Match with the
%inverse benchmark section.} 

\subsubsection{Genz integration and surrogate benchmark}

This benchmark implements a suite of six families of test functions
first proposed in~\cite{Genz_proc_1984} to elucidate the performance
of quadrature algorithms for computing $\mathbb{E}[f(\theta)]$ which
can be computed analytically if $\theta$ is an uncertain parameter
with specific distribution, and $f$ is a specific forward model. It is also useful for benchmarking methods for constructing surrogate models, for example see~\cite{Jakeman_ES_JCP_2015}. The six function families are parameterized by a scalable number $n\ge 1$ of independent and identically distributed uniform variables with non-zero joint density defined on the unit hypercube $[0,1]^n$.

Specifically, the six families of functions are as follows:
{\footnotesize
    \begin{align*} \mathrm{Oscillatory} &&
     f(\theta) &= \cos\left(2\pi w_1 + \sum_{i=1}^n c_i\theta_i\right) \\
    \mathrm{Product Peak} &&
     f(\theta) &= \prod_{i=1}^n \left(c_i^{-2}+(\theta_i-w_i)^2\right)^{-1}\\
    \mathrm{Corner Peak} &&
     f(\theta)&=\left( 1+\sum_{i=1}^n c_i\theta_i\right)^{-(n+1)} \\
    \mathrm{Gaussian Peak} &&
     f(\theta) &= \exp\left( -\sum_{i=1}^n c_i^2(\theta_i-w_i)^2\right) \\
    C_0-\mathrm{Continuous} &&
     f(\theta) &= \exp\left( -\sum_{i=1}^n c_i\lvert \theta_i-w_i\rvert\right) \\
    \mathrm{Discontinuous} &&
     f(\theta) &= \begin{cases}0, & \theta_1>w_1 \;\mathrm{or}\;\\& \theta_2>w_2\\\exp\left(\sum_{i=1}^n c_i\theta_i\right), & \mathrm{otherwise}\end{cases}
     \end{align*}
}

The relative magnitude of the coefficients $c_i>0$ impacts the
difficulty of integrating or approximating these family of
functions. This benchmark implements various types of decay ofthe
coefficients as a function of the index $i$. This introduces
anisotropy into the functions (i.e., it varies each variable's importance) that can be
exploited by adaptive methods. The types of coefficient decay are:

{\footnotesize
     \begin{align*}
         \text{No decay}&&
     \hat{c}_i&=\frac{i+0.5}{n}\\
    \text{Quadratic decay}&&
     \hat{c}_i &= \frac{1}{(i + 1)^2}\\
    \text{Quartic decay}&&
     \hat{c}_i &= \frac{1}{(i + 1)^4}\\
    \text{Exponential decay}&&
     \hat{c}_i&=\exp\left(\log(c_\mathrm{min})\frac{i+1}{n}\right)\\
    \text{Squared-exponential decay}&&
    \hat{c}_i&=10^{\left(\log_{10}(c_\mathrm{min})\frac{(i+1)^2}{n^2}\right)}
    \end{align*}}
    
Here $c_\mathrm{min}=5\times10^{-6}$. With the decay formulas above, the final benchmark coefficients are obtained by normalizing such that
\[c_i = C\frac{\hat{c}_i}{\sum_{i=1}^n \hat{c}_i},\]
where $C>0$ is set by the user and defaults to $1.0$ if not
provided.
Generally, increasing the size of coefficients $\hat c_i$ will make integration or constructing surrogate models of these functions more difficult. 
Finally, while the $0\le w_i \le 1$ parameters were varied in~\cite{Genz_proc_1984}, they do not affect the difficulty
of the integration problem and so we set $w_1=w_2=\ldots=W_n=W$ to a user specified constant $W\in\mathbb{R}$. 

With these definitions, each of the six families of functions maps an uncertain parameter $\theta\in
{\mathbb R}^D$ to $\mathbb R$.
This benchmark only supports evaluation of the forward map.

\begin{table}[h!]
    \centering
    \begin{tabular}{cccp{2.5cm}}
    \toprule
    Config & Type & Default & Description \\
    \midrule
    $n$ & integer & 1 & Function dimension \\
    $C$ & float & 1 & Normalizing constant \\
    $W$ & float & 0.5 & Shift parameter \\
    $T$ & string & \footnotesize{sqexp} & coefficient decay type \\
    $N$ & string & \footnotesize{oscillatory} & integrand family \\
    \bottomrule
    \end{tabular}
\caption{\it Configuration variables for the Genz benchmark.}
\label{tbl:config_genz}
\end{table}

\subsubsection{L2-Sea propagation}
\label{sec:propagation:l2sea}

% \todo[inline]{LT: this is the same benchmark discussed at length in appendix A1, why restate the same thing? I suggest we only leave a link to that. 
% We could also add a reference value (mean and st dev), either here or in Appendix A1}

This benchmark has been presented and discussed in \cref{sec:application_sparse_grids}. 
Here we only specify that the optional configuration parameters
presented when discussing the model in Section~\ref{sec:l2sea_model} are set in the benchmark as follows: 
\textit{sinkoff}=\texttt{'y'}, \textit{trimoff}=\texttt{'y'}.
As a reference value, we also report an approximation of the expected value of the resistance:
more specifically, following the script reported in \Cref{sec:application_sparse_grids}
(note in particular that such example is run with \textit{fidelity}=3) and setting the sparse grid to $w=15$, 
% and the so-called level-to-knots functions requested by \gls{sgmk}
% to \texttt{lev2knots\_lin} for $\Frou$ and to \texttt{lev2knots\_2step} for $\Dr$ (see \cite{sparse_grids_matlab_kit} 
% for details; these settings generate the sparse grid shown at the bottom row of \Cref{fig:results_Lorenzo}, which has 256 points), 
we obtain the sparse grid with 256 points shown at the bottom row of \Cref{fig:results_Lorenzo}, 
and $\mathbb{E}[R] \approx 51.664730931098383$ N.

We also provide a reference value for a second setting, namely assuming 
that $\Frou$ and $\Dr$ are uniformly distributed in their intervals,
rather than having respectively a triangular and a beta random distribution;
we also lower the \textit{fidelity} config parameter to \textit{fidelity}=6.
We then adjust the following lines of the script reported in \Cref{sec:application_sparse_grids}
\begin{lstlisting}[]
fid = 6;
knots_Fr = @(n) knots_leja(n,Fr_a,Fr_b,'sym_line');
lev2knots_Fr = @lev2knots_2step;
knots_T = @(n) knots_leja(n,T_a,T_b,'sym_line');
lev2knots_T = @lev2knots_2step;
w = 9;
S = create_sparse_grid(N,w,{knots_Fr,knots_T},{@lev2knots_Fr,@lev2knots_T});
\end{lstlisting}
and obtain a grid with 181 points that yields $\mathbb{E}[R] \approx 68.398655639653271$ N.

\subsubsection{The cookies problem}
\label{sec:propagation:cookies}

% \todo[inline]{LT: for consistency, give result on st dev here too?}

This benchmark runs a forward \gls{UQ} problem for the cookies model described in \cref{sec:models:cookies}.
The results have been obtained using the Sparse Grids Matlab Kit \cite{sparse_grids_matlab_kit} and the Matlab interface to UM-Bridge. 
More specifically, we assume that the uncertain parameters $\theta_n$ appearing in the definition of the diffusion coefficient \cref{eq:cookies-diffusion-coeff}
are uniform i.i.d. random variables on the range $[-0.99, -0.2]$ and we aim at computing the expected value of the quantity of interest 
$\Psi$ in \cref{eq:cookies-qoi}. The structure of this benchmark therefore identical to the test discussed in \cite{back.nobile.eal:comparison}; 
however, raw numbers are different since in that work the mesh and PDE solver employed were different (standard FEM with piecewise linear basis on a triangular mesh);
in particular, for this benchmark, we fix the splines degree at $p=4$ and the fidelity config parameter to $2$.

As a reference value, we provide the approximation of the expected value $\mathbb{E}[\Psi]$ 
computed with a standard Smolyak sparse grid, based on Clenshaw--Curtis points for level $w=5$; see e.g. \cite{sparse_grids_matlab_kit}
for more details on the sparse grid construction, as well as the benchmark script available on the Um-Bridge project website. 
The resulting sparse grid has 15713 points, and the corresponding approximation of the expected value is $\mathbb{E}[\Psi] \approx 0.064196096847169$.

\subsection{Optimisation benchmarks}\label{sec:optimisation}
UM-Bridge's interface is sufficiently generic to allow more than just \gls{UQ} problems to be solved. As a proof of concept we provide an example optimsation benchmark in this section. 

\subsubsection{L2-Sea optimization}\label{sec:optimisation:l2sea}
% \todo[inline]{Reference Sections~\ref{sec:l2sea_model}
% and \ref{sec:propagation:l2sea} to omit some details. Where the
% benchmark was produced is something that should have been said in one
% of the earlier uses of the model/benchmark.}
% \todo[inline]{This section is far too long. It should describe the
% benchmark briefly; we can leave all of the details to the referenced
% publications. It is not necessary to explain what the optimum actually
% is. It should be possible to describe the benchmark in one page max.}

This optimisation benchmark employs the L2-sea model described in \Cref{sec:l2sea_model}, and was developed within the activities of the NATO-AVT-331 Research Task Group on ``Goal-driven, multifidelity approaches for military vehicle system-level design'' \cite{beran2020}. The goal is to minimize the resistance $R$ of the DTMB 5415 in calm water at fixed speed (i.e. fixed Froude number, namely $\Frou=0.28$), and nominal draft $\Dr=-6.16$, by suitably modifying the shape of the hull of the vessel. 
In other words, the first two components of the parameters vector $\theta$ are fixed, and we optimize on the $N=14$ shape parameters, $\textbf{\thetashape}$, only. The optimization problem thus reads
%
% \begin{eqnarray*}\label{eq:5415prob}
%     \begin{array}{rll}
%         \mathrm{minimize}      & \Delta R (\mathbf{\thetashape}) = \frac{R (\mathbf{\thetashape})}{R_0}-1 \\ % \qquad \mathrm{with} \qquad \thetashape\in\mathbb{R}^N\\
%         \mathrm{subject \,  to}& L_{\rm pp}(\mathbf{\thetashape}) = L_{\rm pp_0}\\
%         \mathrm{and \, to}     & \nabla(\mathbf{\thetashape}) = \nabla_0 \\
%         & |\Delta B(\mathbf{\thetashape})| \leq 0.05B_0 \\
%         & |\Delta T(\mathbf{\thetashape})| \leq 0.05T_0 \\
%         & V(\mathbf{\thetashape})\geq V_0\\
%         & -1\leq {\thetashape}_{i} \leq 1 \quad \mathrm{with} \quad \forall i=1,\dots, N\\
%     \end{array}
% \end{eqnarray*}
\begin{eqnarray*}\label{eq:5415prob}
    \begin{array}{rll}
        \mathrm{minimize}      & f(\mathbf{\thetashape}) \\%\Delta R (\mathbf{\thetashape}) = \frac{R (\mathbf{\thetashape})}{R_0}-1 \\ % \qquad \mathrm{with} \qquad \thetashape\in\mathbb{R}^N\\
        \mathrm{subject \,  to} & h_i(\mathbf{\thetashape}) = 0 & \quad i=1,2\\%L_{\rm pp}(\mathbf{\thetashape}) = L_{\rm pp_0}\\
                                %& \nabla(\mathbf{\thetashape}) = \nabla_0 \\
                                & g_k(\mathbf{\thetashape}) \leq 0 & \quad k=1,\ldots,4\\
                                % & h_2(\mathbf{\thetashape}) \leq 0 \\ 
                                % & h_3(\mathbf{\thetashape}) \leq 0 \\ 
                                % & h_4(\mathbf{\thetashape}) \leq 0 \\ 
                                & -1\leq {\thetashape}_{i} \leq 1 & \quad i=1,\dots, N
    \end{array}
\end{eqnarray*}
where the objective function $f$ is the total resistance $R$, the equality constraints $h_i$ (automatically satisfied by the shape modification method) include ($h_1$) fixed length between perpendiculars of the vessel and ($h_2$) fixed vessel displacement, whereas the inequality constraints $g_1,\ldots,g_4$ encode geometrical constrains on the modified hull in relation to the original one.
In particular, $g_1$ relates to the beam, $g_2$ to the draft (both allowing $\pm5\%$ of maximum variation), and $g_3, g_4$ relate to the minimum cylindrical
dimensions to contain the sonar in the bow dome, see \cite{grigoropoulos2017mission} for details:
these are the four additional outputs of the L2-sea model that were hinted at the end of \Cref{sec:l2sea_model}.

A reference optimum has been found in \cite{serani:l2sea} with a multi-fidelity method (i.e., the config parameter \textit{fidelity} is not fixed a-priori but chosen adaptively by the optimizer), where an objective improvement of 12.5\% is achieved with an $\mathrm{NCC}=628$. Specifically, the minimum of the resistance is at 
$\mathbf{\thetashape}=(-1.000000, -0.906250, -0.781250, \allowbreak -0.242188, 0.533203, -0.906250, \; 0.128418, \; 0.812500,\allowbreak -0.070313,-0.878296, \allowbreak 0.422363, 0.031250, 0.365234,\allowbreak -0.175659)$ and the corresponding value computed at \textit{fidelity}=1 is 36.3116 N. 
Note that NCC is the normalized computational cost and it is evaluated as the ratio between the CPU time needed for the generic $j$-th fidelity and the highest-fidelity. To allow comparison between methods, the standard NCC to be used as the cost of function evaluation for each fidelity level is \{1.00, 0.46, 0.20, 0.11, 0.07, 0.04, 0.03\}, from highest to lowest.
These results have been obtained by fixing the config parameters \textit{sinkoff} and \textit{trimoff} to \texttt{'y'}.

% \todo[inline]{LT: Figures  \ref{fig_sea:meshes} and \ref{fig_sea:GridStudy} have appeared in a previous work
% by Serani Pellegrini Diez (authors here too) and three more
% persons not authors here (the second one is a slightly different version). Do we need to do something special about attribution?
% Do we just leave it? Do we drop them?}

% \begin{table}[h!]
% \caption{\it Properties of the L2-Sea optimization benchmark.}
% \label{tbl:properties_l2sea-opt}
%     \centering
%     \begin{tabular}{ccp{3cm}}
%     \toprule
%     Mapping & Dimensions & Description \\
%     \midrule
%     input  & [16] & The first input is the Froude number (fixed to 0.28); the second is the draft (fixed to -6.16); the other 14 are the $x$ design variables for the shape modification with $-1\leq x_i \leq 1$  (with parent hull has $x_i = 0$) for $i=1,\dots,14$\\
%     output & [5]  & The first output is the model scale total resistance in Newton, whereas the other four are the optimization geometrical constraints (negative to be satisfied). \\
%     \bottomrule
%     \end{tabular}
% \end{table}

%\printbibliography
%\end{refsection}
%\end{comment}
\end{document}